\documentclass[aps,prl,showpacs,floatfix,amsmath,amssymb,amsfonts,twocolumn,superscriptaddress]{revtex4-1}
\usepackage{graphicx}
\usepackage{color}
\usepackage{ulem}
\usepackage{bm}
\usepackage[colorlinks=true,allcolors=blue]{hyperref}
\usepackage{latexsym}
\usepackage{dsfont}
\usepackage{epstopdf}
\usepackage{CJK}
\usepackage{float}
\usepackage{titlesec}
\newcommand{\mgcro}{MgCr$_2$O$_4$}
\newcommand{\be}{\begin{equation}}
\newcommand{\ee}{\end{equation}}
\newcommand{\bea}{\begin{eqnarray}}
\newcommand{\eea}{\end{eqnarray}}

\newcommand{\red}[1]{{\color{red} #1}}

\def\cha{\cos(2\pi \!\cdot\! h/4)}
\def\chb{\cos(2\pi \!\cdot\! h/2)}
\def\chc{\cos(2\pi \!\cdot\! 3h/4)}

\def\cka{\cos(2\pi \!\cdot\! k/4)}
\def\ckb{\cos(2\pi \!\cdot\! k/2)}
\def\ckc{\cos(2\pi \!\cdot\! 3k/4)}

\def\cla{\cos(2\pi \!\cdot\! \ell/4)}
\def\clb{\cos(2\pi \!\cdot\! \ell/2)}
\def\clc{\cos(2\pi \!\cdot\! 3\ell/4)}

\newcommand{\of}[1]{\left( #1 \right)}

\begin{document}
\title{Magnetic excitations of the classical spin liquid \mgcro}   
\author{X. Bai}
    \affiliation{School of Physics, Georgia Institute of Technology, Atlanta, GA 30332, USA}  
\author{J.~A.~M. Paddison} 
    \affiliation{School of Physics, Georgia Institute of Technology, Atlanta, GA 30332, USA}
    \affiliation{Churchill College, University  of  Cambridge, Storey's Way, Cambridge CB3 0DS, UK} 
\author{E. Kapit} 
    \affiliation{Rudolf Peierls Centre for Theoretical Physics, University of Oxford, Parks Road, Oxford OX1 3NP, UK}
     \affiliation{Department of Physics, Colorado School of Mines, Golden, CO, 80401, USA}   
\author{S.~M.~Koohpayeh}
    \affiliation{Institute for Quantum Matter and Department of Physics and Astronomy, The Johns Hopkins University, Baltimore, MD 21218, USA}
\author{J.-J. Wen}
    \altaffiliation{Present Address: Stanford Institute for Materials and Energy Sciences, SLAC National Accelerator Laboratory, Menlo Park, CA 94025, USA.}
    \affiliation{Institute for Quantum Matter and Department of Physics and Astronomy, The Johns Hopkins University, Baltimore, MD 21218, USA}
 \author{S.~E.~Dutton}
    \altaffiliation{Present Address: Cavendish Laboratory, Department of Physics, University of Cambridge, JJ Thomson Ave., Cambridge CB3 0HE, UK} 
    \affiliation{Department of Chemistry, Princeton University, Princeton, New Jersey 08544, USA}
\author{A.~T.~Savici}
    \affiliation{Neutron Scattering Division, Oak Ridge National Laboratory, Oak Ridge, TN 37831, USA}
\author{A.~I.~Kolesnikov}
    \affiliation{Neutron Scattering Division, Oak Ridge National Laboratory, Oak Ridge, TN 37831, USA}
\author{G.~E.~Granroth}
    \affiliation{Neutron Scattering Division, Oak Ridge National Laboratory, Oak Ridge, TN 37831, USA}     
\author{C.~L.~Broholm}
    \affiliation{Institute for Quantum Matter and Department of Physics and Astronomy, The Johns Hopkins University, Baltimore, MD 21218, USA}
\author{J.~T.~Chalker}
    \affiliation{Rudolf Peierls Centre for Theoretical Physics, University of Oxford, Parks Road, Oxford OX1 3NP, UK}
\author{M.~Mourigal}
    \email{mourigal@gatech.edu}
    \affiliation{School of Physics, Georgia Institute of Technology, Atlanta, GA 30332, USA}
    \affiliation{Institute for Quantum Matter and Department of Physics and Astronomy, The Johns Hopkins University, Baltimore, MD 21218, USA}
\date{October 28, 2018}
\begin{abstract}
We report a comprehensive inelastic neutron-scattering study of the frustrated pyrochlore antiferromagnet MgCr$_2$O$_4$ in its cooperative paramagnetic regime. Theoretical modeling yields a microscopic Heisenberg model with exchange interactions up to third-nearest neighbors, which quantitatively explains all the details of the dynamic magnetic response. Our work demonstrates that the magnetic excitations in paramagnetic MgCr$_2$O$_4$ are faithfully represented in the entire Brillouin zone by a theory of magnons propagating in a highly-correlated paramagnetic background. Our results also suggest that MgCr$_2$O$_4$ is proximate to a spiral spin-liquid phase distinct from the Coulomb phase, which has implications for the magneto-structural phase transition in  MgCr$_2$O$_4$.
\end{abstract}
\maketitle

The classical pyrochlore Heisenberg antiferromagnet is a canonical model of frustrated magnetism. With only nearest-neighbor (NN) exchange interactions, it does not exhibit magnetic ordering down to zero temperature and instead hosts a liquid-like state of strongly correlated spins. In real space, this cooperative paramagnet is a system of underconstrained spins on a network of corner-sharing tetrahedra. The energy is minimized if the vector sum of spins is zero on every tetrahedron, giving rise to an extensive ground-state degeneracy. Mapping spin variables to flux variables on the bonds of the dual diamond lattice transforms this spin constraint to a divergence-free condition on the flux fields. Consequently, spin correlations decay algebraically in real space, and sharp features---known as pinch points---are present in reciprocal space. This exotic magnetic state of matter is termed a Coulomb phase~\citep{Reimers_1992,Moessner_1998,henley2010coulomb}. 

The best candidate materials to realize the Coulomb phase include the spin ices~\cite{Bramwell_2001,Fennell_2009,Morris_2009} and the cubic $AB_2$O$_4$ spinels and Na$A^\prime B_2$F$_7$ fluorides~\cite{Krizan_2014,Ross_2016,plumb2017continuum}, in which a transition-metal ion $B$ occupies a pyrochlore lattice. Canonical spinel examples are CdCr$_2$O$_4$ \citep{Chung}, ZnCr$_2$O$_4$ \citep{Lee2002}, and \mgcro~\citep{Suzuki2007,Tomiyasu2008a}, which are all highly-frustrated antiferromagnets that ultimately order magnetically at temperatures $T_{\mathrm{N}}$ much smaller than the scale of exchange interactions. Contrary to expectations, neutron-scattering experiments on these materials do not reveal sharp pinch points; instead, only broad ring-like diffuse scattering patterns are observed. These experimental observations have been explained in terms of decoupled hexagonal spin clusters---loops of six spins with alternating directions \cite{Lee2002}. While phenomenological model has been remarkably successful in explaining magnetic scattering features \cite{Lee2002,Suzuki2007,Tomiyasu2008a,Tomiyasu2013a}. It leaves three key questions unaddressed. First, what is the microscopic origin of cluster-like scattering in terms of the underlying magnetic interactions? Second, how does frustration relate to the complex ordered structures that $B$-site spinels often exhibit below $T_{\mathrm{N}}$? And, third, what is the origin of the broad magnetic excitation spectrum observed in the cooperative paramagnetic state? This final question is of particular importance because three explanations have been proposed: (i) scattering is broad in energy, because excitations have a short lifetime; (ii) scattering is broad because the excitations are fractionalized; (iii) scattering is broad in momentum, because the excitations are riding on a disordered background.

\begin{figure*}[th!]
\includegraphics[width=1.9\columnwidth]{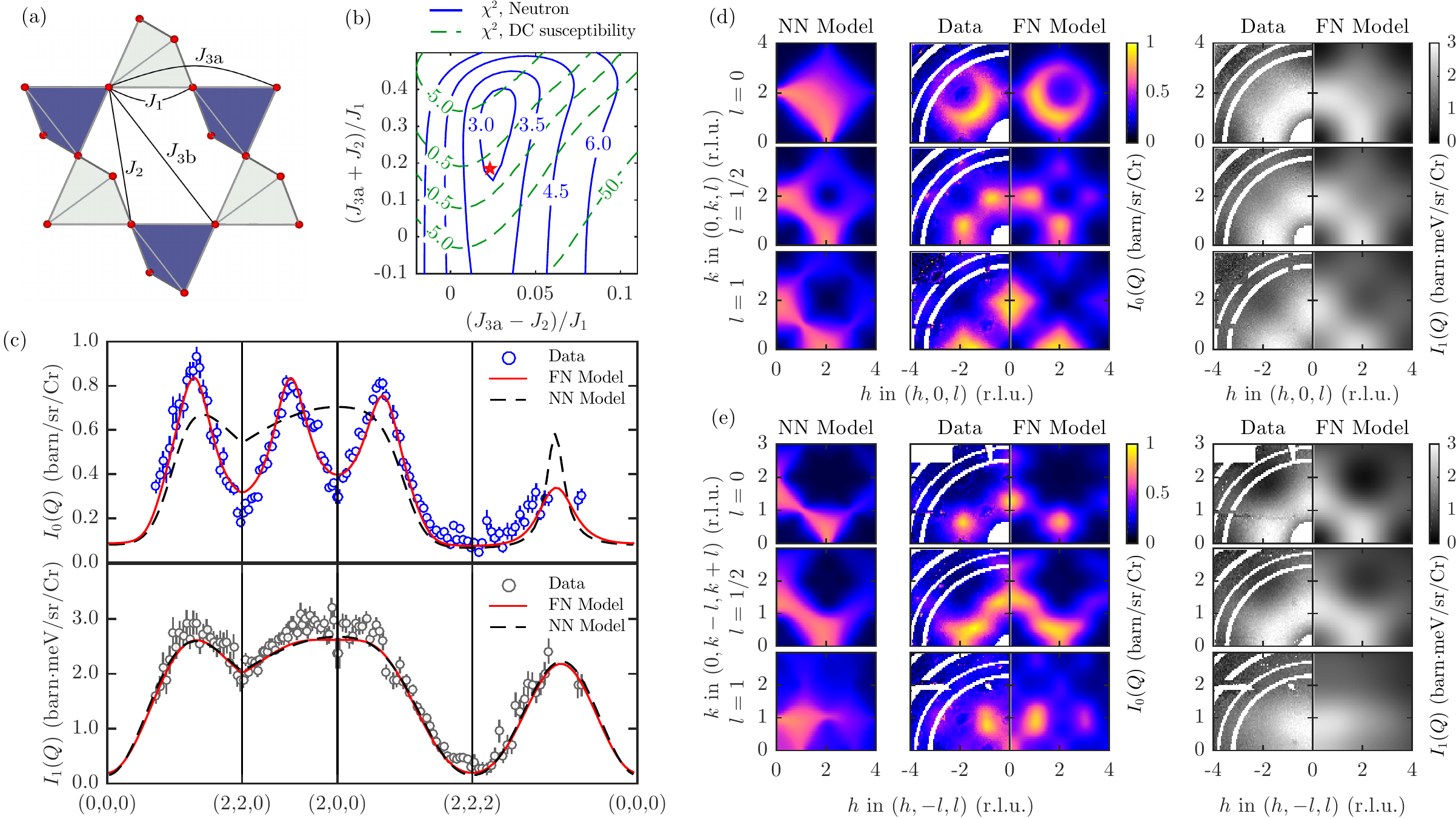}
\caption{(a) The pyrochlore lattice of Cr$^{3+}$ ions (red spheres) in \mgcro\ and definition of exchange interactions up to third neighbors. Note that $J_{3\text{a}}$ and $J_{3\text{b}}$ span the same distance but are not equivalent by symmetry. (b) Contour plot of the goodness of fit $\chi^2$ between calculations and neutron (blue solid lines) and bulk susceptibility (green dashed lines) measurements. FN exchange interactions $J_2$ and $J_{3\text{a}}$ are fixed on a grid with $J_1$ and $J_{3\text{b}}$ fitted at each grid point. The choice of $J_2 \pm J_{3\text{a}}$ as plotting axes highlights the nearly equivalent spin structure factors obtained for $J_2=J_{3\text{a}}$. Spin correlations are calculated using the self-consistent Gaussian approximation (SCGA) at $T=20$~K. The red star is the best overall fit. (c) Momentum dependence of $I_0({\bf Q})=F(|{\bf Q}|)\mathcal{S}({\bf Q})$ and $I_1({\bf Q})=F(|{\bf Q}|)\mathcal{K}({\bf Q})$ along several paths of the Brillouin zone (BZ) at $T=20$~K, and comparisons with SCGA predictions for NN (dashed black line) and FN (solid red line) models. For the NN model, $J_1\!=\!38$~K. (d--e) Selected slices across $I_0({\bf Q})$ and $I_1({\bf Q})$ for fixed momentum transfer along the $(0,0,l)$ and $(0,-l,l)$ directions, respectively, and comparison between NN and FN models calculated using the SCGA. Throughout, white rings are masked aluminum background lines. In (c)--(e), only $E_i\!=\!80$~meV data are shown, but both $40$~meV and $80$~meV data were included in fits. \vspace{-0.5cm}}
\label{fig1}
\end{figure*} 

In this Letter, we use a combination of neutron spectroscopy and modeling to determine the spin Hamiltonian of MgCr$_2$O$_4$ and the nature of its magnetic excitations in the correlated paramagnetic regime at temperature $T\!=\!20$\,K. We study this material because it is a paradigmatic example of a frustrated antiferromagnetic spinel that shows cluster-like scattering above $T_\mathrm{N}$ and exotic magnetic order below $T_\mathrm{N}$. Our results significantly advance previous studies by measuring and explaining the entire four-dimensional (4D) magnetic response of MgCr$_2$O$_4$ as a function of energy and momentum. We use quantitative modeling to determine a set of exchange interactions that best reproduce our experimental data. Remarkably, we find that linear spin-wave theory accurately captures all the details of the correlated paramagnetic response in MgCr$_2$O$_4$, revealing the harmonic nature of excitations in this classical spin liquid. Furthermore, we find that our model remains highly frustrated despite the presence of further-neighbor (FN) interactions. We explain this result by showing that MgCr$_2$O$_4$ is proximate to a highly-degenerate spiral-spin-liquid phase distinct from the Coulomb phase. Our results suggest competition between nearly-degenerate states drives the complex low-temperature states often observed in frustrated $B$-site spinels.

\begin{figure*}[th!]
\includegraphics[width=1.9\columnwidth]{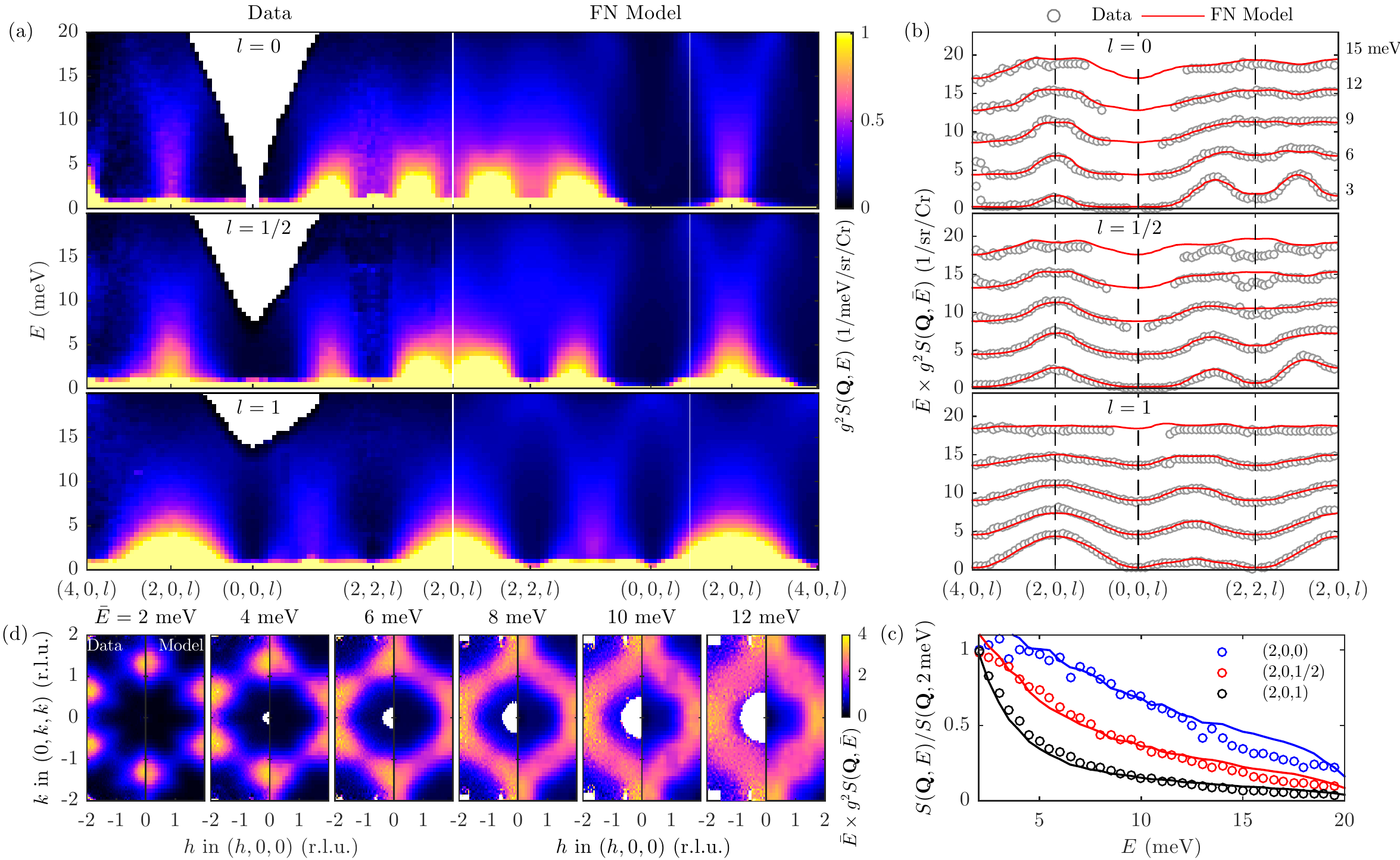}
\caption{Magnetic excitation spectra of \mgcro\ at $T\!=\!20$~K measured with incident neutron energy $E_i\!=\!40$~meV, and comparison with linear spin-wave theory (LSWT) calculations for our FN model. (a) Momentum-energy slices through $g^2 S ({\bf Q},E)$ along different paths, comparing data (left column) and FN model (right column). (b) Cuts at constant energies $\bar{E}\pm0.2$~meV through the data (gray circles) and FN model (red lines), where $\bar{E}$ is labeled on each plot. The intensity is multiplied by $\bar{E}$ and offset by $4 $/sr/Cr for clarity.  (c) Energy dependence of the experimental (colored circles) and modeled (colored lines) dynamical structure factor at selected momenta, normalized to the energy transfer $E_0=2$~meV. (d) Slices at constant energies $\bar{E}\pm0.2$~meV through the data (left column) and the FN model (right column) in the $(h,k,k)$ plane. Throughout, blank space is due to kinematic constraints on the scattering, and the extra intensity at $(4,0,0)$ arises from a strong nuclear Bragg peak and its associated acoustic phonon.\vspace{-0.5cm}}
\label{fig2}
\end{figure*} 

The crystal structure of \mgcro\ at $T=20$~K is cubic (space group $Fd\bar{3}m$, $a\!=\!8.33$~\AA\/). Magnetic Cr$^{3+}$ ions interact magnetically with their nearest neighbors (NN) primarily via direct exchange ($d_{\rm Cr-Cr}\!=\!2.95$~\AA) and with further neighbors (FN) via superexchange [Fig.~\ref{fig1}(a)]. Thermo-magnetic measurements~\citep{blasse1963neel,rudolf2007spin,Dutton2011a,Koohpayeh2013a} reveal net antiferromagnetic interactions with a Weiss constant ranging from $\theta_{\text{W}}\!=\!-346$K \citep{rudolf2007spin,blasse1963neel} to $-433$K \citep{Dutton2011a,Koohpayeh2013a}, and are compatible with spin-only magnetic moments for Cr$^{3+}$ ($S\!=\!3/2$ and $g\!\approx\!2.05$)~\cite{Dutton2011a}. Below $\sim 40$~K ($\approx 0.1 \theta_{\text{W}}$), the magnetic susceptibility markedly departs from the Curie-Weiss law which contrasts with predictions for the NN model~\cite{Moessner_1999}. Futhermore a cooperative paramagnetic regime appears with cluster-like scattering~\cite{Suzuki2007,Tomiyasu2008a,Tomiyasu2013a,gao2018manifolds}. This regime persists down to $T_{\text{N}}\!\approx\!13$~K \citep{klemme2000heat,Dutton2011a,Koohpayeh2013a}, where the onset of long-range magnetic ordering \citep{blasse1963neel,Dutton2011a,Koohpayeh2013a} is accompanied by a structural distortion to tetragonal or lower symmetry \citep{Ehrenberg2002,Ortega2008,Kemei2013a} due to spin-lattice coupling \citep{Xiang_2011,Oleg2002,Lee2002,Nilsen2015}. Magnetic Bragg peaks observed below $T_{\rm N}$ are indexed by two inequivalent propagation vectors, $\boldsymbol{k}_{L,1}\!=\!\left(\frac{1}{2},\frac{1}{2},0\right)$ and $\boldsymbol{k}_{L2}\!=\!\left(1,0,\frac{1}{2}\right)$~\citep{Shaked1970} with respect to the cubic cell; the magnetic structure of this so-called ``$L$ phase" is not fully solved~\citep{Shaked1970,gao2018manifolds}. Moreover, an additional partially-ordered magnetic phase (``$H$ phase") with a single propagation vector $\boldsymbol{k}_{H}\!=\!(0,0,1)$ is observed for some samples between $T_{\rm N}$ and $T_H\!\approx\!16$~K~\citep{Shaked1970,Suzuki2007}.

To understand the nature of the magnetic excitations in \mgcro\, we performed neutron-scattering experiments that expose its magnetic excitation spectrum as a function of neutron momentum transfer ${\hbar\bf Q}\!=\!\hbar\bm{k}_i\!-\!\hbar\bm{k}_f$ and energy transfer $E\!=\!E_i\!-\!E_f$ to the sample. Large single crystals of \mgcro\ were grown using the floating-zone technique following systematic sample-quality studies~\cite{Dutton2011a,Koohpayeh2013a}. Our 10 best crystals were co-aligned on an aluminum holder for a total sample mass $m\!\approx\!13.5$~g and overall mosaic $\leq3^\circ$ \red{[see Sec.~S1]}. Inelastic neutron-scattering data were collected on the SEQUOIA instrument \citep{Granroth2010a,Stone2014a} at the Spallation Neutron Source, Oak Ridge National Laboratory (USA). Incoming neutron energies of $E_i\!=\!40$ and $80$ meV were used, yielding elastic energy resolutions of 0.8(4) and 1.6(8) meV, respectively. The sample mount was cooled to $T\!=\!20$~K using a closed-cycle refrigerator and rotated about a vertical axis in steps of $1^\circ$ over a range $>\!90^\circ$. The data were converted to absolute units in {\scshape Mantid} \citep{arnold2014mantid} using measurements of a vanadium standard, analyzed in {\scshape Horace} \citep{ewings2016horace} where background contributions and Bragg peaks from the sample were masked, and symmetrised in the $m\bar{3}m$ Laue class \red{[see Sec.~S2]}. The normalized magnetic intensity can be written $I({\bf Q},{E})\!=\! (\frac{1}{2}\gamma r_0)^2[gf(|{\bf Q}|)]^2\,\mathcal{S} ({\bf Q},E)$, where $(\frac{1}{2}\gamma r_0)^2\!=\!0.07265\!\times\!10^{-24}$~cm$^2$ \cite{Xu2013a}, $f(|{\bf Q}|)$ is the magnetic form factor, and $\mathcal{S} ({\bf Q},E)$ is the magnetic scattering function. We obtained energy-integrated quantities $I_{\alpha}({\bf Q})\!\equiv\!\int_0^{E^\prime}\!\!\text{d}E\,E^{\alpha}\,(1+e^{- E/{k_\mathrm{B}T}}) \mathcal{S}({\bf Q},E)$, where $\alpha\in\{0,1\}$, and $E^\prime\!=\!20$~meV is chosen to encompass the magnetic excitation bandwidth. The quantities $I_{0}({\bf Q})$ and $I_{1}({\bf Q})$ are proportional to the instantaneous magnetic structure factor $\mathcal{S}({\bf Q})$ and the first moment $\mathcal{K}({\bf Q})$, respectively, with the constant of proportionality $F(|{\bf Q}|)=(\frac{1}{2}\gamma r_0)^2[gf(|{\bf Q}|)]^2$.

To model the magnetism of \mgcro, we use the Heisenberg model $\mathcal{H}\!=\!\frac{1}{2}\sum_{ij}J_{ij}\,{\bf S}_{i}\cdot{\bf S}_j$, where ${\bf S}_{i}$ represents the spin at one of the $N$ sites ${\bf R}_i$ of the pyrochlore lattice, and the four interactions $J_{ij}\in\{J_1,J_2,J_{3\text{a}},J_{3b}\}$ extend to third-nearest neighbors [Fig.~\ref{fig1}(a)]. We will show that it is crucial to model the two inequivalent third-neighbor pathways $J_{3\text{a}}$ and $J_{3b}$ separately. Our choice of a Heisenberg model is motivated by the small orbital contribution to the magnetic moment ($g\approx 2.05$) and a preliminary reverse Monte Carlo analysis \cite{McGreevy_1988,Paddison_2012,Paddison_2017,Paddison_2018} that revealed an isotropic distribution of spin orientations \red{[see Sec.~S3]}.  For a Heisenberg paramagnet, the structure factor is the Fourier transform of instantaneous two-spin correlators, $\mathcal{S}({\bf Q})\!=\!{\frac{2}{3N}}\sum_{ij}\!\!\left\langle {\bf S}_{i}\!\cdot\!{\bf S}_{j}\right\rangle\cos\left({\bf Q}\!\cdot\!{\bf r}_{ij}\right)$, where ${\bf r}_{ij} = {\bf R}_{i}\!-\!{\bf R}_{j}$ is the vector between the spin pair. The first moment contains correlators weighted by the corresponding interactions~\citep{Hohenberg1974,Stone_2001}, $\mathcal{K}({\bf Q})\!=\!-{\frac{1}{3N}}\sum_{ij}\!J_{ij}\left\langle {\bf S}_{i}\!\cdot\!{\bf S}_{j}\right\rangle[1-\cos({\bf Q}\!\cdot\!{\bf r}_{ij})]$. As $J_{3\text{a}}$ and $J_{3\text{b}}$ are symmetry inequivalent but associated with the same lattice harmonics, it is impossible to determine their values by a simple ratio between Fourier coefficients of the structure factor and the first moment. Therefore, we employ the self-consistent Gaussian approximation (SCGA) \citep{Conlon2010a} to calculate $\mathcal{S}({\bf Q})$ and $\mathcal{K}({\bf Q})$ from the magnetic interaction matrix; this method is in excellent quantitative agreement with classical Monte Carlo simulations \red{[see Sec.~S5]}. 

Determining the magnetic interactions of \mgcro\ is a challenging problem, because the spin correlations of the model are essentially degenerate along the line $J_2\!=\!J_{3\text{a}}$ in interaction space \citep{chern2008partial}. Consequently, we used three complementary approaches. First, we performed a global fit to $\mathcal{S}({\bf Q})$ and $\mathcal{K}({\bf Q})$ for a grid of values of $J_2$ and $J_{3\text{a}}$, with $J_1$ and $J_{3\text{b}}$ fitted at each grid point. The corresponding goodness-of-fit $\chi^2$, shown  in Fig.~\ref{fig1}(b), reveals a shallow valley of possible minima \red{[see Sec.~S4]}. Second, we calculated the goodness-of-fit to the temperature dependence of bulk magnetic susceptibility data between 20~K and 400~K for all the parameter sets $\{J_{1},J_{2},J_{3\text{a}},J_{3\text{b}}\}$ extracted from the $\mathcal{S}({\bf Q})+\mathcal{K}({\bf Q})$ fits. The intersection of $\chi^2$ minima for these two results yields $J_1\!=\!38.05(3)$~K, $J_2/J_1\!=\!0.0815$, $J_{3\text{a}}/J_1\!=\!0.1050$ and $J_{3\text{b}}/J_1\!=\!0.0085(1)$ [red star in Fig.~\ref{fig1}(b)]. Finally, we validated these parameters using fits to the energy-resolved $S(\mathbf{Q},\omega)$, as discussed below [Fig.~\ref{fig2}]. 

In Fig.~\ref{fig1}, we compare the experimental $I_\alpha({\bf Q})$ with SCGA calculations for our optimized exchange parameters. The FN interactions are small, with a maximum of $J_{3\text{a}}\!\approx\!0.1J_1$, and are all antiferromagnetic, in contrast to first-principles estimates~\citep{yaresko2008electronic}. Crucially, however, these interactions quantitatively explain the cluster-like scattering \cite{Lee2002,Tomiyasu2008a,Tomiyasu2013a,gao2018manifolds} [Fig.~\ref{fig1}(d-e)]. Compared to the NN model, our model correctly captures the suppressed intensity at the $(2,0,0)$ and $(2,2,0)$ pinch-point positions [Fig.~\ref{fig1}(c)], indicating the destruction of the Coulomb phase by FN interactions~\cite{Conlon2010a}. In real space, the spin correlators as a function of distance show an alternation in sign, which explains the apparent success of the decoupled hexagonal spin-cluster model \red{[see Sec.~S5]}. However, our FN Heisenberg model enables three key advances. First, it allows a complete microscopic description of the spin dynamics; second, it allows the frustration to be understood in terms of degeneracies of the model; and, third, it enables the nature of the low-temperature ordered phases to be predicted in absence of magnetoelastic effects. We discuss these results in turn below.

\begin{figure}[t!]
\includegraphics[width=1\columnwidth]{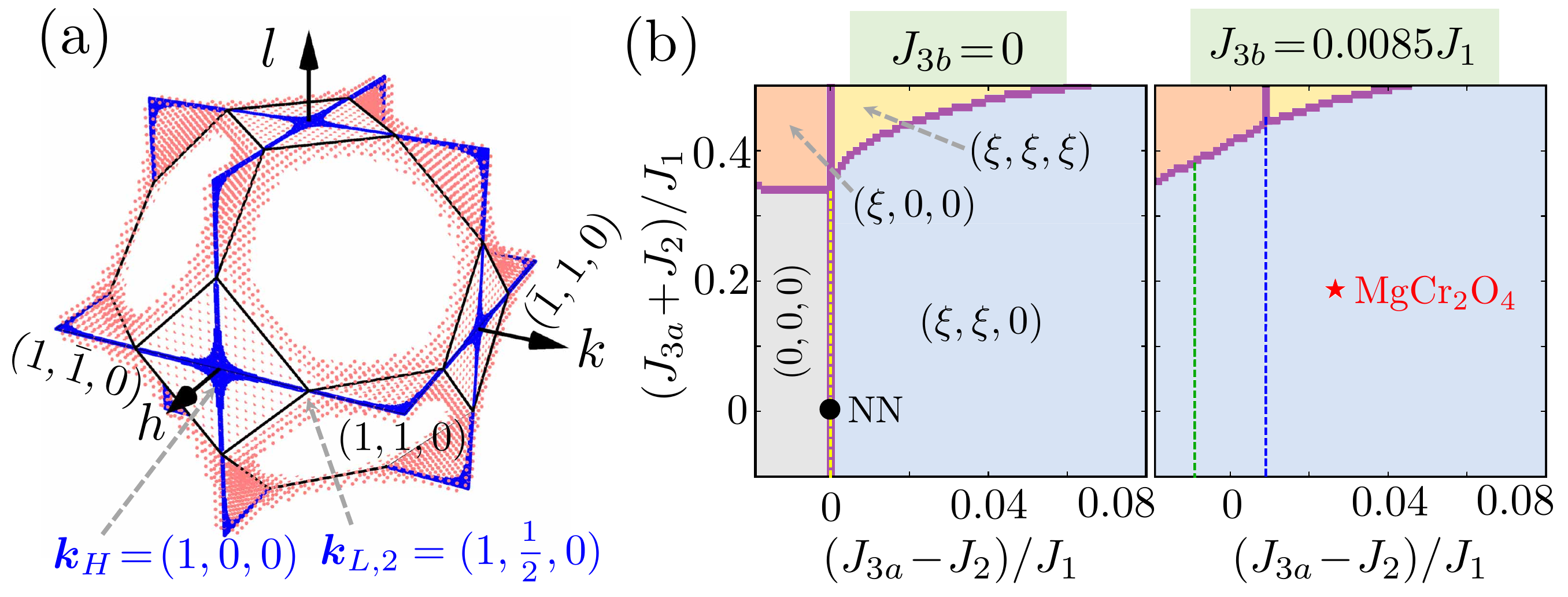}
\caption{(a) Sets of ordering wavevectors $\boldsymbol{\kappa}$ with nearly-degenerate energies represented in the Brillouin zone as colored surfaces. The pink surface shows wavevectors with energies within $\sim$$0.5\%$ of the global energy minimum for the FN interaction parameters of \mgcro. The blue lines show wavevectors with energy equal to the global energy minimum for the macroscopically-degenerate phases represented by a blue dashed line in the phase diagram. (b) Mean-field phase diagrams of our FN Heisenberg model as a function of $J_2$ and $J_{3\text{a}}$, showing results for $J_{3\text{b}}=0$ (left) and $J_{3\text{b}}=0.0085J_1$ (right). Phases with different ordering wavevectors $\boldsymbol{\kappa}$ are shown in different colors. Special phases (dashed lines) correspond to a macroscopic number of ordering wavevectors with degenerate energies. They emerge from $J_{3\text{a}}-J_2=0$ (yellow line), which corresponds to the NN model, to form two half-planes corresponding to $J_{3\text{a}}- J_{2}=\mp J_{3\text{b}}$ (green and blue lines).}
\label{fig3}
\end{figure} 

Magnetic excitation spectra of our sample are presented in Fig.~\ref{fig2}. The excitations are gapless with a bandwidth of $\approx\!20$~meV ($\sim\!4J_1S$), although the dominant contribution to the spectral weight resides below $\approx\!5$~meV ($\sim\!J_1S$) [Fig.~\ref{fig2}(a)]. Close to the suppressed pinch point at $(2,0,0)$, excitations are relatively sharp and dispersive along the $(\xi,0,0)$ direction [Fig.~\ref{fig2}(b)], a feature also observed in NaCaNi$_2$F$_7$~\cite{plumb2017continuum}. Along other directions, such as $(2,\xi,0)$ and $(\xi,\xi,0)$, excitations form a broad continuum [Fig.~\ref{fig2}(a)] whose energy dependence is Lorentzian with a {\bf Q}-dependent relaxation rate [Fig.~\ref{fig2}(c)]. A simple factorization of the dynamic response as $\mathcal{S}({\bf Q},E)\!=\!\mathcal{S}({\bf Q})f(E)$, which implies spatially incoherent excitations, is not possible for \mgcro\ [Fig.~\ref{fig2}(d)], in contrast to theoretical predictions for the lowest branch of excitations in the NN model~\citep{Conlon2009a}. 

To examine the nature of excitations, we calculated $\mathcal{S}({\bf Q},E)$ in the paramagnetic regime using linear spin-wave theory (LSWT) in a framework previously used to model metallic spin-glasses~\cite{Walker_1977,Walker_1980}. For a given set of interactions, we use Monte Carlo simulations to generate ensembles of spin configurations at low but finite temperature to avoid ordering, calculate harmonic fluctuations of each spin configuration, and average $\mathcal{S}({\bf Q},E)$ over these samples \red{[see Sec.~S6]}. We compared LSWT calculations---performed for several sets of interactions near the shallow $\chi^2$ minimum of Fig.~\ref{fig1}(b)---with the entire 4D momentum-energy dependence of our experimental data \red{[see Sec.~S7]}. The best match is obtained for our previously-determined FN model, with LSWT calculations in striking agreement with the experimental observations [Fig.~\ref{fig2}]. 

Our microscopic model also explains the persistence of a classical spin-liquid in \mgcro\, despite FN interactions. In classical spin liquids, the lowest-energy eigenvalues of the interaction matrix are degenerate throughout large regions of the Brillouin zone, which suppresses magnetic ordering. We find that for the FN parameters of \mgcro\/, ordering wavevectors $\boldsymbol{\kappa}$ with energies within $0.5$\% of the global energy minimum describe a large surface near the zone boundary [Fig.~\ref{fig3}(a)]. This result is surprising because FN interactions are generically expected to lift the degeneracy of the NN model. To explain it, we calculated the phase diagram of ordered states as a function of $J_{2}$, $J_{3\text{a}}$, and $J_{3b}$ [Fig.~\ref{fig3}(b)]. Crucially, we uncover planes in interaction space along which the degeneracy of possible ordered states is exact and macroscopic. Our FN parameters place \mgcro\ in proximity to such a phase, for which wavevectors of the type $\boldsymbol{\kappa}\!=\!(1,h,0)$ are degenerate [blue lines in Fig.~\ref{fig3}(a)]. The corresponding states are a degenerate set of coplanar spirals \red{[see Sec.~S8]}, analogous to the ``spiral spin liquid" states previously known only for the $J_{1}$-$J_{2}$ model on the diamond lattice \cite{Bergman2007}. This result explains the similarity of cluster-like scattering in \mgcro\ to neutron-scattering data for diamond-lattice systems such as MnSc$_2$S$_4$~\cite{Gao2016}. 

Our analysis sets a benchmark for the comprehensive determination of magnetic interactions in materials where the traditional approach of spin-wave dispersion modeling is not available---either because the system does not order at an accessible temperature, or because the nature of this ordering is controlled by a magnetic Hamiltonian that is distinct from that of the paramagnetic phase due to magnetoelastic effects. The latter is the case in frustrated spinels such as \mgcro\ and ZnCr$_2$O$_4$. Furthermore the presence of several symmetry-unrelated ordering wavevectors makes magnetic structure solution very challenging. However, our results present a key insight: the degeneracy of our spiral spin liquid state encompasses two of the ordering wave-vectors, $\boldsymbol{\kappa}_H$ and $\boldsymbol{\kappa}_{L,2}$, that are observed experimentally below $T_{\rm N}$ in \mgcro\ and ZnCr$_2$O$_4$. This result suggests that the complex magnetic orderings observed in these frustrated spinels is a consequence of the near-degeneracy of competing ordered states shown in Fig.~\ref{fig3}(a). While the exact ground state is likely selected by magneto-structural effects beyond the Heisenberg model, we anticipate that our paramagnetic Hamiltonian will provide a valuable starting-point to develop a microscopic theory of magnetic ordering in these complex materials.

It is remarkable that, within the resolution of our experiment, the spin dynamics of \mgcro\ at $T=20$~K can be entirely described by spins precessing around their local mean field, with no evidence of quantum effects~\cite{plumb2017continuum}. Crucially, this excludes fractionalization and short lifetime as the physical origin for the broad momentum-energy response; rather, it indicates that scattering is broad in wave-vector because excitations propagate in a spatially disordered background.

{{\it Note added at the time of submission:} During the completion of this manuscript a paper making similar observations for NaCaNi$_2$F$_7$ appeared on the arXiv~\cite{Zhang_2018}. In conjunction these papers on pyrochlore antiferromagnets with different ranges of interactions and spin quantum numbers indicate the robustness of our theoretical results.
\begin{acknowledgments}
We thank Oleg Tchernyshov for many useful discussions during the earlier stages of this project. The work at Georgia Tech and at the Johns Hopkins-Princeton Institute for Quantum Matter was supported by the U.S. Department of Energy, Office of Basic Energy Sciences, Materials Sciences and Engineering Division under awards number DE-SC-0018660 and DE-FG02-08ER46544, respectively. The work at Oxford University was supported by the EPSRC under grant EP/I032487/1.  The research at Oak Ridge National Laboratory's Spallation Neutron Source was sponsored by the U.S. Department of Energy, Office of Basic Energy Sciences, Scientific User Facilities Division. J.A.M.P. acknowledges financial support from Churchill College, Cambridge (UK). M.M. and J.T.C. acknowledge the Kavli Institute for Theoretical Physics (KITP) where part of this research was started. KITP is supported by the National Science Foundation under Grant No. NSF-PHY-1748958.
\end{acknowledgments}


%

\clearpage
\onecolumngrid
\setcounter{figure}{0}
\setcounter{table}{0}
\setcounter{equation}{0}
\renewcommand{\thefigure}{S\arabic{figure}}
\renewcommand{\thetable}{S\arabic{table}}
\renewcommand{\theequation}{S\arabic{equation}}
\renewcommand{\thesection}{S\arabic{section}}   
\titleformat*{\section}{\bf}

\begin{center}
{\large \bf Supplementary Information}\\[10px]
{\large  \bf Magnetic excitations of the classical spin liquid \mgcro}   

\end{center}

\section{S1. Single-crystal sample}

\begin{figure}[h]
\includegraphics[width=0.6\columnwidth]{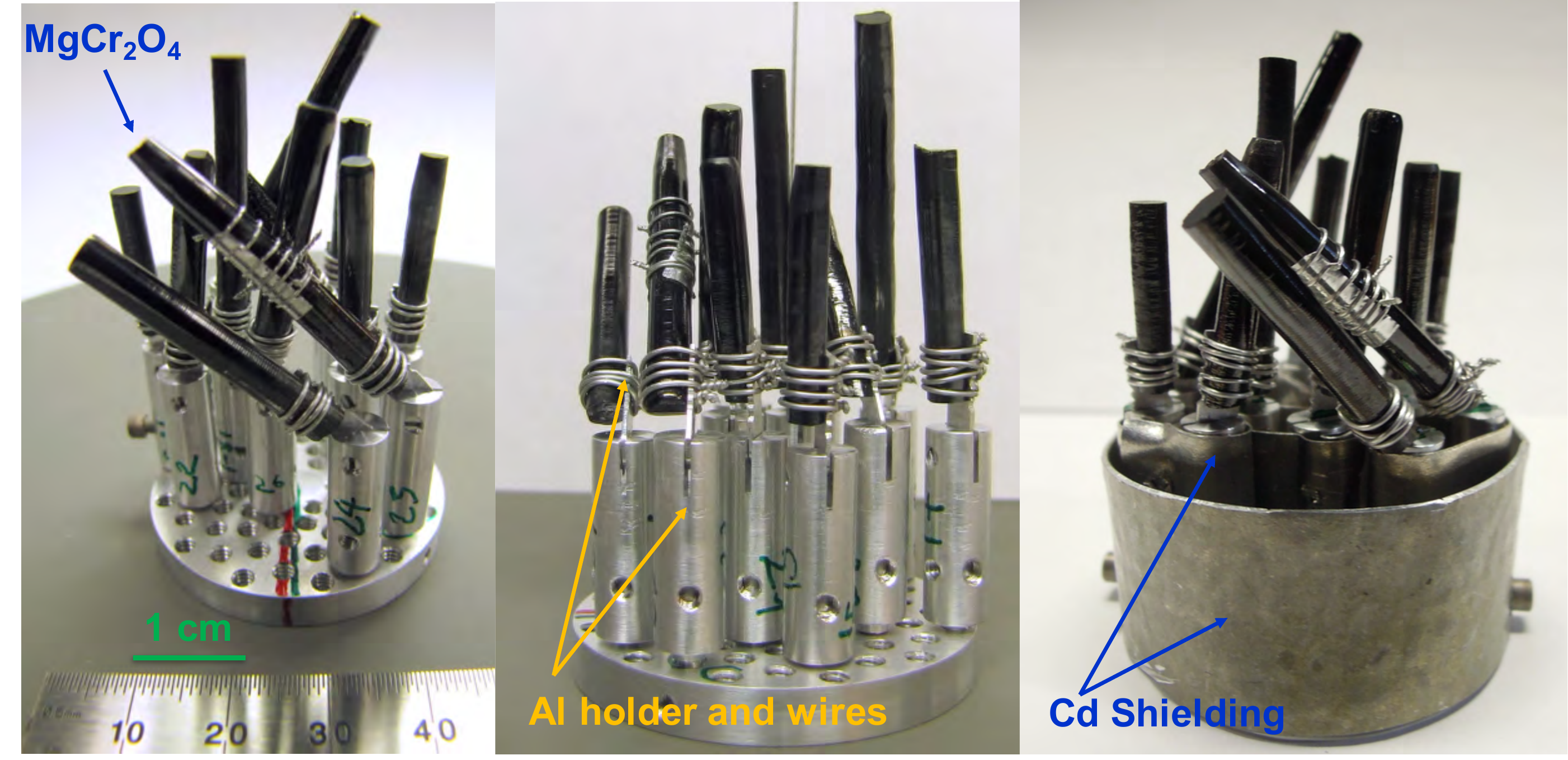}
\caption{Picture of the single-crystal mount used in our experiments}
\end{figure}

\begin{figure}[h]
\includegraphics[width=0.89\columnwidth]{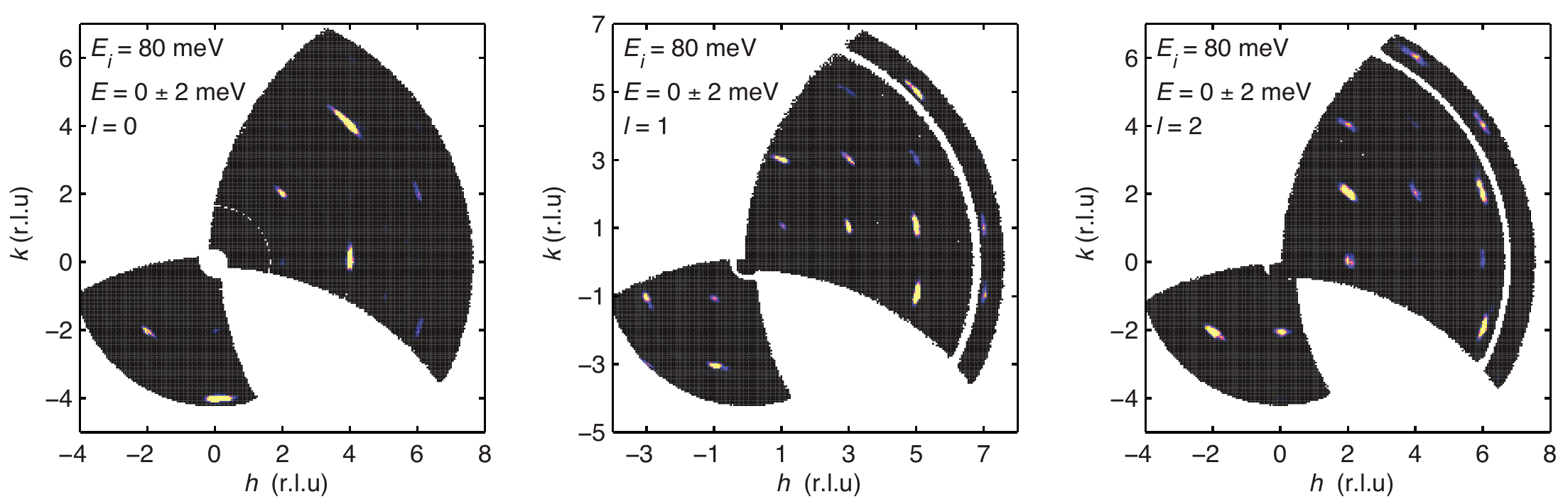}
\caption{Elastic line of our single-crystal mount showing nuclear Bragg peaks with positions and intensities compatible with the Fd$\bar{3}$m space-group.}
\end{figure} 

\clearpage

\section{S2. Data folding and symmetrization}

\begin{figure}[h]
\includegraphics[width=0.7\columnwidth]{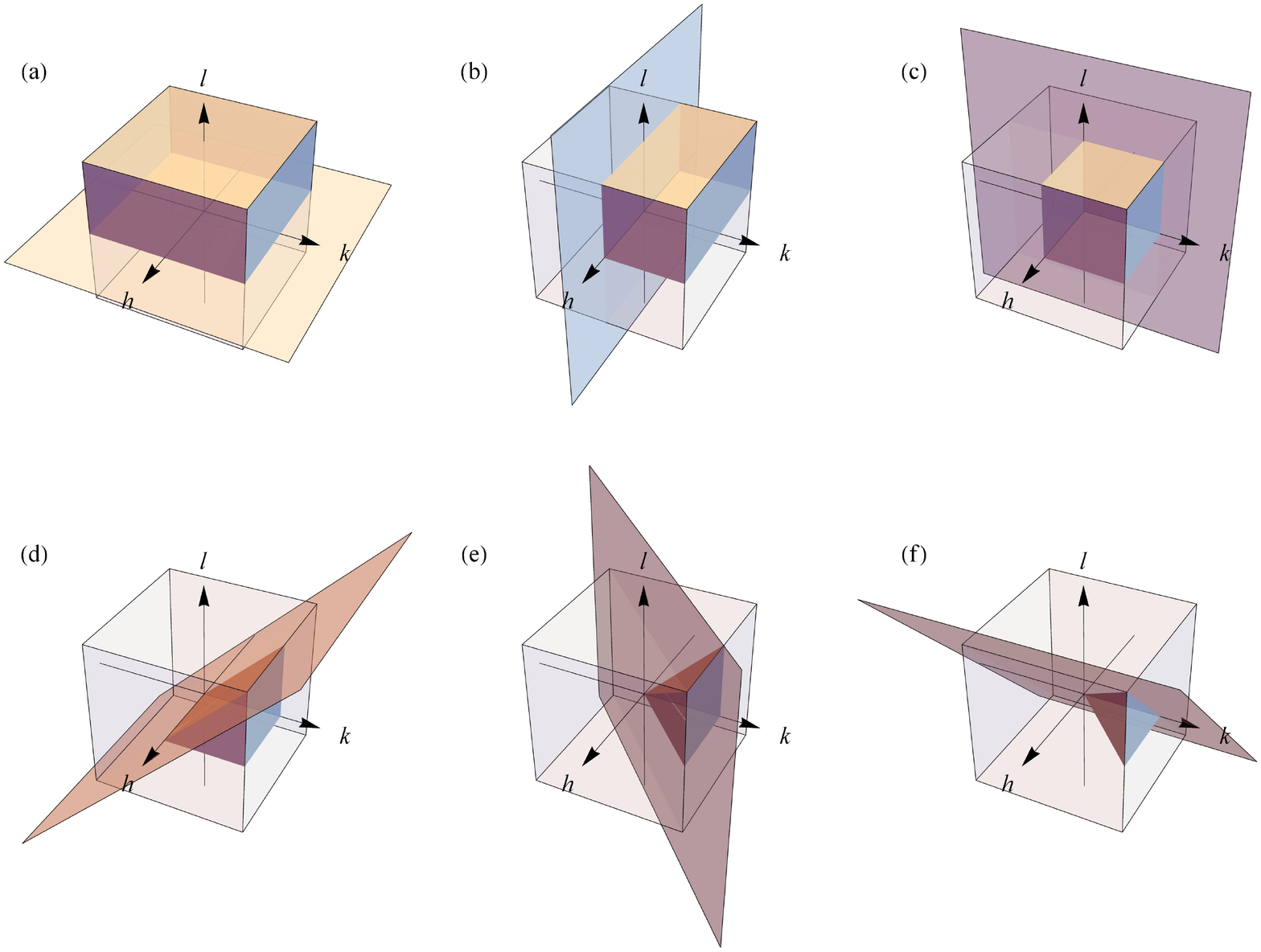}
\caption{Graphical representation of the momentum-space folding procedure used to symmetrize our data.}
\end{figure} 

\section{S3. Reverse Monte-Carlo analysis of spin-space anisotropy}
We used a reverse Monte Carlo (RMC) approach \cite{McGreevy_1988}
to analyze our magnetic diffuse-scattering data. The RMC approach
fits spin configurations directly to experimental data without using
a model of the magnetic interactions. For our refinements, we fitted
the energy-integrated single-crystal data measured on SEQUOIA at 20\,K;
the two datasets with incident energies of 40 and 80\,meV were fitted
simultaneously. Our spin configurations contained $8\times8\times8$
conventional unit cells (8192 vector spins). Our single-crystal RMC
refinement algorithm has been described previously \cite{Paddison_2018}.
An overall intensity scale factor and flat background level were refined
for each dataset. Fifteen independent refinements were performed and
the results averaged to improve their statistical accuracy. Each refinement
was performed for $300$ proposed moves per spin, after which no significant
improvements in the fit was apparent. 

Because the RMC approach is data-driven and independent of an interaction
model, it allows the assumptions of our interaction model to be tested
in an unbiased way. Arguably the most important assumption of our
interaction model is that the interactions are described by a Heisenberg
form without anisotropy terms. To test this assumption, we look for
the presence of anisotropy in the distribution function of spin orientations
determined from RMC refinement,
\[
p(\theta,\phi)=\frac{n(\theta,\phi)}{N{\rm d}(\cos\theta)\,{\rm d}\phi},
\]
where $\ensuremath{n(\theta,\phi)}$ is the number spins with orientations
within the range $\ensuremath{{\rm d}(\cos\theta),\mathrm{d}\phi}$,
and $N$ is the total number of spins. It has been shown previously
that RMC refinement is sensitive to anisotropy in pyrochlore magnets,
if it is indeed present~\cite{Paddison_2017}. However, the function
$\ln(p)$ for MgCr$_{2}$O$_{4}$ shown in Fig.~\ref{fig:rmc_anisotropy}(a)
reveals no evidence for anisotropy, beyond statistical fluctuations
that are also present in entirely random spin configurations of the
same size {[}Fig.~\ref{fig:rmc_anisotropy}(b){]}.
\begin{figure}

\centering{}\includegraphics{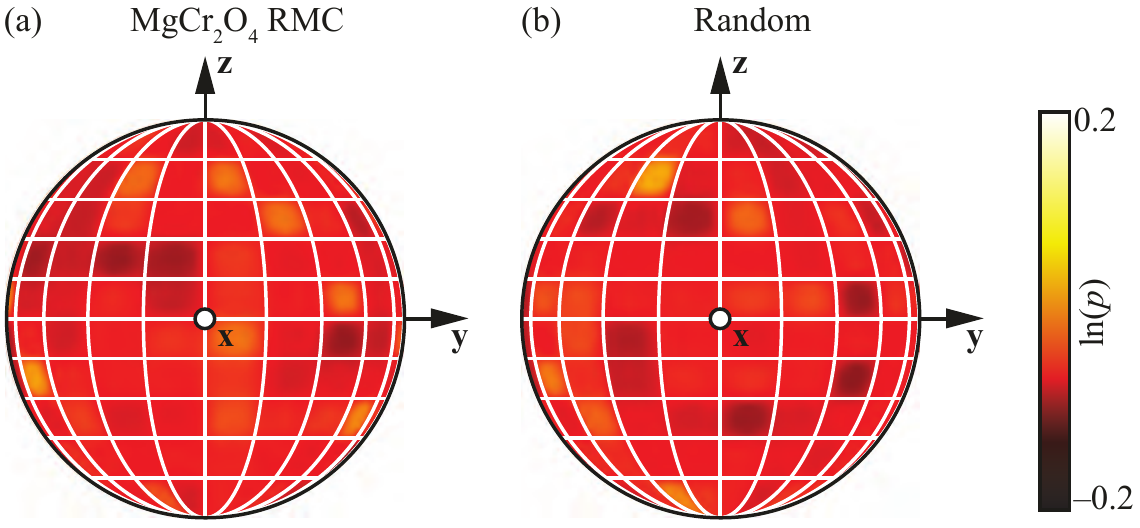}\caption{\label{fig:rmc_anisotropy}Stereographic projections of the distribution
function of spin orientations for (a) MgCr$_{2}$O$_{4}$ spin configurations
determined from RMC refinement to single-crystal diffuse-scattering
data, and (b) entirely random spin configurations. The distribution
functions reveal no evidence for spin anisotropy in MgCr$_{2}$O$_{4}$.
The $\mathbf{x}$, $\mathbf{y}$, and $\mathbf{z}$ axes are defined
locally such that $\text{\textbf{z }}\in\frac{1}{\sqrt{3}}\left\langle 111\right\rangle $
is parallel to the local three-fold axis of the Cr$^{3+}$ site, and
the local $\mathbf{x}\in\frac{1}{\sqrt{6}}\left\langle 112\right\rangle $
and $\mathbf{y}\in\frac{1}{\sqrt{2}}\left\langle 110\right\rangle $
axes are mutually perpendicular and in the plane perpendicular to
$\mathbf{z}$.}
\end{figure}

\section{S4. Results of SCGA fits to energy-integrated and susceptibility data}
 The SCGA fits to energy-integrated quantities $I_0({\bf Q})$ and $I_1({\bf Q})$ of both 40 and 80 meV datasets are performed for a grid of values of $J_2$ and $J_{3\text{a}}$. In addition to $J_1$ and $J_{3\text{b}}$, overall scale factors, $s_{40}$ and $s_{80}$, are introduced in the fits for each dataset to compensate for the discrepancy in the absolute normalization. Constant background parameters $I_{0,\text{bk}}^{40}$, $I_{1,\text{bk}}^{40}$, $I_{0,\text{bk}}^{80}$ and $I_{1,\text{bk}}^{80}$ are also included to account for the incoherent scattering signal and instrumental background. Fig.~\ref{fig1} gives an overview of values of all the fitting parameters. The red star is determined from the goodness of fit for the neutron and bulk susceptibility data. The corresponding values are listed in Tab.~\ref{SI_chifit}.

\begin{figure}[h]
\includegraphics[width=0.7\columnwidth]{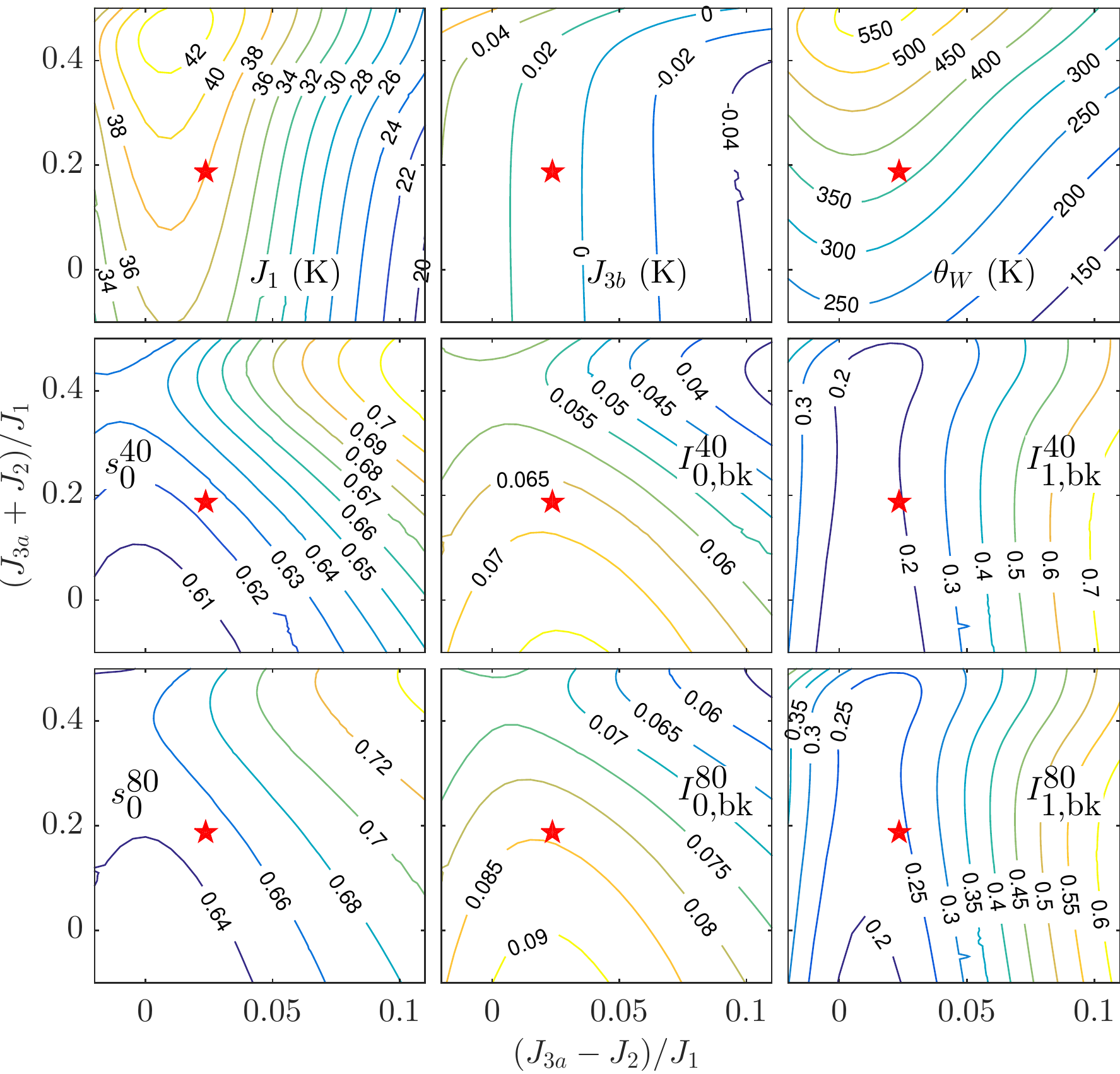}
\caption{Contour plots of fitting parameters as a function of $J_2$ and $J_{3\text{a}}$.  The red star indicates the best fit, determined from the goodness of fit for the 40 and 80 meV neutron data at 20~K and bulk susceptibility data between 20~K and 400~K [Fig.~\ref{fig1}b].}
\label{SI_grid}
\end{figure}

\begin{table*}[h]
\centering
\begin{tabular}{|c|c|c|c|c|c|c|c|c|c|c|c|}
\hline
$J_1$(K) & $J_2/J_1$ & $J_{3\text{a}}/J_1$ & $J_{3\text{b}}/J_1$& $\theta_{\text{W}}$ (K)& $T$ (K) & $s_{40}$  & $I_{0,\text{bk}}^{40}$ & $I_{1,\text{bk}}^{40}$& $s_{80}$  &$I_{0,\text{bk}}^{80}$ & $I_{1,\text{bk}}^{80}$\\
\hline
38.05(3) & 0.0815 & 0.1050 & 0.0085(1)  &  364.3 & 20 &  0.6241(5)&	0.0675(2) &	0.1946(13)	&0.6503(5)	&0.0843(2)	&0.2399(11)\\
\hline

\end{tabular}
\caption{{Values of the best fitting parameters for \mgcro.} The Weiss temperature is computed from $\theta_{\text{W}}=(6J_1+12J_2+6J_{3\text{a}}+6J_{3\text{b}})S(S+1)/3$ with $S=3/2$.}\label{FitParam}
\label{SI_chifit}
\end{table*}

\begin{figure}[h]
\includegraphics[width=0.5\columnwidth]{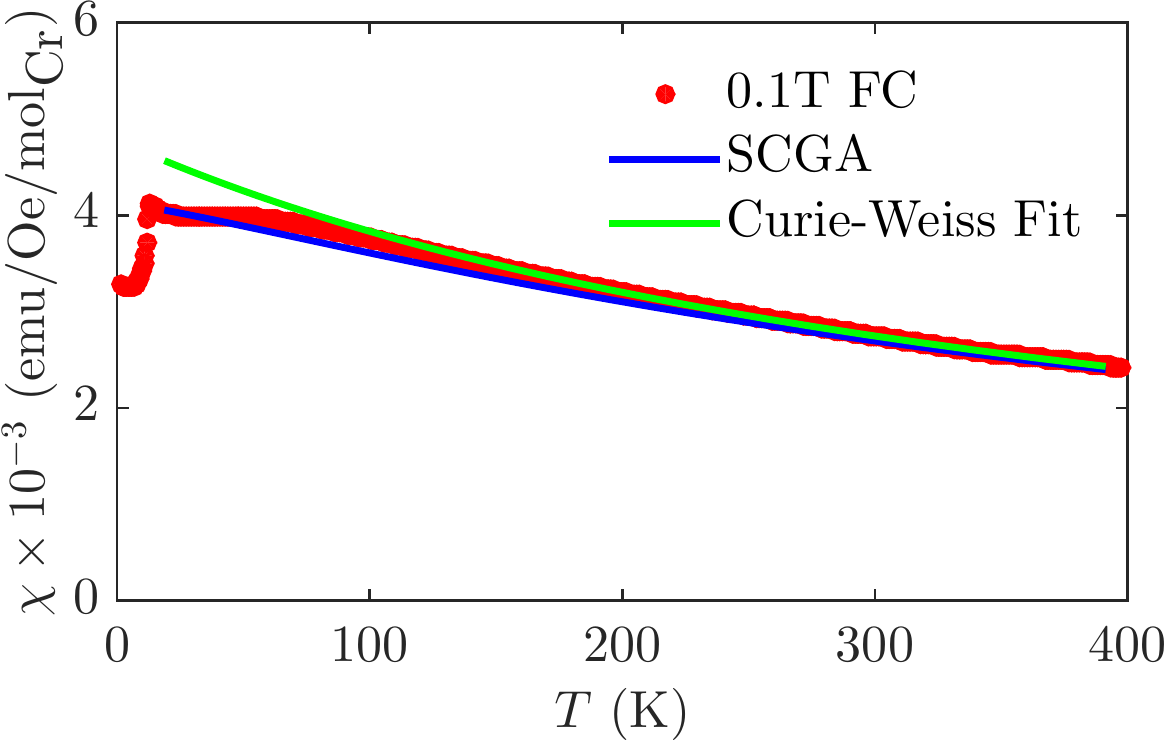}
\caption{{Temperature dependence of bulk magnetic susceptibility of \mgcro. The red dots are data collected at 0.1~Tesla during cool down. The blue curve is the calculation using SCGA for the best fitting parameters listed in Tab.~\ref{SI_tabcorr}. The green curve is the Curie-Weiss fit of the data between 200~K and 400~K, yielding a Weiss constant of 405~K.}} 
\end{figure} 

\clearpage
\section{S5. Comparison of spin correlations for the FN model (calculated using the SCGA and Monte Carlo) and hexagonal spin-cluster model}

The structure factor of the hexagonal spin-cluster model is
\begin{align}
\mathcal{S}({\bf Q})=&\dfrac{4}{9}\left[\left(1-\cos ^2\left(\frac{\pi  k}{2}\right)\right) \left(\cos \left(\frac{\pi  \ell}{2}\right)-\cos \left(\frac{\pi  h}{2}\right)\right)^2\right.\notag\\
&+\left(1-\cos ^2\left(\frac{\pi  h}{2}\right)\right) \left(\cos \left(\frac{\pi  k}{2}\right)-\cos \left(\frac{\pi  \ell}{2}\right)\right)^2\notag\\
&\left.+\left(1-\cos ^2\left(\frac{\pi  \ell}{2}\right)\right) \left(\cos \left(\frac{\pi  h}{2}\right)-\cos \left(\frac{\pi  k}{2}\right)\right)^2\right]\,.
\end{align}
This can be rewritten in terms of the Fourier expansion
\begin{align}
\mathcal{S}({\bf Q})=\dfrac{2}{3}\left(\left\langle  {\bf S}_{0}\cdot {\bf S}_{0}\right\rangle f_0 +\left\langle  {\bf S}_{0}\cdot {\bf S}_{1} \right\rangle f_1+\left\langle  {\bf S}_{0}\cdot {\bf S}_{2} \right\rangle f_2+\left\langle  {\bf S}_{0}\cdot {\bf S}_{3} \right\rangle f_3 \right)
\end{align}
where $\left\langle  {\bf S}_{0}\cdot {\bf S}_{0} \right\rangle =  1$, $\left\langle  {\bf S}_{0}\cdot {\bf S}_{1} \right\rangle =  -1/3$, $\left\langle  {\bf S}_{0}\cdot {\bf S}_{2} \right\rangle =  1/6$  and $\left\langle  {\bf S}_{0}\cdot {\bf S}_{3} \right\rangle=\left(\left\langle  {\bf S}_{0}\cdot {\bf S}_{3\text{a}} \right\rangle+\left\langle  {\bf S}_{0}\cdot {\bf S}_{3\text{b}} \right\rangle\right)/2 =  -1/12$. All the other correlators vanish identically. The Fourier functions $f_i$ are given in Tab.~\ref{SI_tabcorr}. In particular, two types of the third nearest neighbors are distinguished by the lattice symmetry, while their corresponding Fourier basis take the same form. 

\begin{table}[h!]
\begin{tabular}{|c|rl|} \hline
$i$   & & $f_i$  \\ \hline
0        & & 1 \\ \hline
1        & 2 $\times$ & $\big[ \cha \big\{ \cka + \cla \big\} + \cka \cla \big] $\\ \hline
2        & 4 $\times$ & $\big[ \chb \cka \cla$  \\ && $+ \cha \big\{ \ckb \cla + \cka \clb \big\} \big] $ \\ \hline
3 & 4 $\times$ & $\big[ \chb \big\{ \ckb+ \clb \big\}  + \ckb \clb \big]  $ \\ \hline
4        & 2 $\times$ & $\big[ \chc \big\{ \cka + \cla \big\}  + \ckc \big\{ \cha + \cla \big\} $ \\
                     && $ + \clc \big\{ \cha + \cka \big\} \big] $ \\ \hline
5 & $...$ & $...$ \\ \hline
\end{tabular}
\caption{Lattice harmonics entering the instantaneous structure factor, $\mathcal{S}({\bf Q}) = 2/3\sum_{{i}}\left\langle {\bf S}_{0}\cdot{\bf S}_{i}\right\rangle f_i $, where $f_i = \sum_{{\bf R}_{i}}\cos\left({\bf Q}\cdot({\bf R}_{0} - {\bf R}_{i})\right)$.  The level of neighbors is indexed by $i$. 
}
\label{SI_tabcorr}
\end{table}

\begin{figure}[h]
\includegraphics[width=0.45\columnwidth]{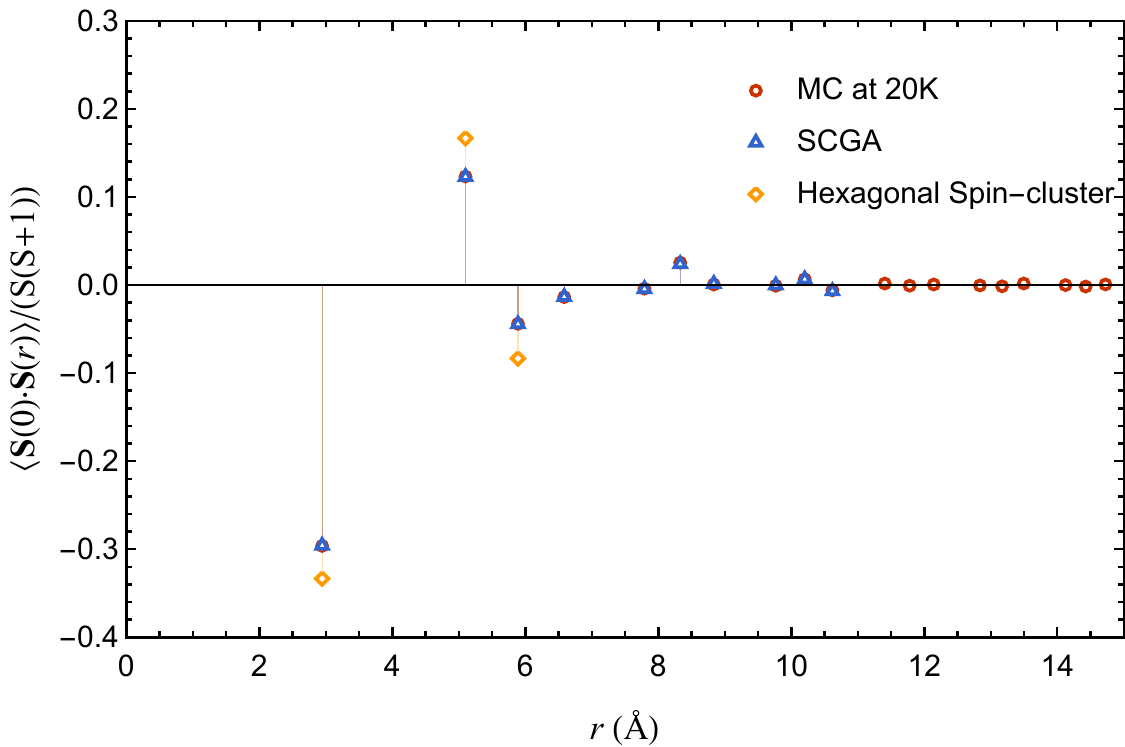}
\caption{{Spin correlations as a function of distance. The red circles are obtained from Monte Carlo simulations at 20~K for the best fitting parameters. The blue triangles are calculated using SCGA up to 10th neighbors, which show excellent agreement with MC results. The yellow diamonds are correlators for the hexagonal spin-cluster model. All the correlators beyond the third nearest neighbor are zero. Both signs and relative strength of the first three correlators from this model resemble those of our microscopic FN model. }}
\label{SI_corr}
\end{figure} 

\clearpage
\section{S6. Linear Spin-Wave Theory calculations}
Understanding of the spin wave excitations \mgcro\ was enabled by comparing neutron scattering results to semiclassical simulations of $\mathcal{S}\of{\mathbf{Q},E}$ for the pyrochlore lattice, with $\mathbf{Q}, E, T$ and the values of further neighbor interactions taken as the input parameters.  

Our numerical modeling proceeded as follows: we studied spins on a pyrochlore lattice with $6 \times 6 \times 6$ cubic unit cells and periodic boundary conditions, containing 3456 spins in total, governed by a Heisenberg Hamiltonian with nearest neighbor $J$ and longer ranged interactions $J_{2}, J_{3a}$ and $J_{3b}$. To calculate $\mathcal{S}\of{\mathbf{Q},E}$, we first computed an approximate classical ground state using the Monte Carlo technique. In the case of pure nearest neighbor interactions (where the ground state is massively degenerate) we reached the $T \to 0$ limit by following the Monte Carlo iterations with a steepest descent method to arrive at an effectively exact ground state. In the case of further neighbor interactions, the true ground state is ordered (with a very small $T_{\rm N}$); since our study was interested in the spin wave structure of the disordered paramagnetic phase, we thermalized the system at approximately 10~K to ensure the ordered state was never reached. Having derived a base classical spin configuration, we then numerically constructed the quantum harmonic spin wave Hamiltonian as in Walker and Walstedt~\cite{Walker_1977,Walker_1980}, exactly diagonalized the Hamiltonian using a Bogoliubov transformation to obtain the full single-excitation spectrum, and then used the resulting eigenstates to calculate the dynamical structure factor $\mathcal{S}\of{\mathbf{Q},E}$, with Bose factors added to incorporate finite temperature. This method produces the leading order term in the $1/S$ expansion; at this level the eigenfrequencies of the quantum large-$S$ problem and the normal modes of small oscillations about the classical ground state are identical.

To improve our numerical results, we repeated the above process ten times for each set of interactions chosen, and then averaged the resulting distributions of $\mathcal{S}\of{\mathbf{Q},E}$ over classical ground states. For the nearest neighbor case, averaging over ground states mitigates finite size effects that result from studying the excitations about a single ground state chosen from a massively degenerate manifold. For case of longer ranged interactions, our choice to study the system at finite temperature to prevent ordering led to a fraction of the lowest energy modes ($E \sim T$ or below) being unstable, as they described oscillations about a configuration which was not the system's true ground state. Such modes have complex eigenfrequencies and cannot be properly normalized in the Bogoliubov formalism. However, as far as we were able to ascertain, the momentum space distribution of the unstable modes is an effectively random fraction of the total spectral weight at that energy, so we were able to reconstruct the low-energy excitations by averaging over thermal ``ground state" spin configurations to sample from the stable modes which had well-defined normalization, simply omitting any contribution from unstable modes.

Finally, to study the energy-integrated spectral weight $\mathcal{S}\of{\mathbf{Q}}$ we employed the self-consistent Gaussian approximation outlined in Ref.~\cite{Conlon2010a}. Unlike the harmonic approach detailed above, this method implicitly accounts for interactions between spin waves and does not suffer from normalization issues due to unstable modes, but it is a time-independent method and thus does not provide energy-resolved data. As its computational cost is significantly lower than the harmonic approach, we used it to extract the longer ranged interactions $J_{2}, J_{3a}$ and $J_{3b}$ from a numerical fit to the neutron scattering data, and then used those parameters in the harmonic calculation to obtain finite-energy results.

\section{S7. Spin dynamics for different exchange models and results of LSWT fits}
To make quantitative comparisons between interaction models and experimental data, we fit results of LSWT calculations to the experimental data with three fitting parameters, the overall energy scale $W$, the intensity scale $I_0$ and a constant background $I_{\text{bk}}$. The fitting is performed simultaneously for constant-momentum slices with $L=0,0.5,1$ and $1.5$ r.l.u. and energy transfer between 1.5~meV and 25~meV. The simulated data is interpolated on a $\bf Q$-grid that matches the experimental data and the overall intensity is normalized according to the sum rule, prior to the fitting. All the exchange models and corresponding fitting results are summarized in Fig.~\ref{SI_exchanges} and Tab.~\ref{SI_ALLPARAMTABLE}. Detailed inelastic spectra are presented in Fig.~\ref{SI_Inelastic1} to \ref{SI_Inelastic2}. 

\begin{figure}[h]
\includegraphics[width=0.5\columnwidth]{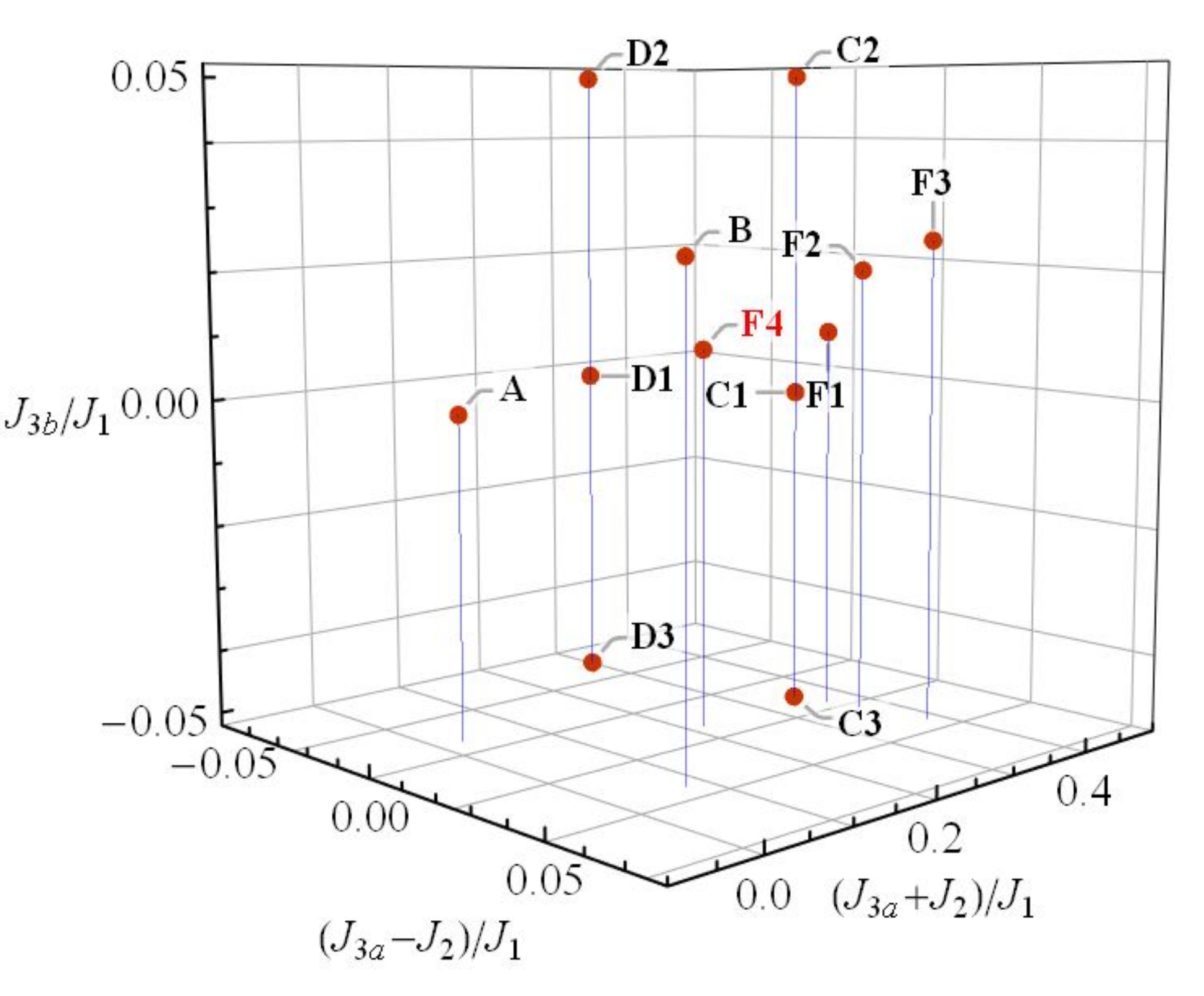}
\caption{{Overview of all the exchange parameters used in LSWT calculations.}}
\label{SI_exchanges}
\end{figure} 

\begin{table}[h]
\begin{tabular}{|c|c|c|c|c|c|c|c|c|c|c|c|c|}
\hline 
      & A & B     & C1   & C2   & C3    & D1   & D2   & D3    & F1     & F2    & F3    & {\red{ F4} }     \\ \hline
$J_2/J_1$ & 0 & 0     & 0.15 & 0.15 & 0.15  & 0.15 & 0.15 & 0.15  & 0.15   & 0.15  & 0.15  & \red{0.08145} \\ \hline
$J_{3\text{a}}/J_1$ & 0 & 0.05  & 0.17 & 0.17 & 0.17  & 0.12 & 0.12 & 0.12  & 0.1774 & 0.185 & 0.2   & \red{0.105}   \\ \hline
$J_{3\text{b}}/J_1$ & 0 & 0.025 & 0    & 0.05 & -0.05 & 0    & 0.05 & -0.05 & 0.01   & 0.02  & 0.025 & \red{0.00878} \\ \hline
$W$     &35.918   & 19.3418  & 17.6436  & 18.0734  & 16.9698  & 16.0742  & 17.1556  & 12.5     & 17.636   & 18.3618  & 18.3784  & \red{20.442 }  \\ \hline
$I_{\text{0}}$     &0.74193  & 1.213    & 1.4196   & 1.3546   & 1.3229   & 1.3707   & 1.3857   & 0.48817  & 1.4227   & 1.3598   & 1.354    & \red{1.0531}   \\ \hline
$I_{\text{bk}}$     &0.034101 & 0.027938 & 0.038731 & 0.026772 & 0.04146  & 0.044522 & 0.033162 & 0.099382 & 0.03265  & 0.027036 & 0.020982 & \red{0.0317 }  \\ \hline
$\chi^2$  &0.037663 & 0.031606 & 0.034808 & 0.033442 & 0.044838 & 0.044944 & 0.037591 & 0.077936 & 0.032907 & 0.032382 & 0.033653 & \red{0.031522} \\ \hline
\end{tabular}
\caption{{Values of the exchange parameters and fitting results. The reduced $\chi^2$ confirms that the set of parameters, F4, obtained from fitting the energy-integrated quantities and bulk magnetic susceptibility also best reproduces inelastic neutron data. Moreover, the intensity scale $I_0$ is close to 1, indicating that the calculation for this set of parameters correctly captures the ratio of inelastic to elastic spectral weight.} }\label{SI_ALLPARAMTABLE}
\end{table}

\begin{figure}[h]
\includegraphics[width=0.41\columnwidth]{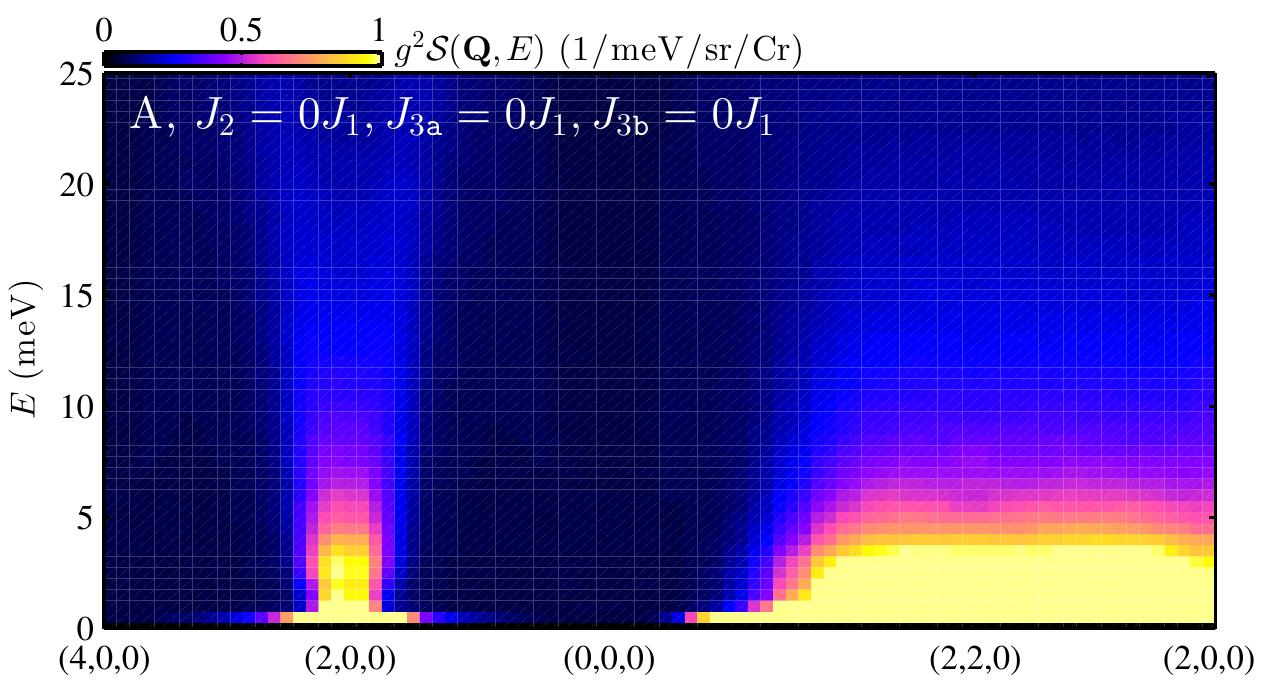}
\includegraphics[width=0.41\columnwidth]{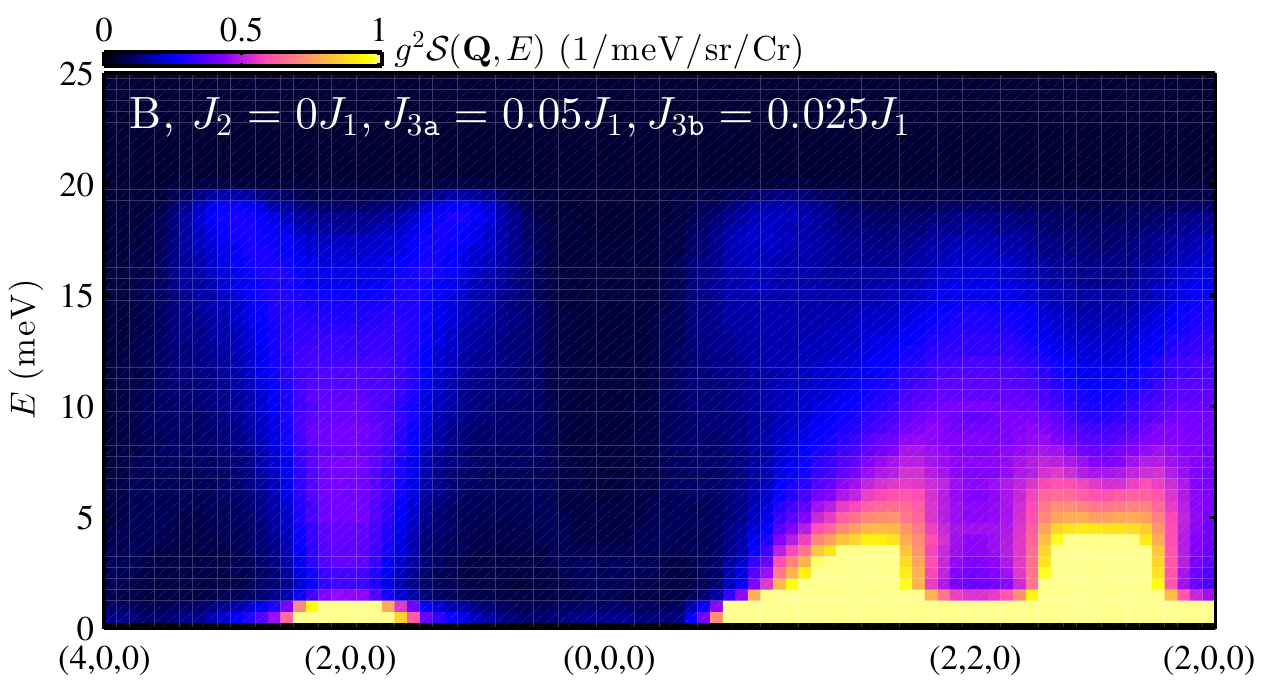}\\
\includegraphics[width=0.41\columnwidth]{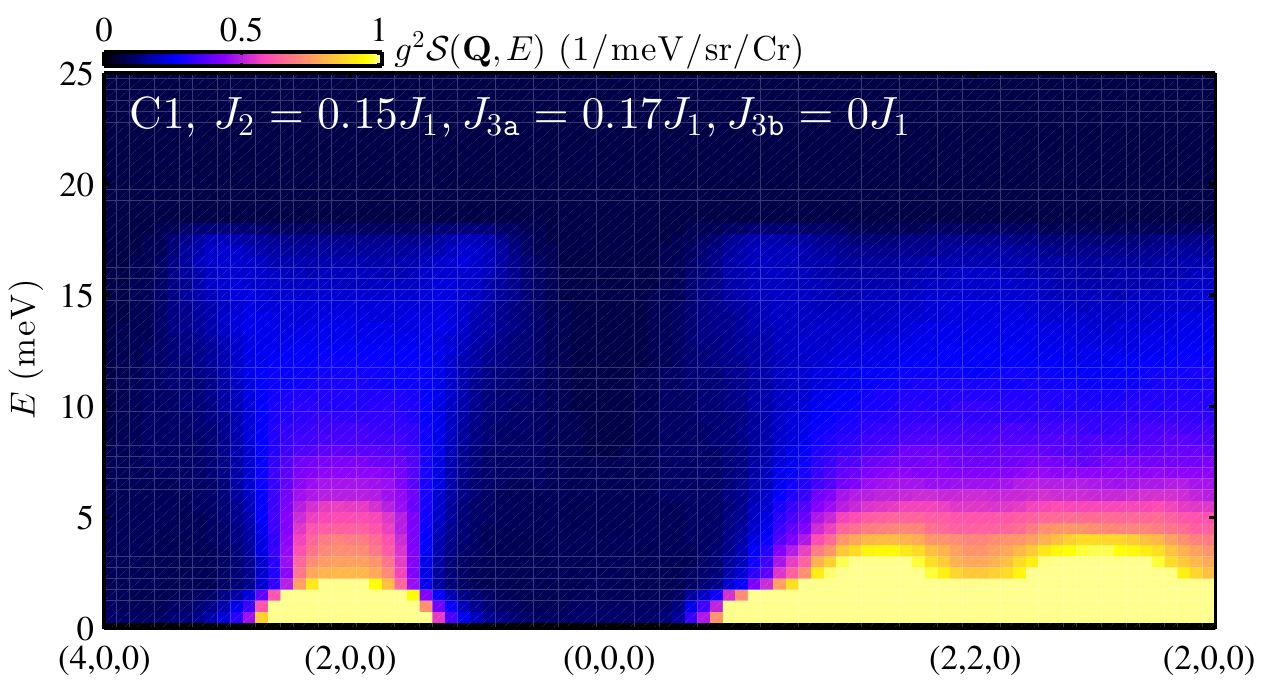}
\includegraphics[width=0.41\columnwidth]{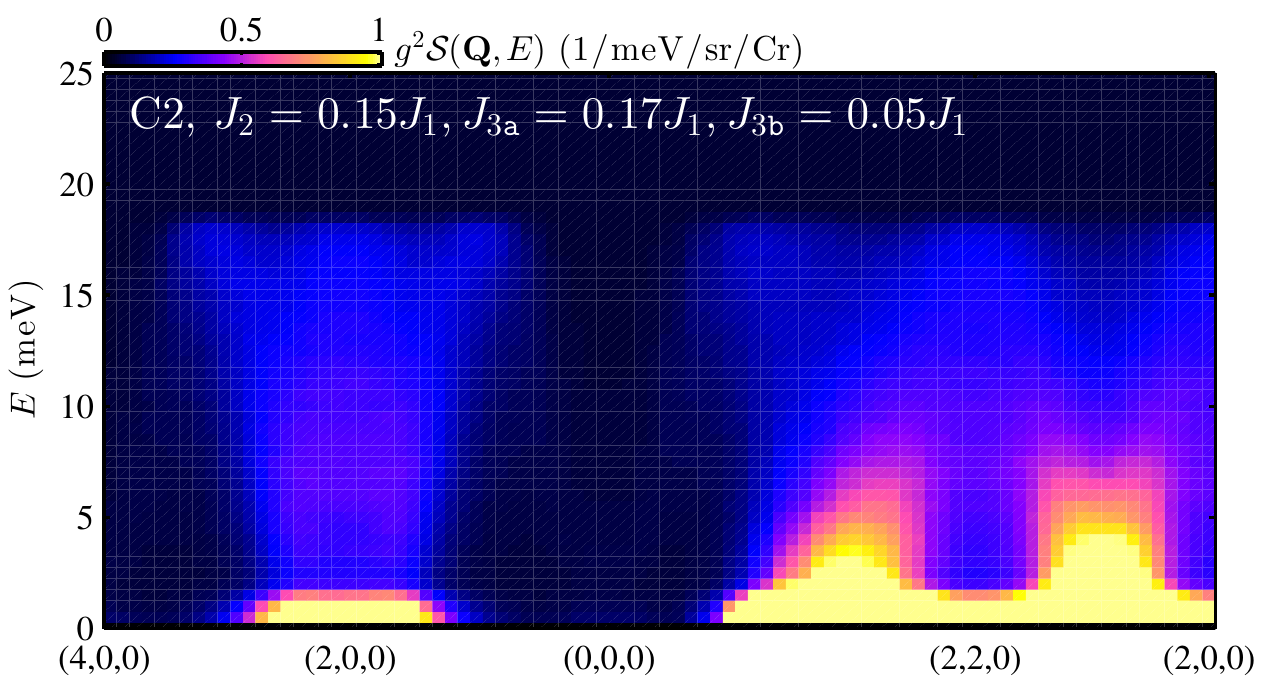}\\
\includegraphics[width=0.41\columnwidth]{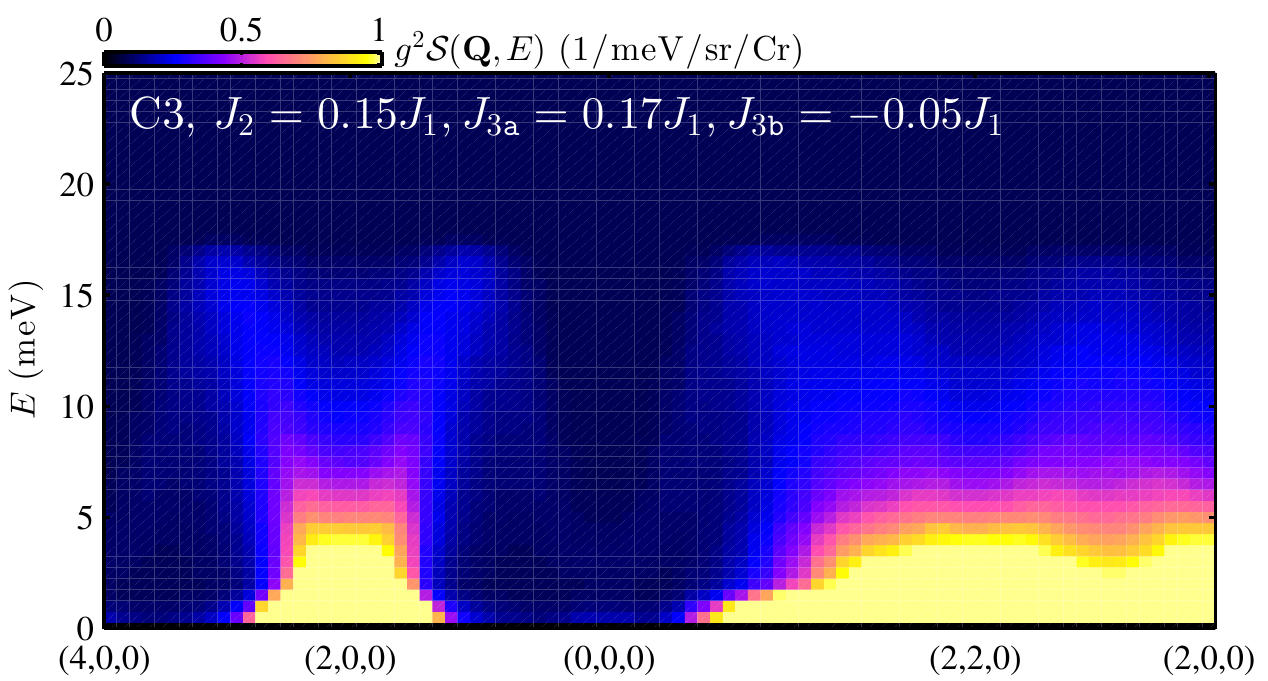}
\includegraphics[width=0.41\columnwidth]{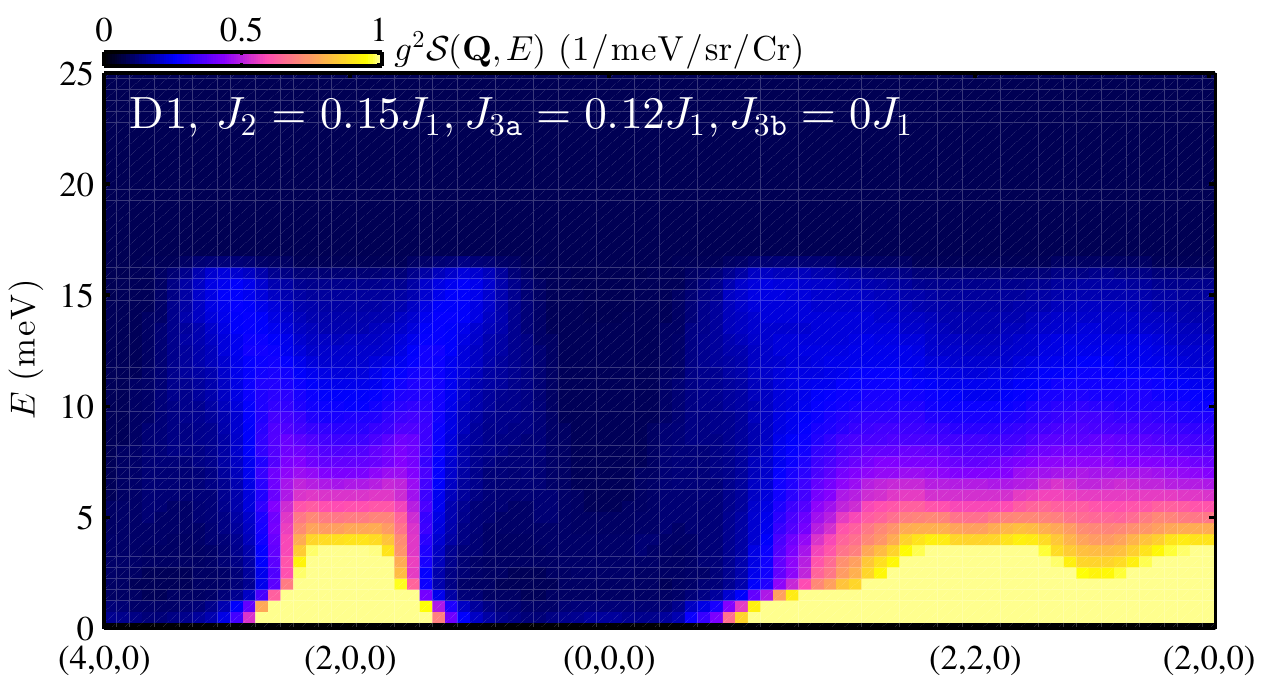}\\
\includegraphics[width=0.41\columnwidth]{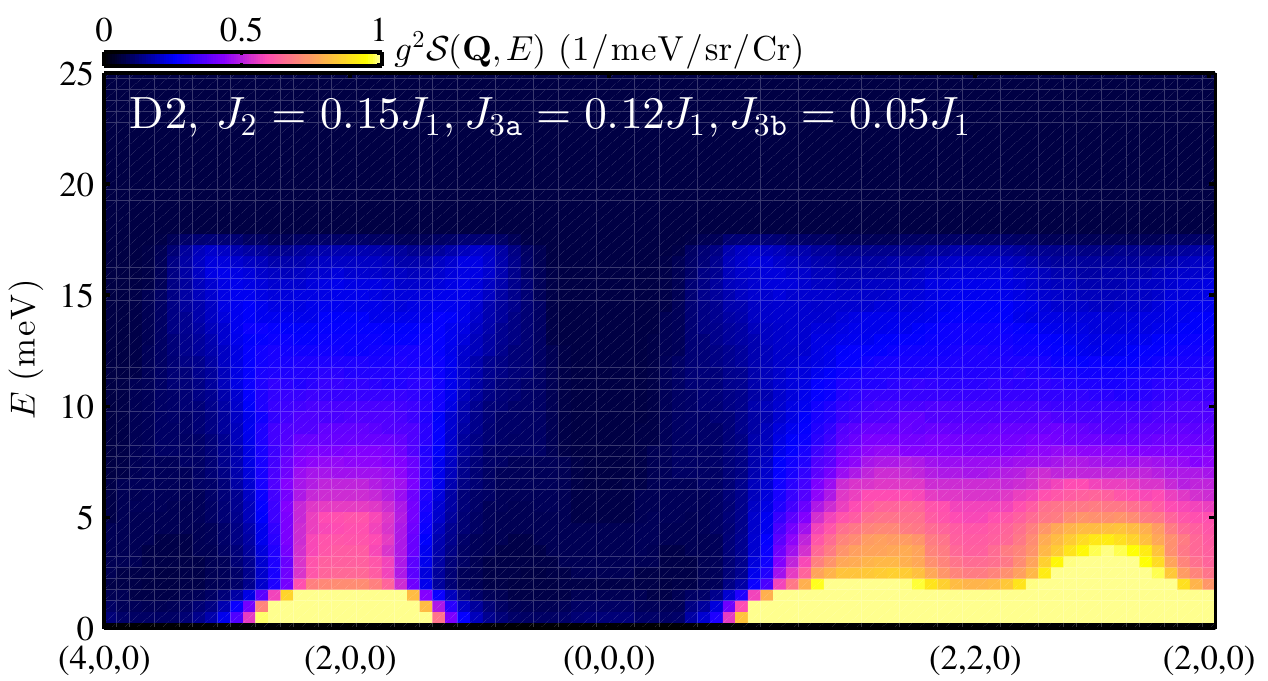}
\includegraphics[width=0.41\columnwidth]{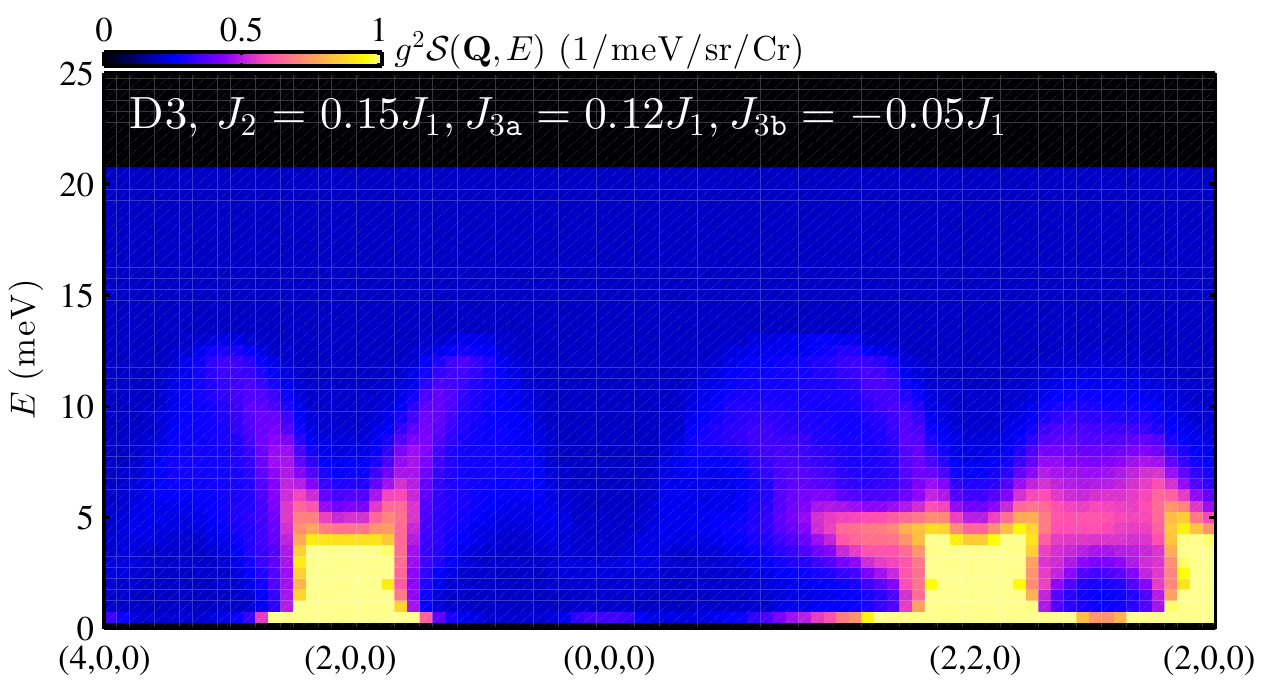}\\
\includegraphics[width=0.41\columnwidth]{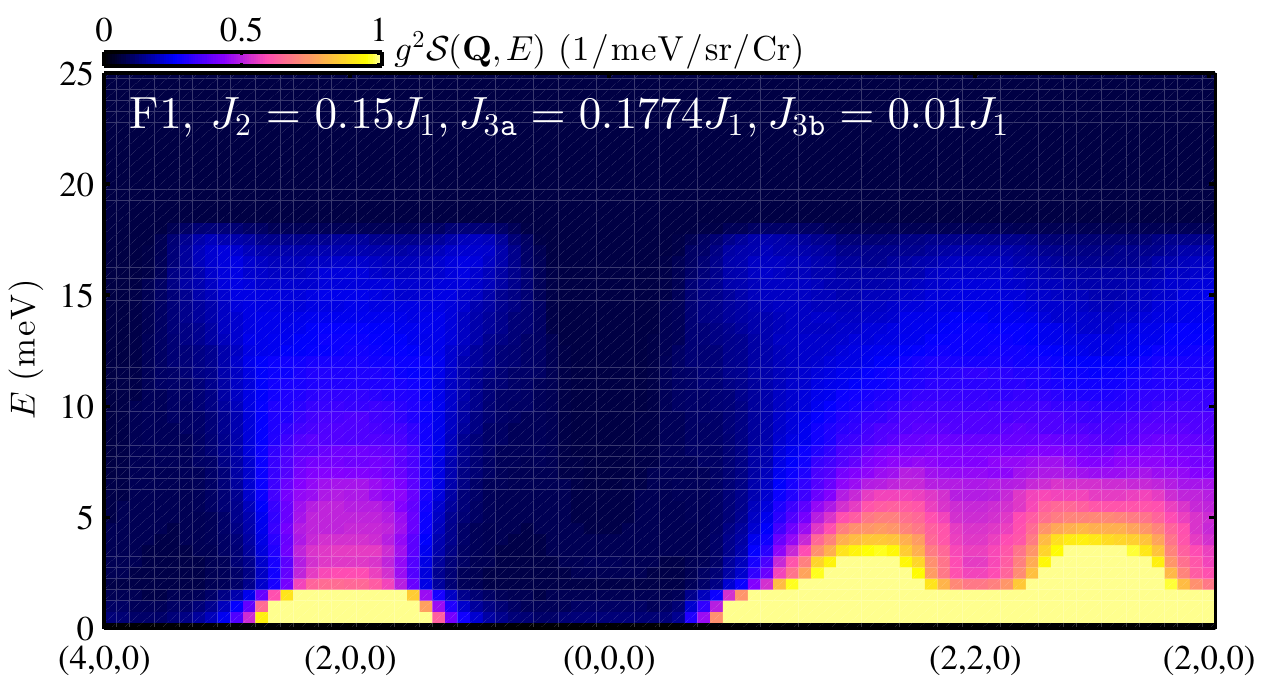}
\includegraphics[width=0.41\columnwidth]{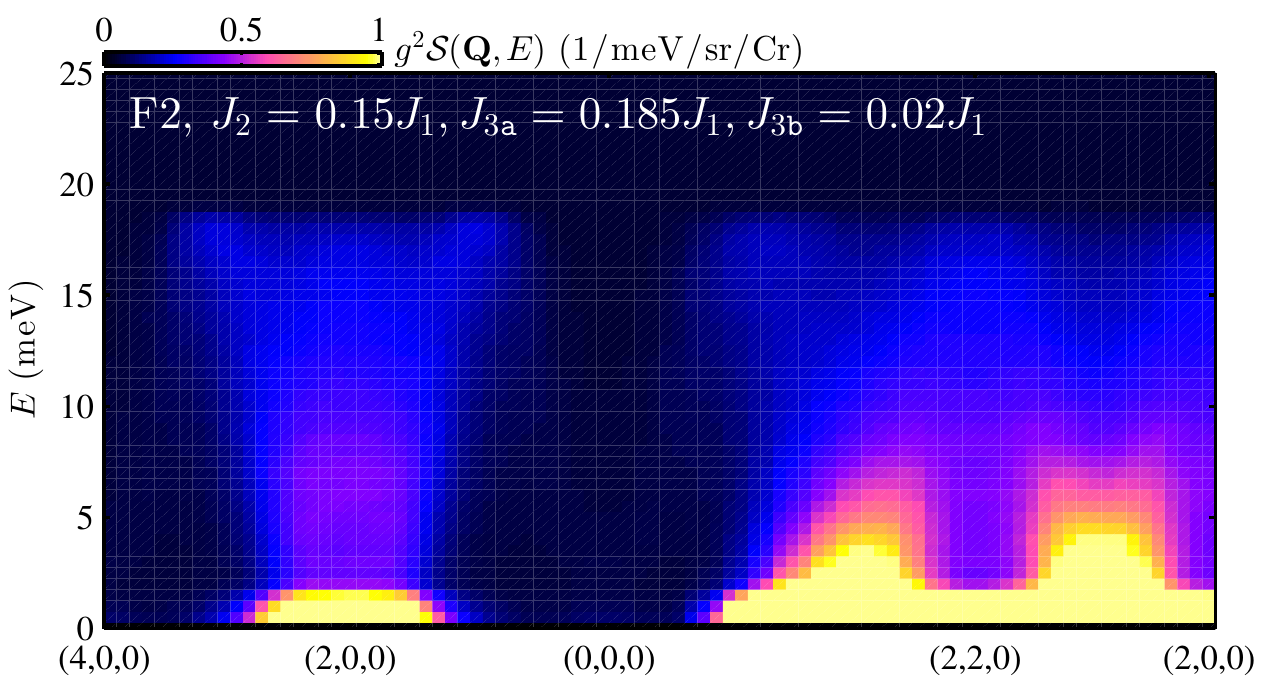}\\
\includegraphics[width=0.41\columnwidth]{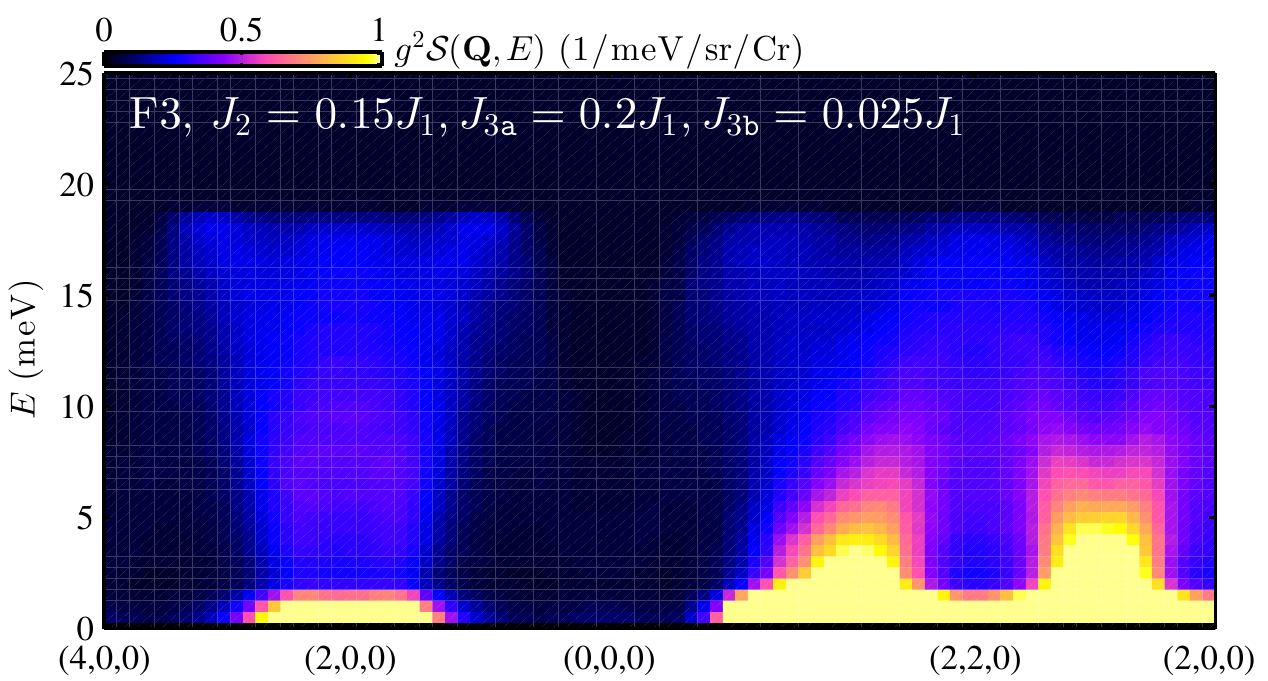}
\includegraphics[width=0.41\columnwidth]{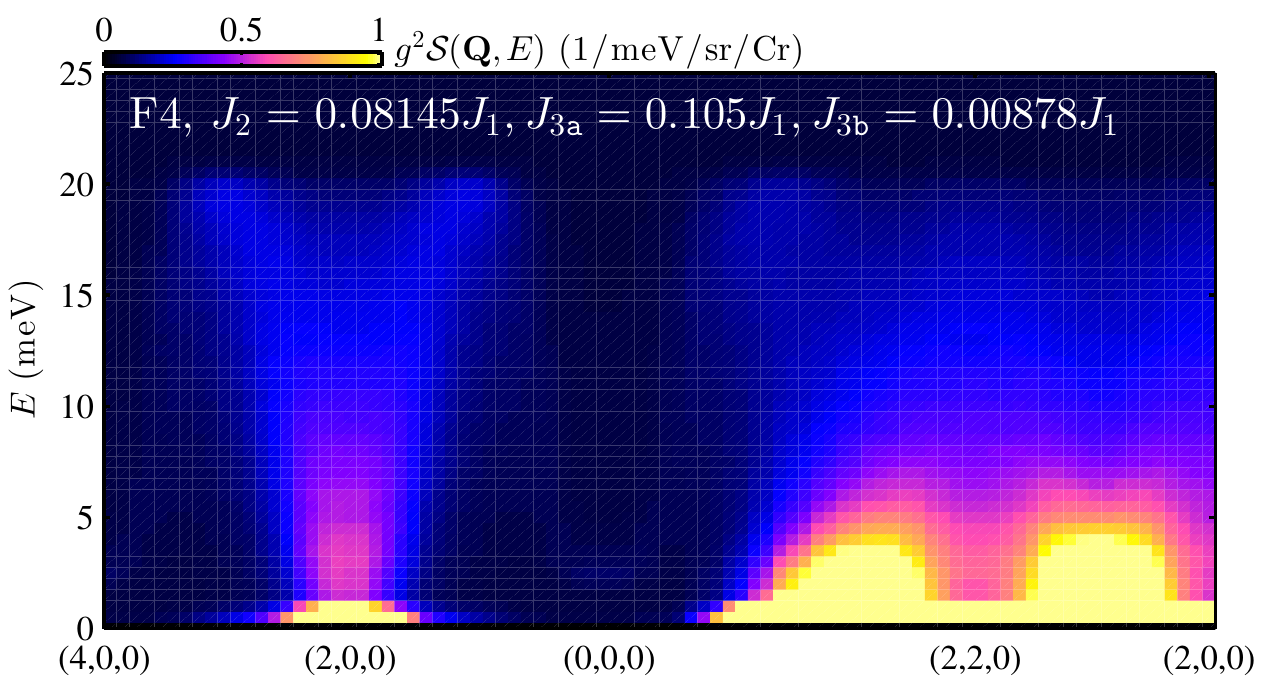}
\caption{{Calculated inelastic spectra along the path $(4,0,0)\rightarrow(2,0,0)\rightarrow(0,0,0)\rightarrow(2,2,0)\rightarrow(2,0,0)$.}}
\label{SI_Inelastic1}
\end{figure} 

\clearpage

\begin{figure}[h]
\includegraphics[width=0.41\columnwidth]{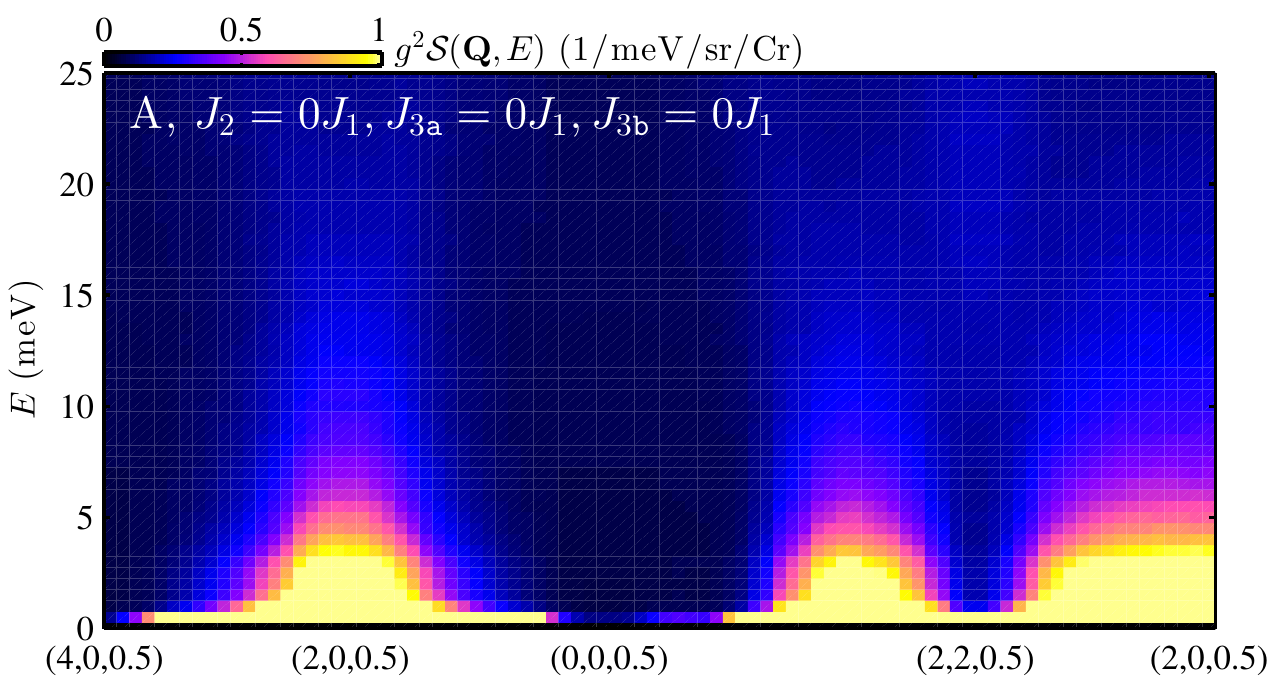}
\includegraphics[width=0.41\columnwidth]{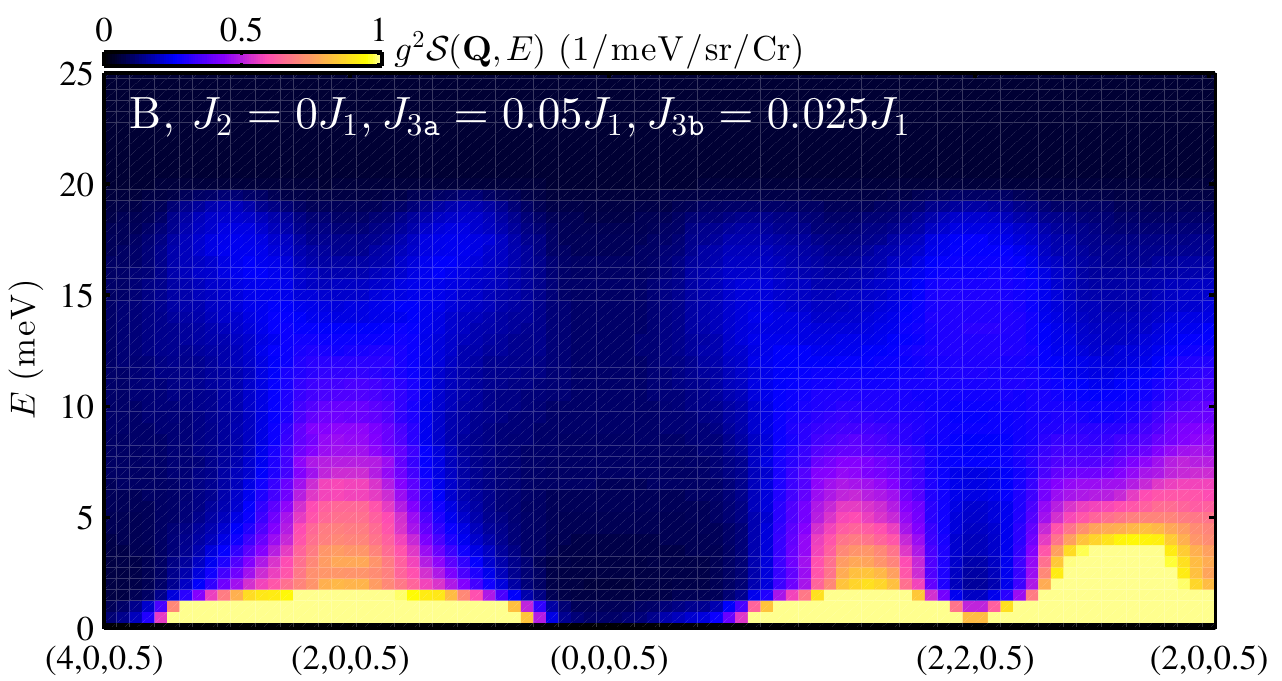}\\
\includegraphics[width=0.41\columnwidth]{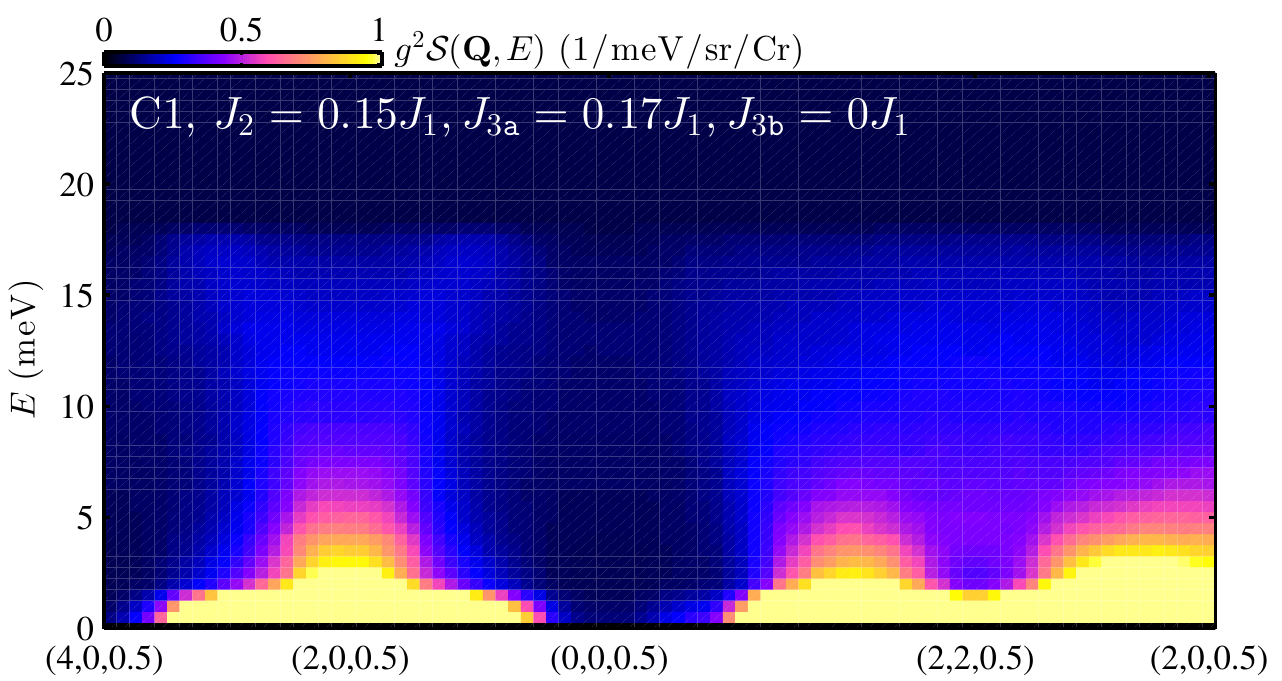}
\includegraphics[width=0.41\columnwidth]{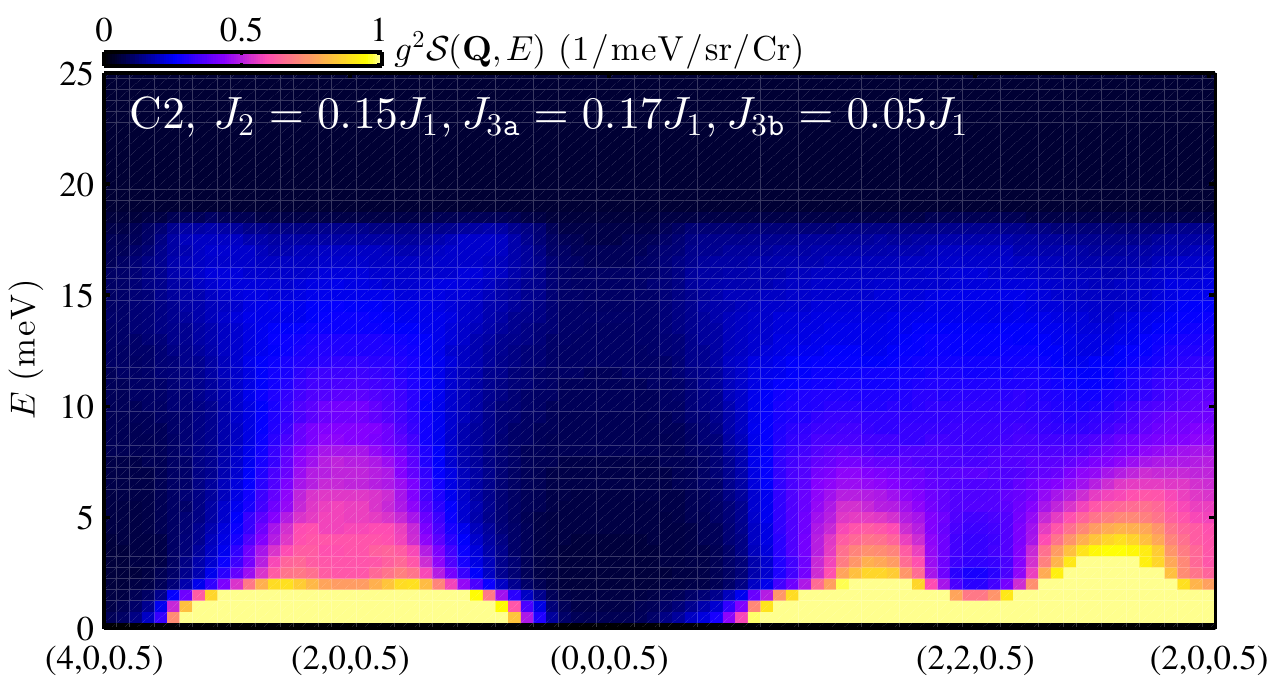}\\
\includegraphics[width=0.41\columnwidth]{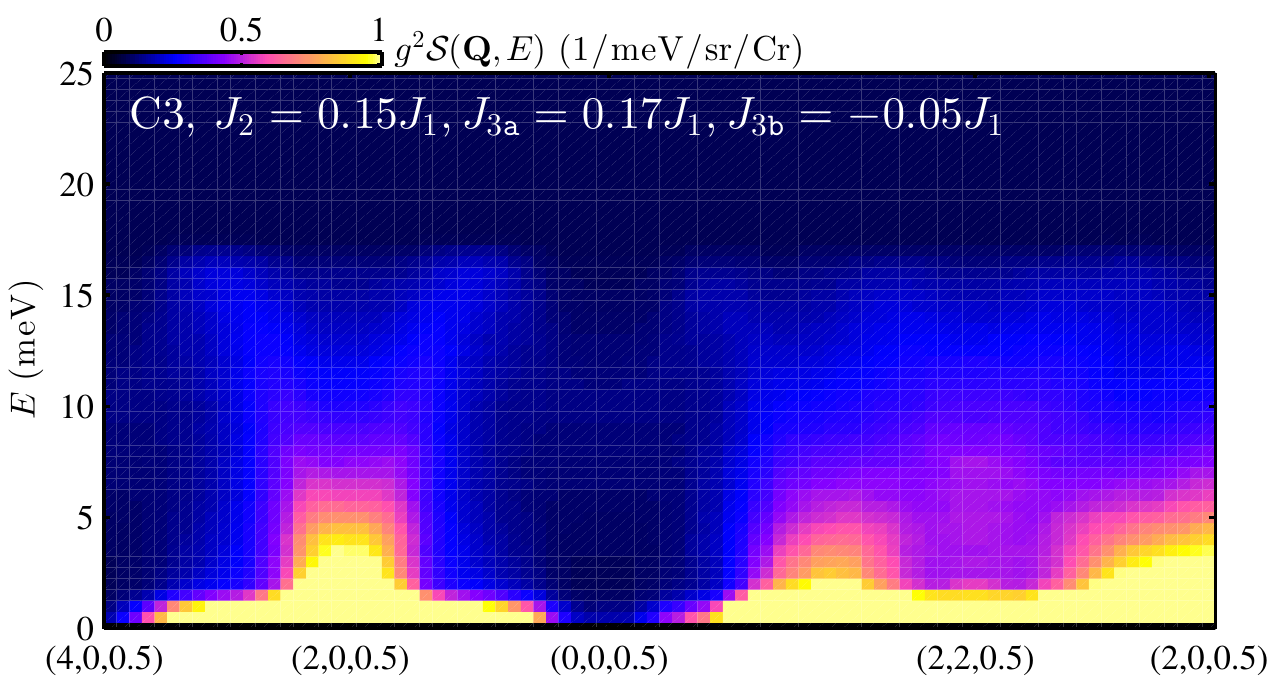}
\includegraphics[width=0.41\columnwidth]{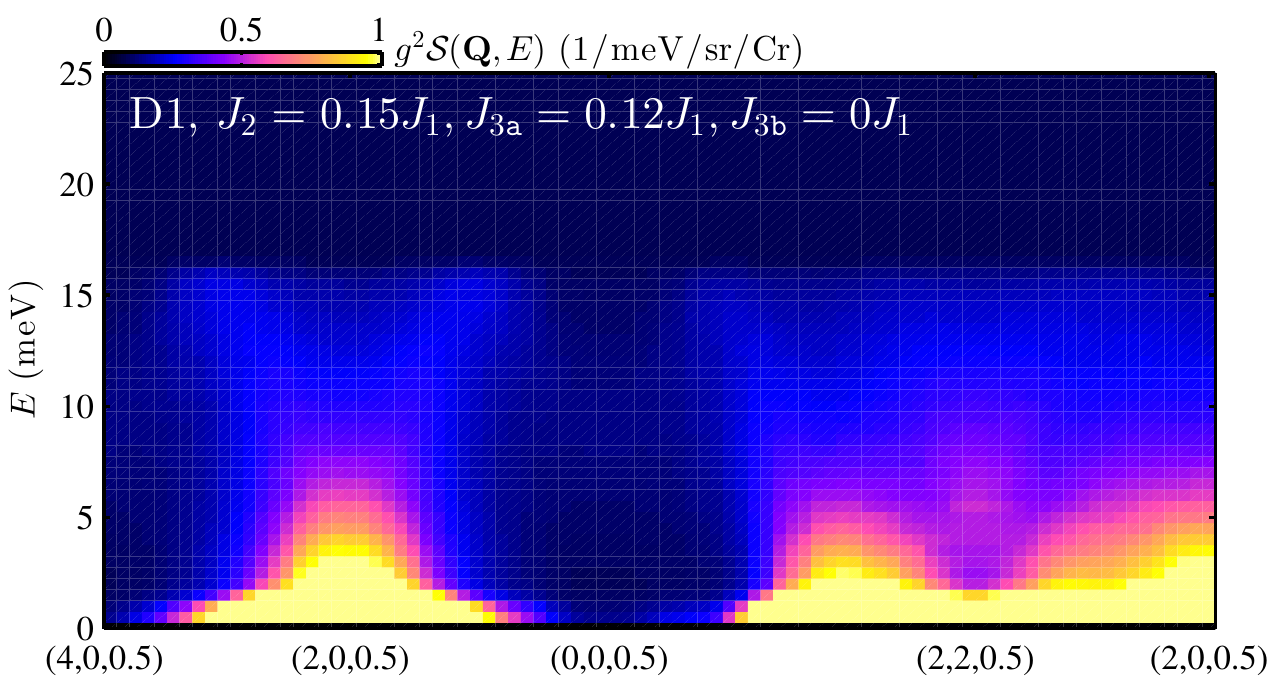}\\
\includegraphics[width=0.41\columnwidth]{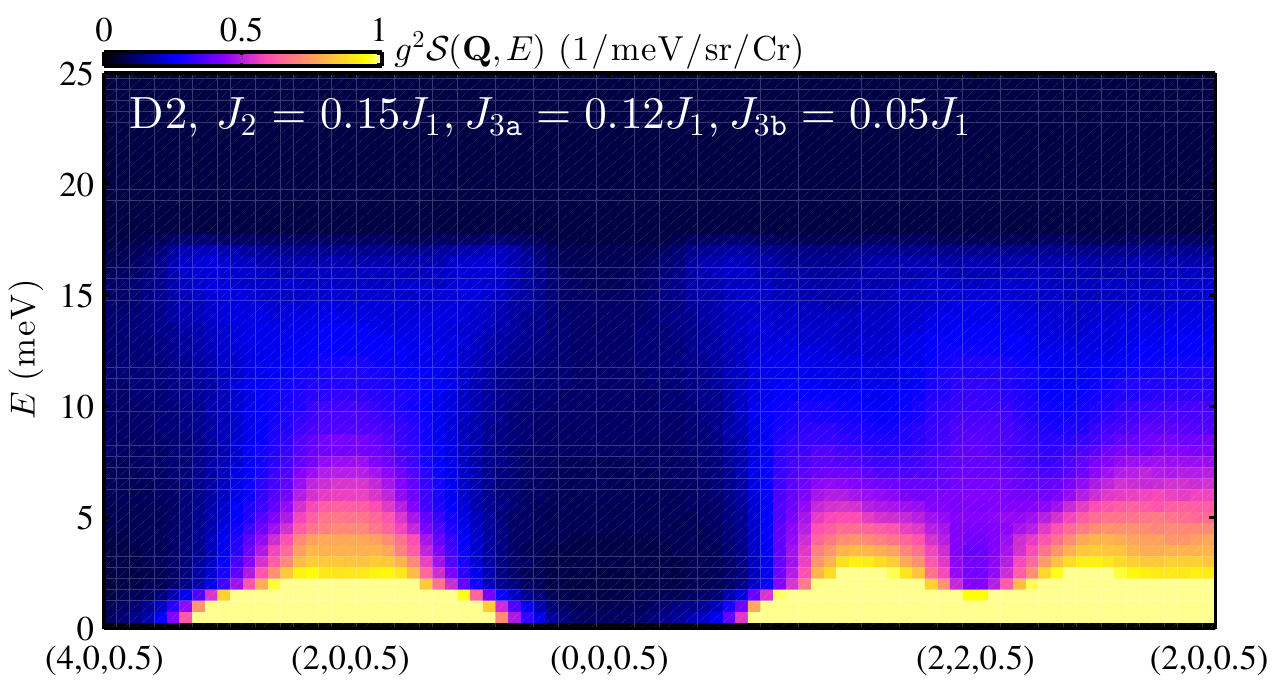}
\includegraphics[width=0.41\columnwidth]{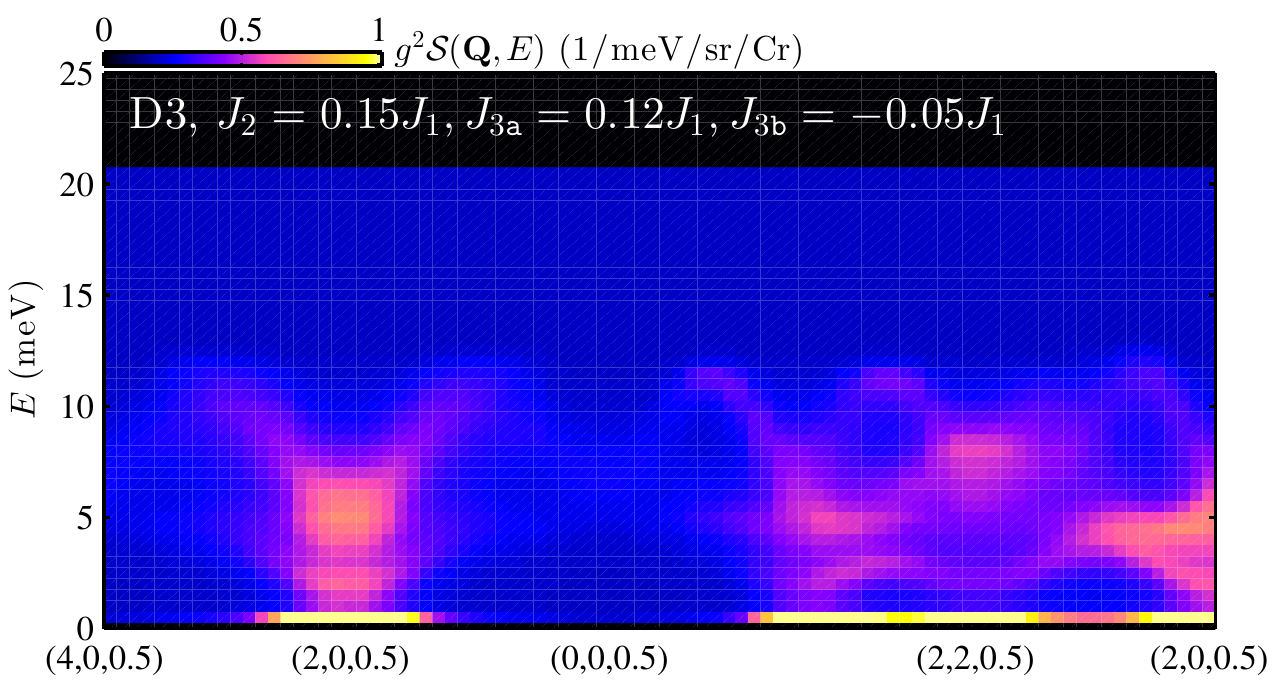}\\
\includegraphics[width=0.41\columnwidth]{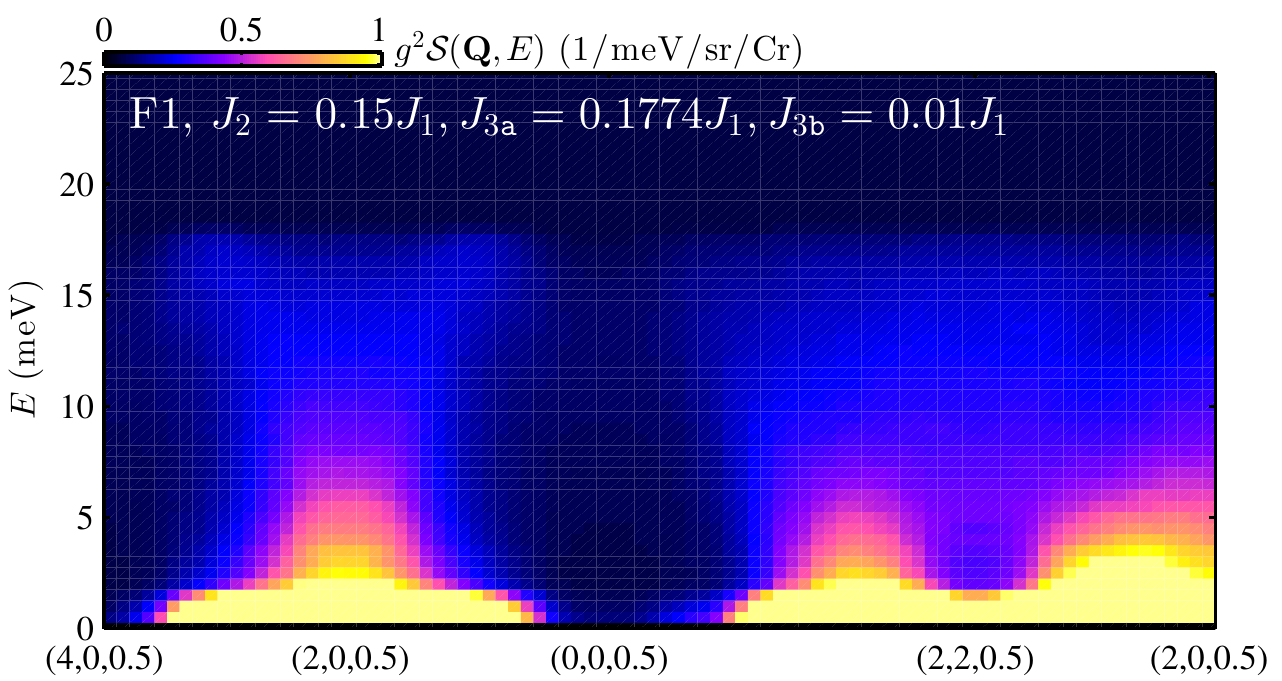}
\includegraphics[width=0.41\columnwidth]{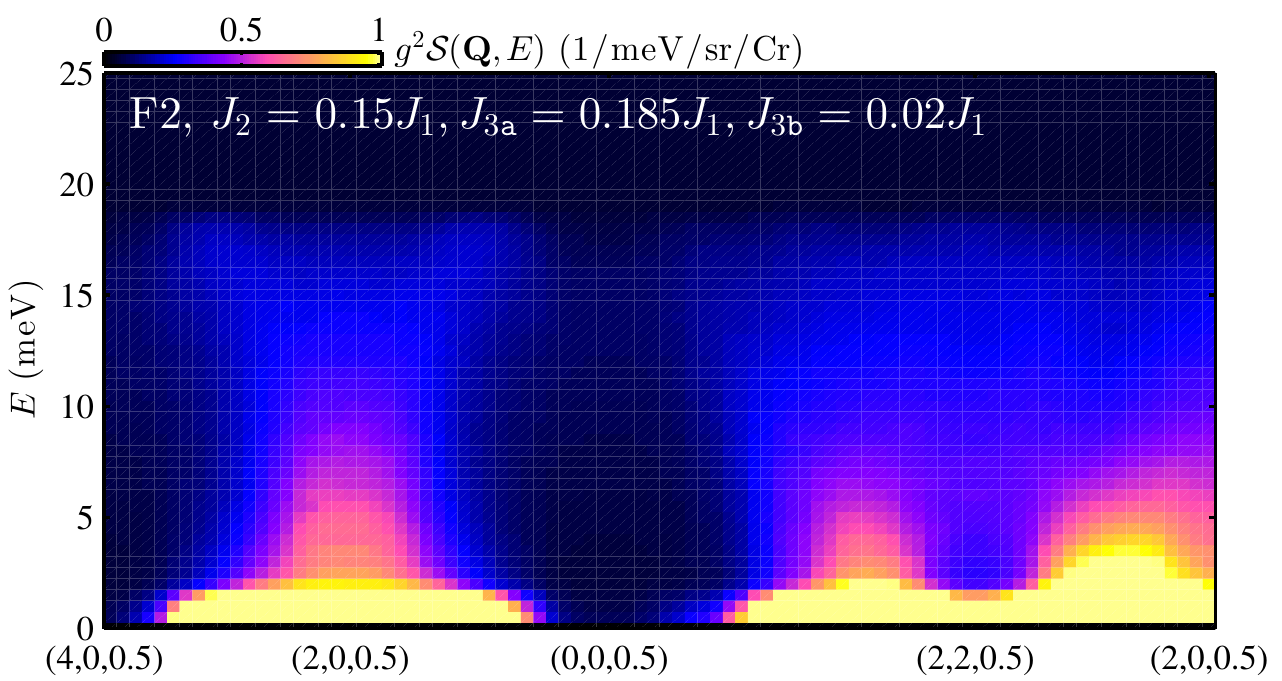}\\
\includegraphics[width=0.41\columnwidth]{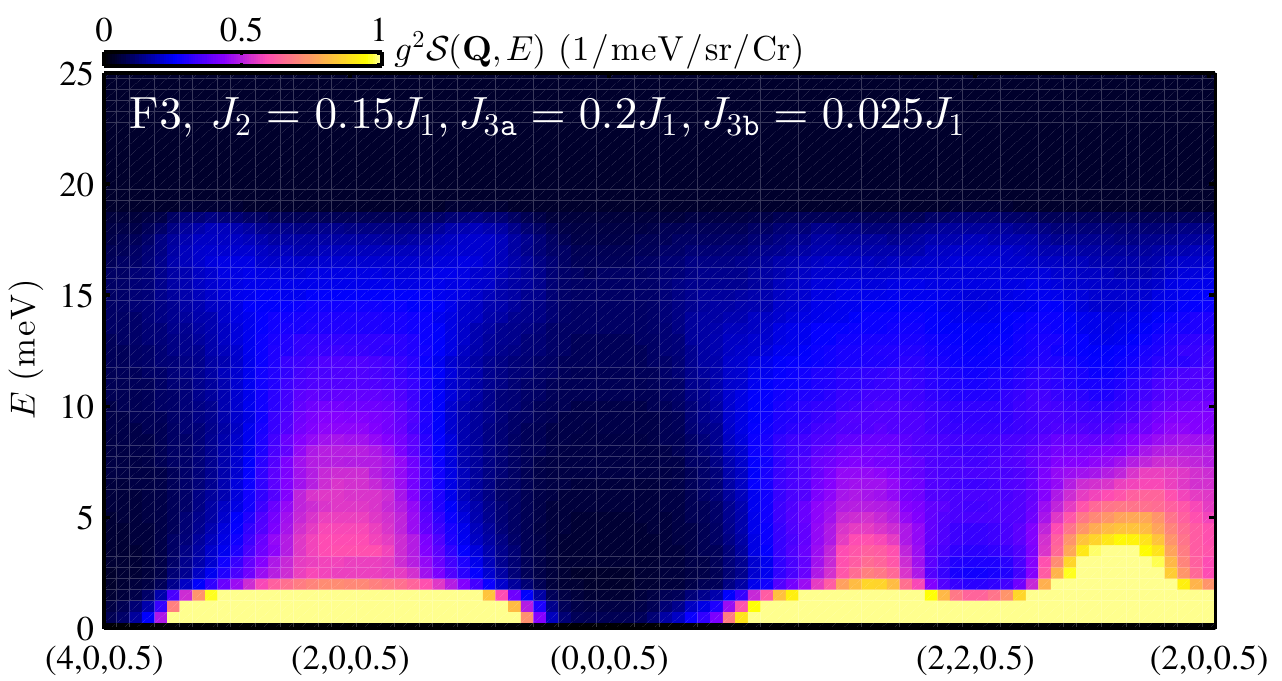}
\includegraphics[width=0.41\columnwidth]{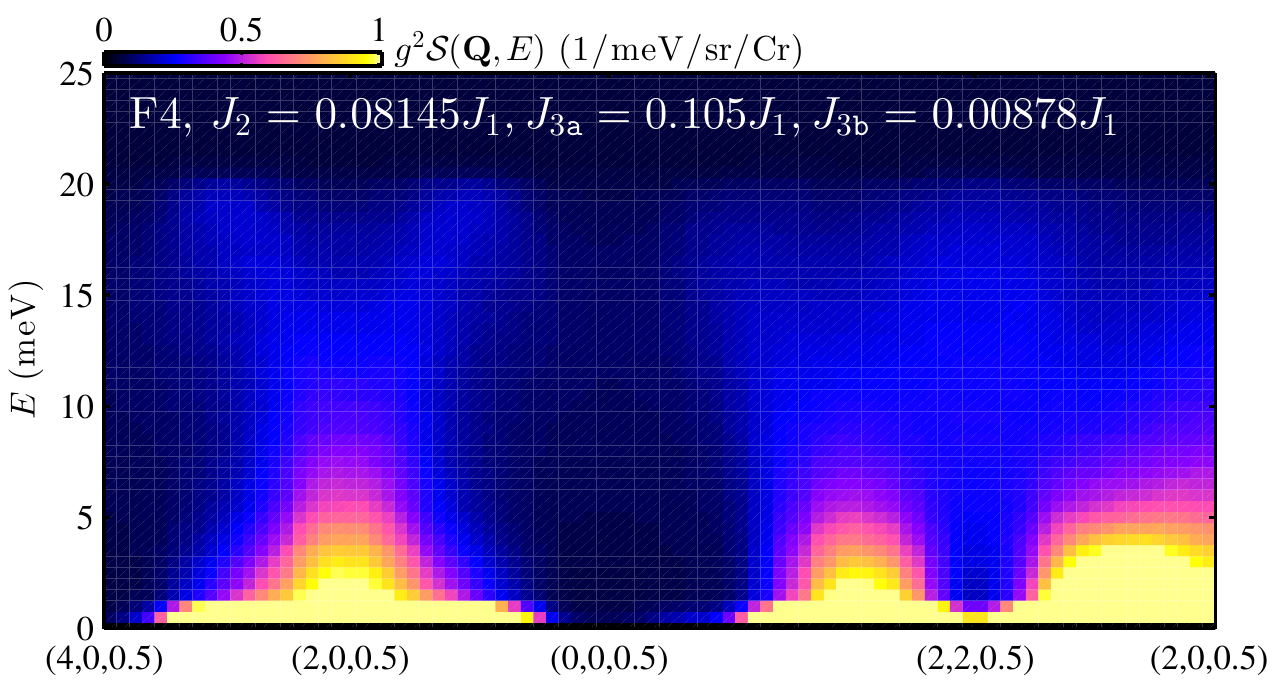}
\caption{{Calculated inelastic spectra along the path $(4,0,0.5)\rightarrow(2,0,0.5)\rightarrow(0,0,0.5)\rightarrow(2,2,0.5)\rightarrow(2,0,0.5)$.}}
\label{SI_Inelastic2}
\end{figure} 
\clearpage

\begin{figure}[h]
\includegraphics[width=0.41\columnwidth]{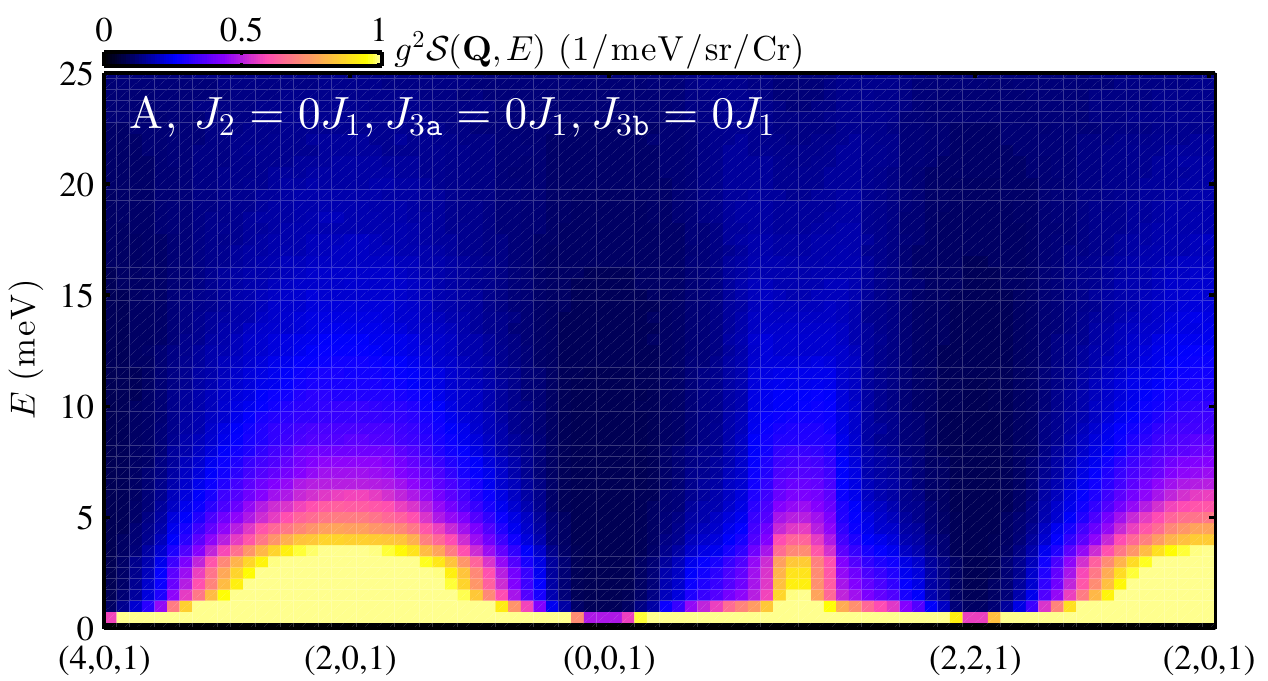}
\includegraphics[width=0.41\columnwidth]{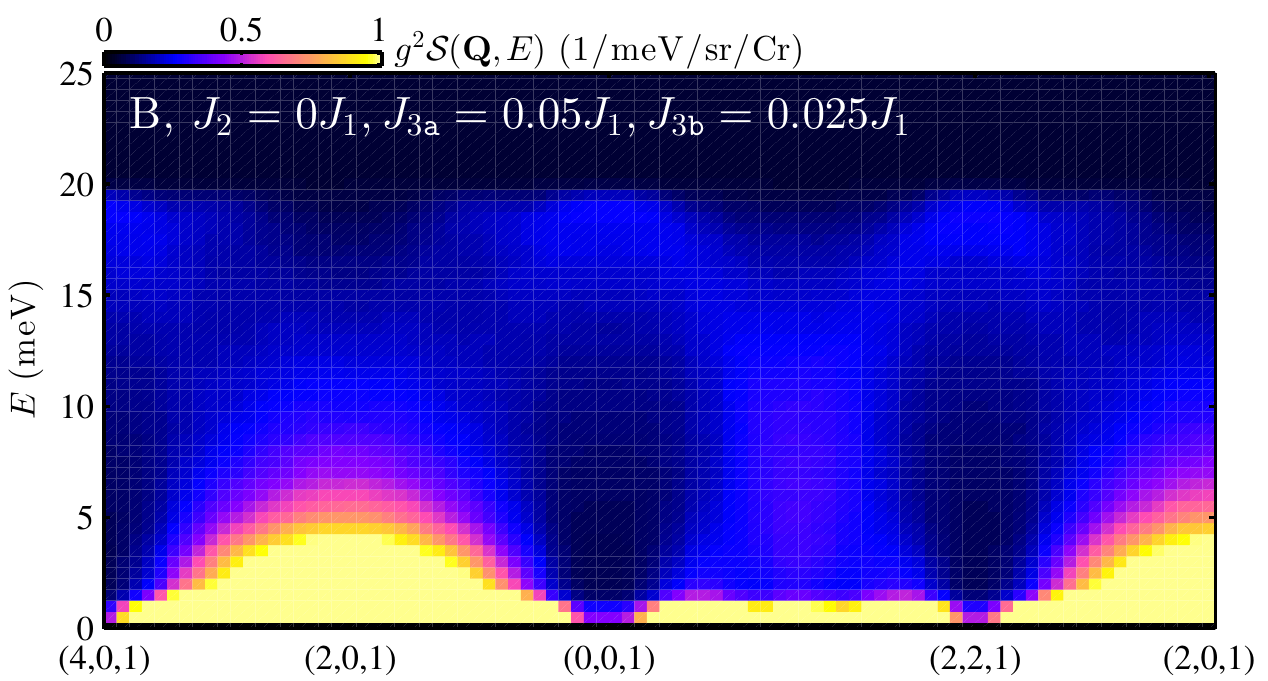}\\
\includegraphics[width=0.41\columnwidth]{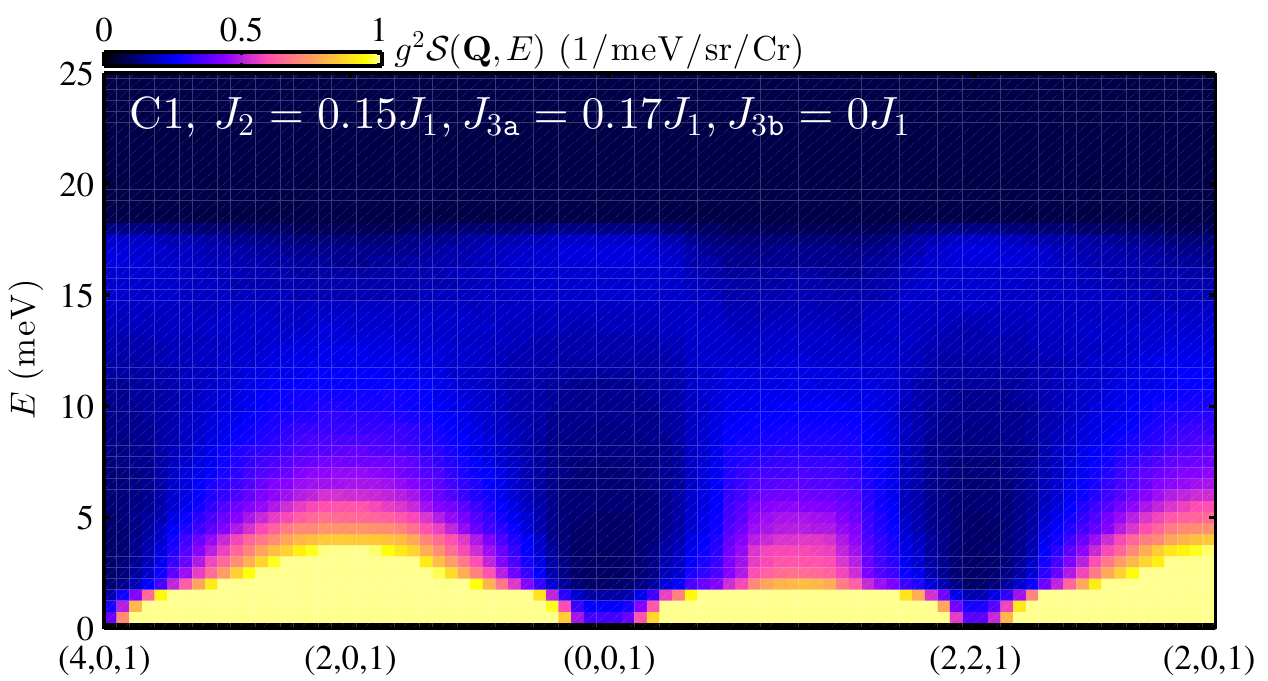}
\includegraphics[width=0.41\columnwidth]{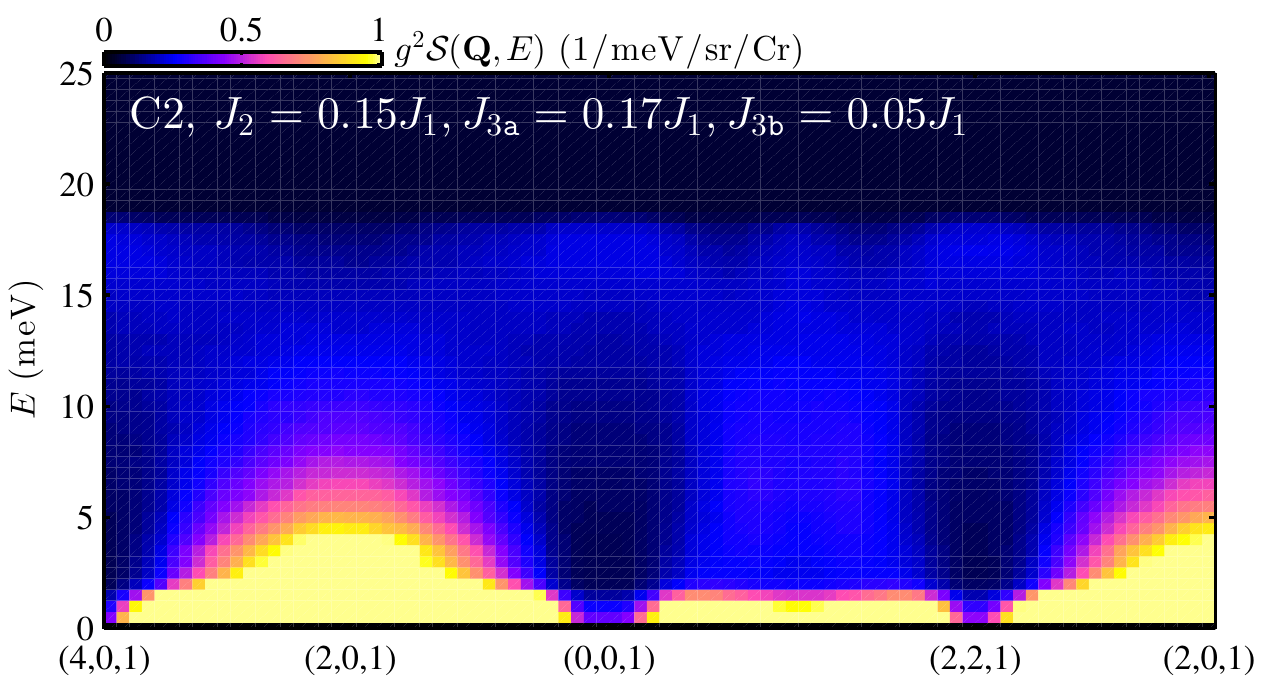}\\
\includegraphics[width=0.41\columnwidth]{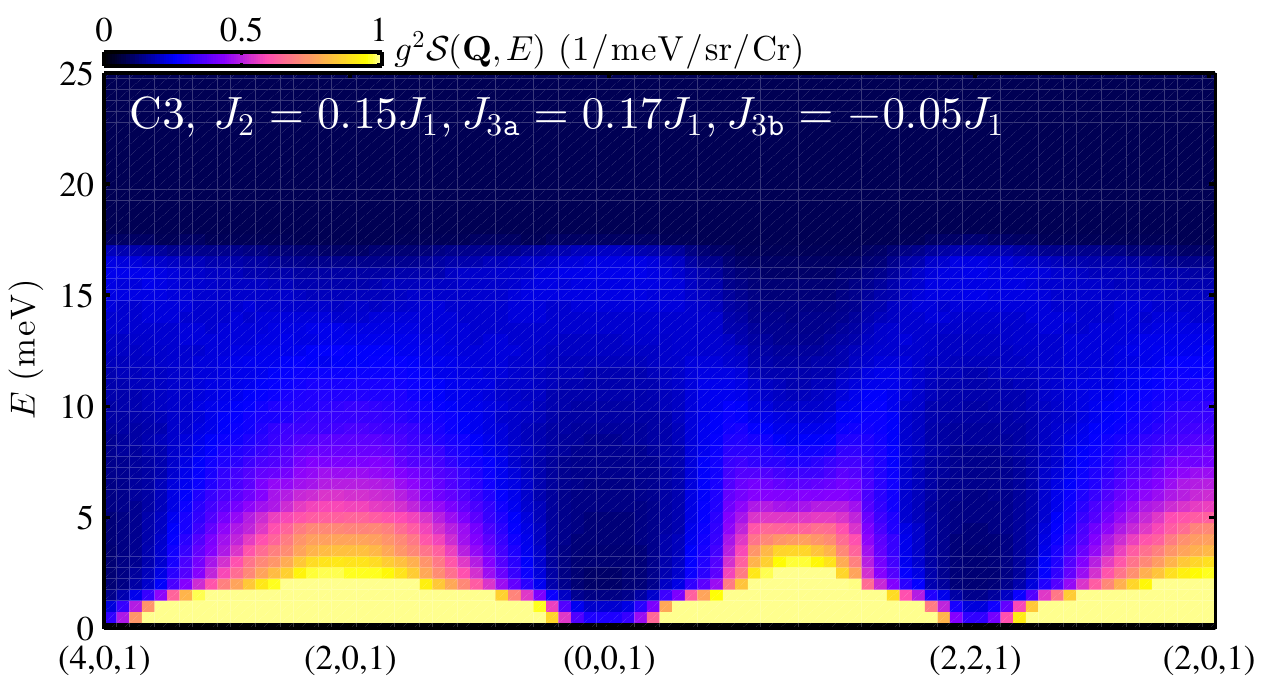}
\includegraphics[width=0.41\columnwidth]{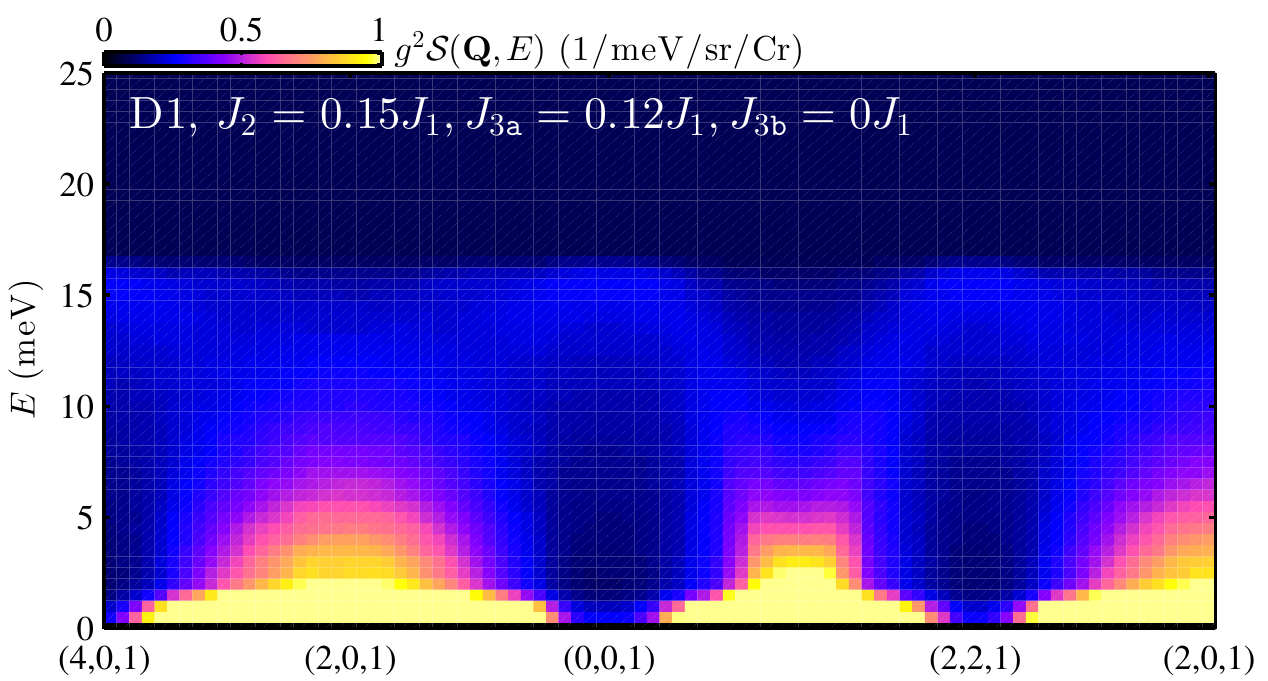}\\
\includegraphics[width=0.41\columnwidth]{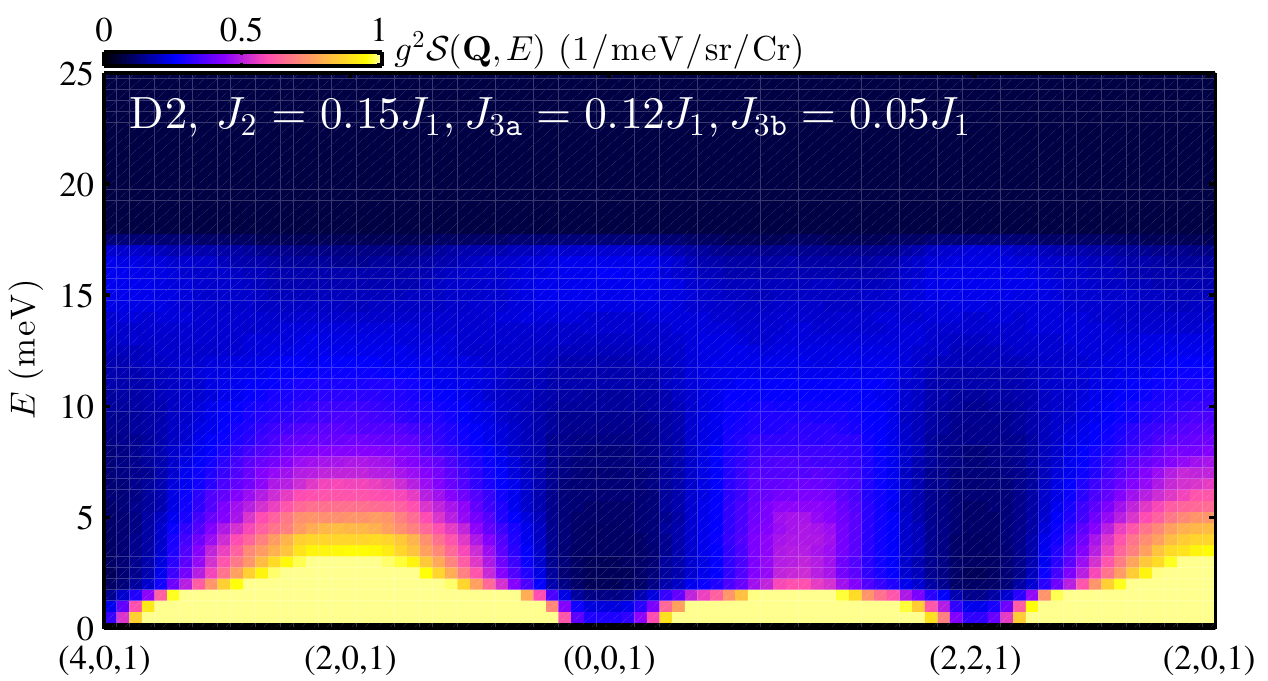}
\includegraphics[width=0.41\columnwidth]{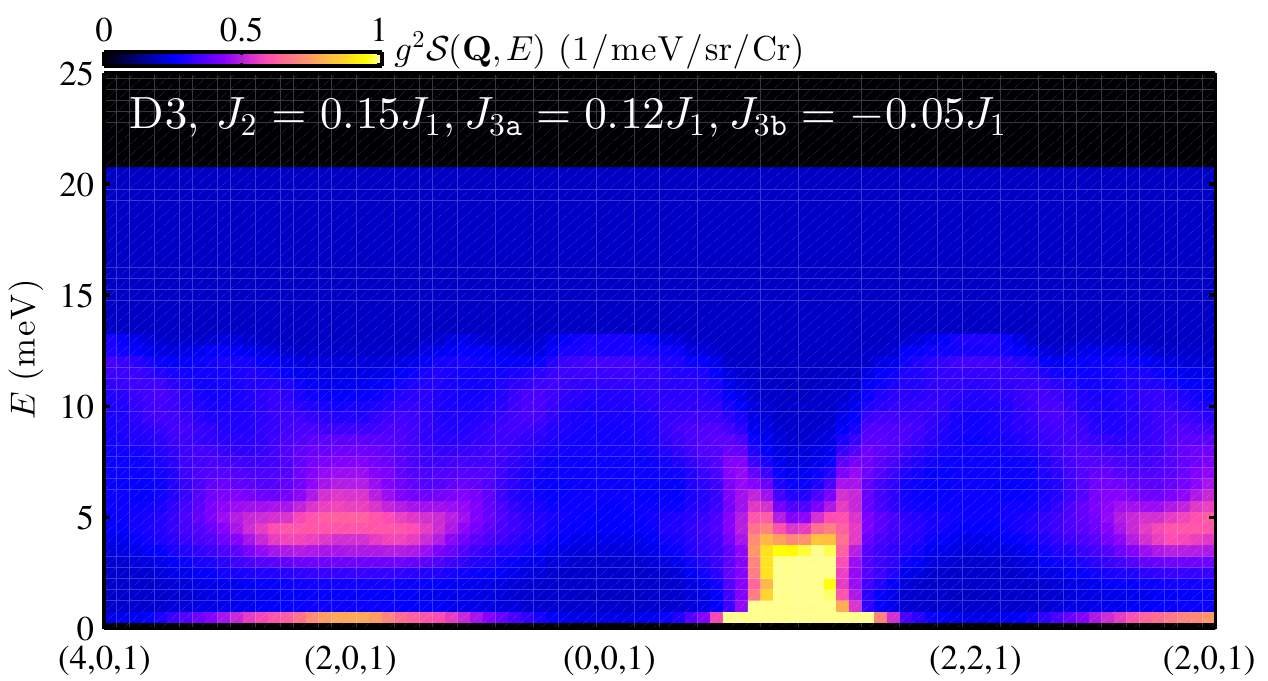}\\
\includegraphics[width=0.41\columnwidth]{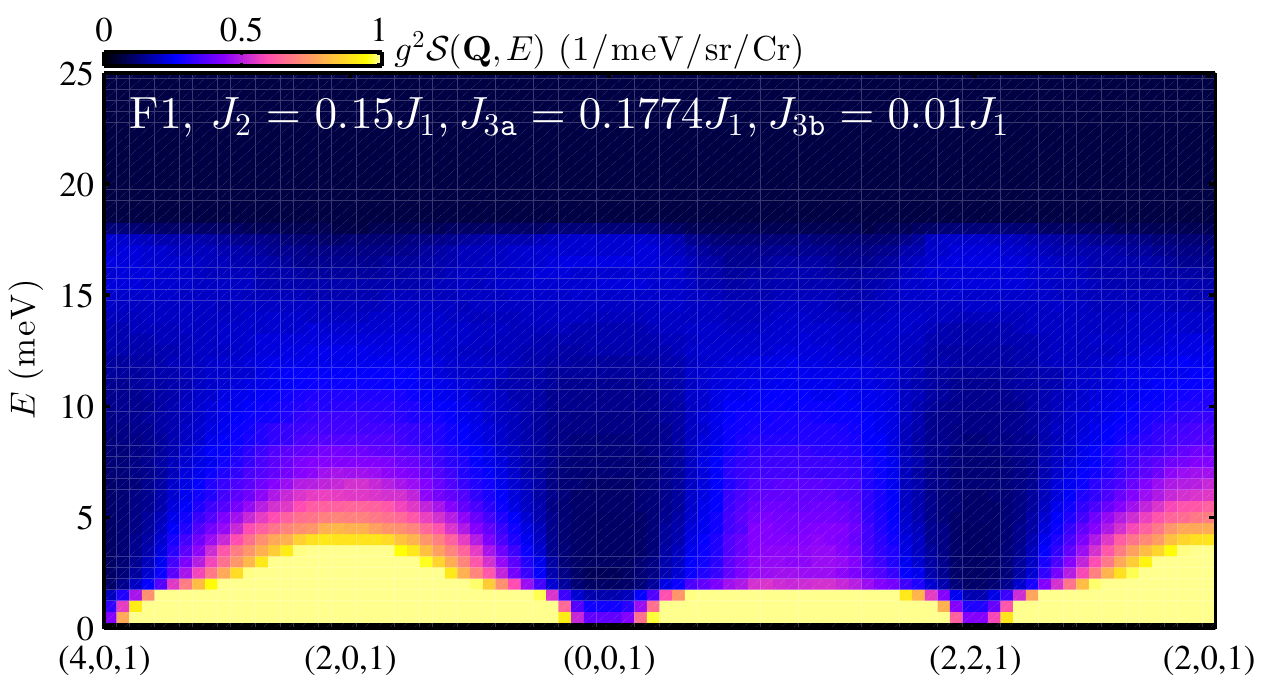}
\includegraphics[width=0.41\columnwidth]{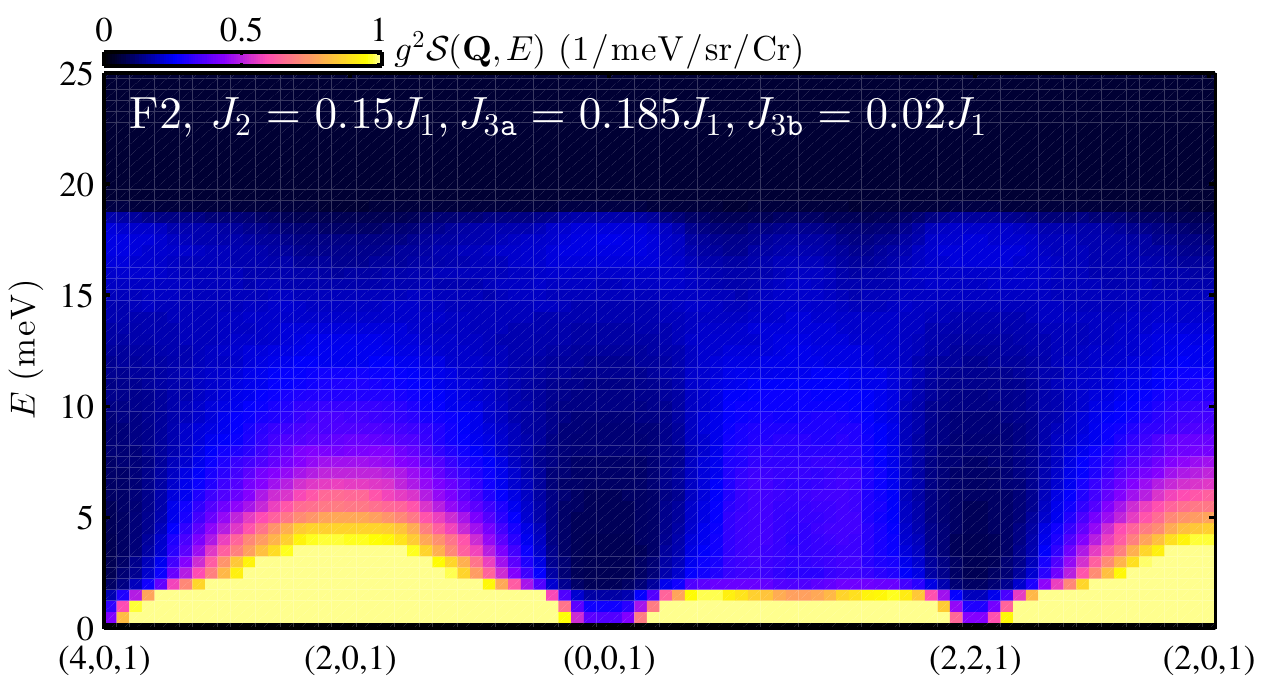}\\
\includegraphics[width=0.41\columnwidth]{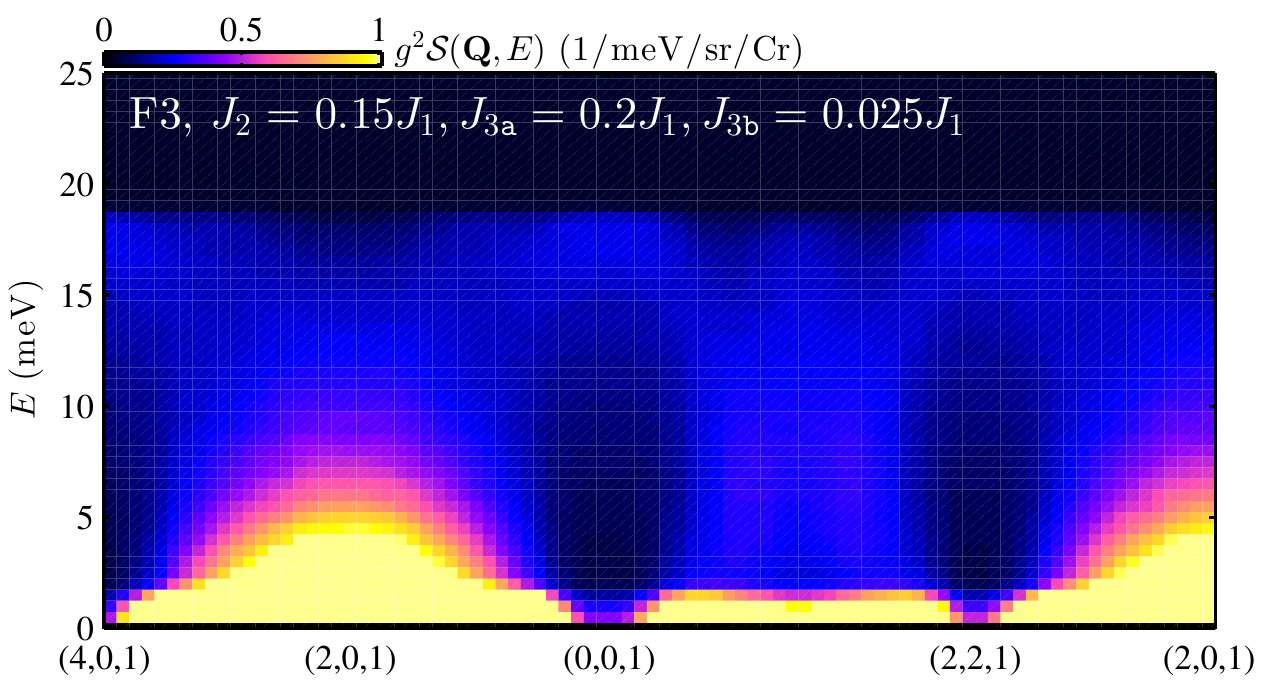}
\includegraphics[width=0.41\columnwidth]{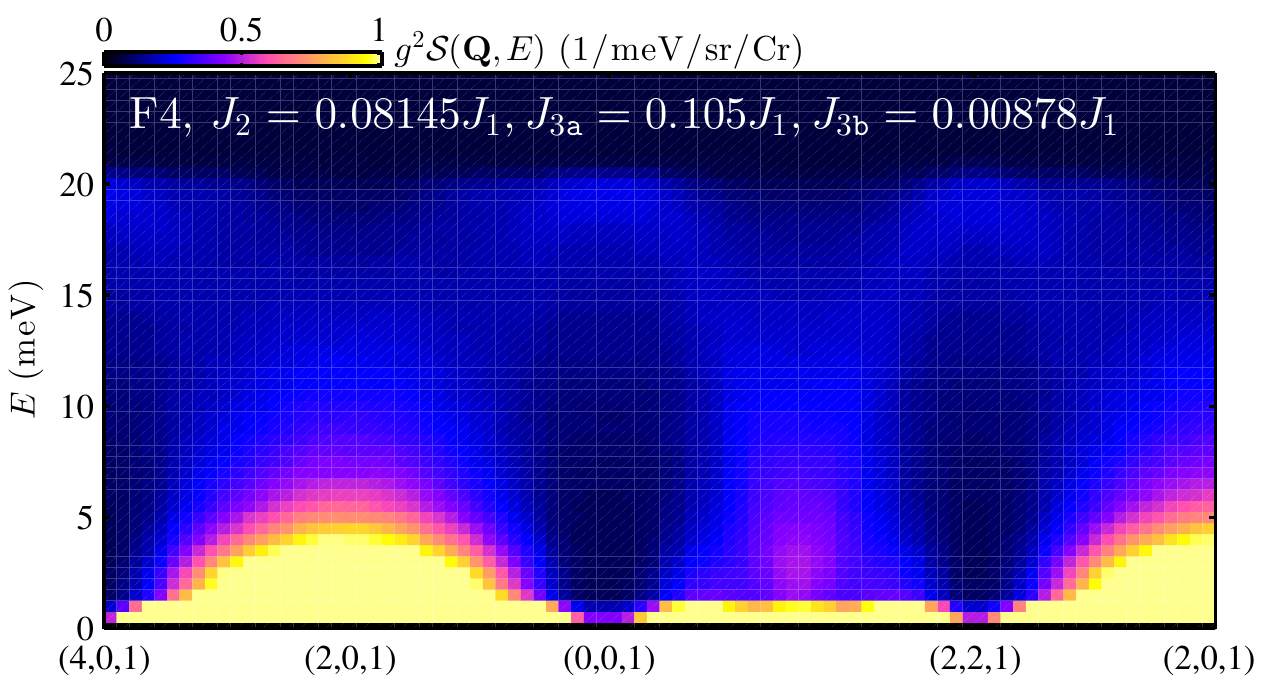}
\caption{{Calculated inelastic spectra along the path $(4,0,1)\rightarrow(2,0,1)\rightarrow(0,0,1)\rightarrow(2,2,1)\rightarrow(2,0,1)$.}}
\label{SI_Inelastic3}
\end{figure} 

\begin{figure}[h]
\includegraphics[width=0.41\columnwidth]{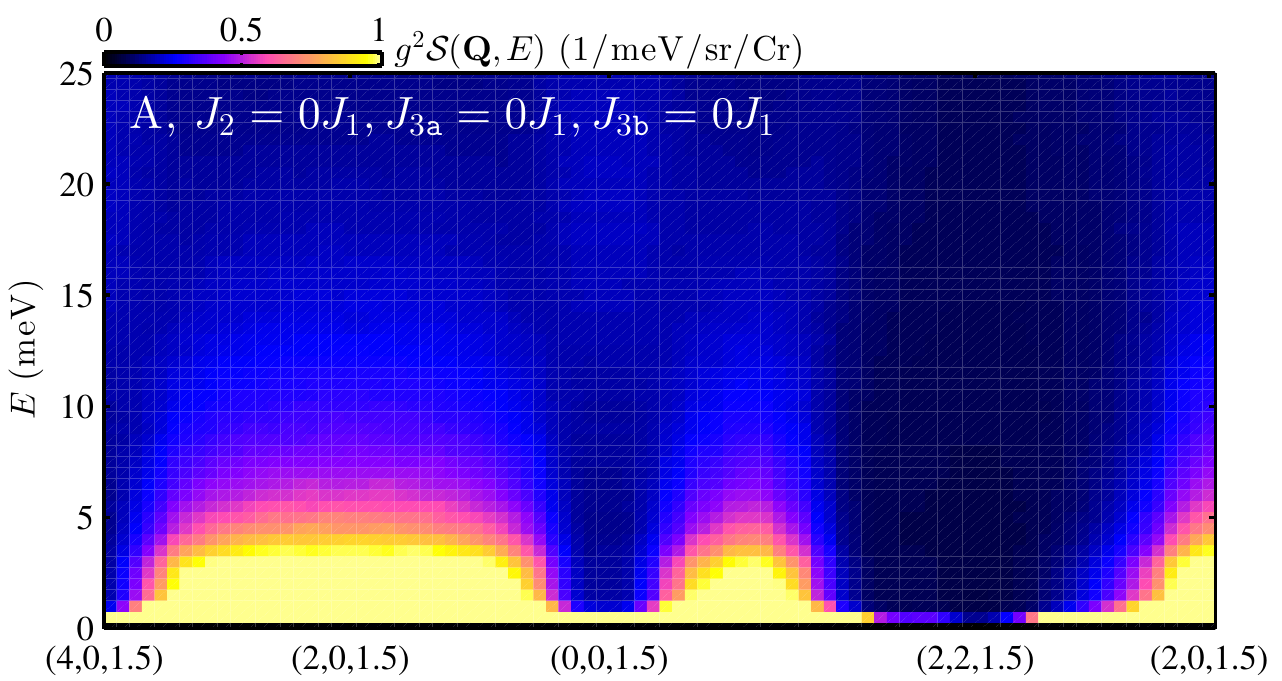}
\includegraphics[width=0.41\columnwidth]{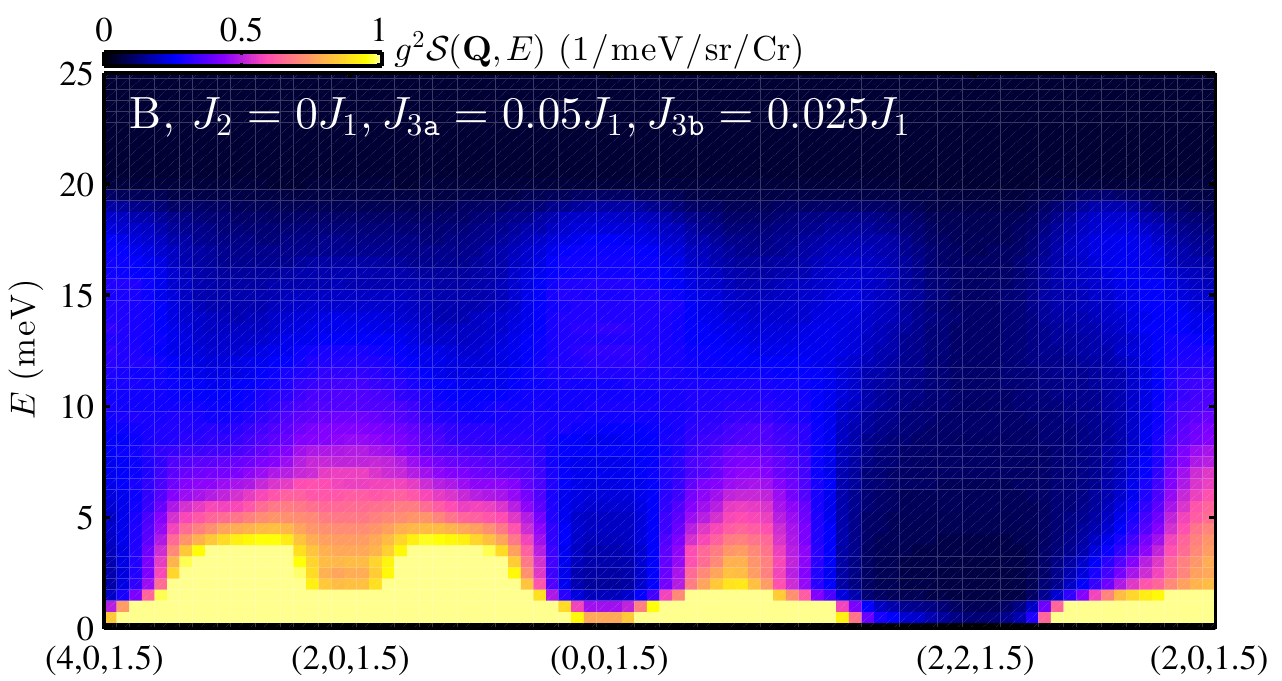}\\
\includegraphics[width=0.41\columnwidth]{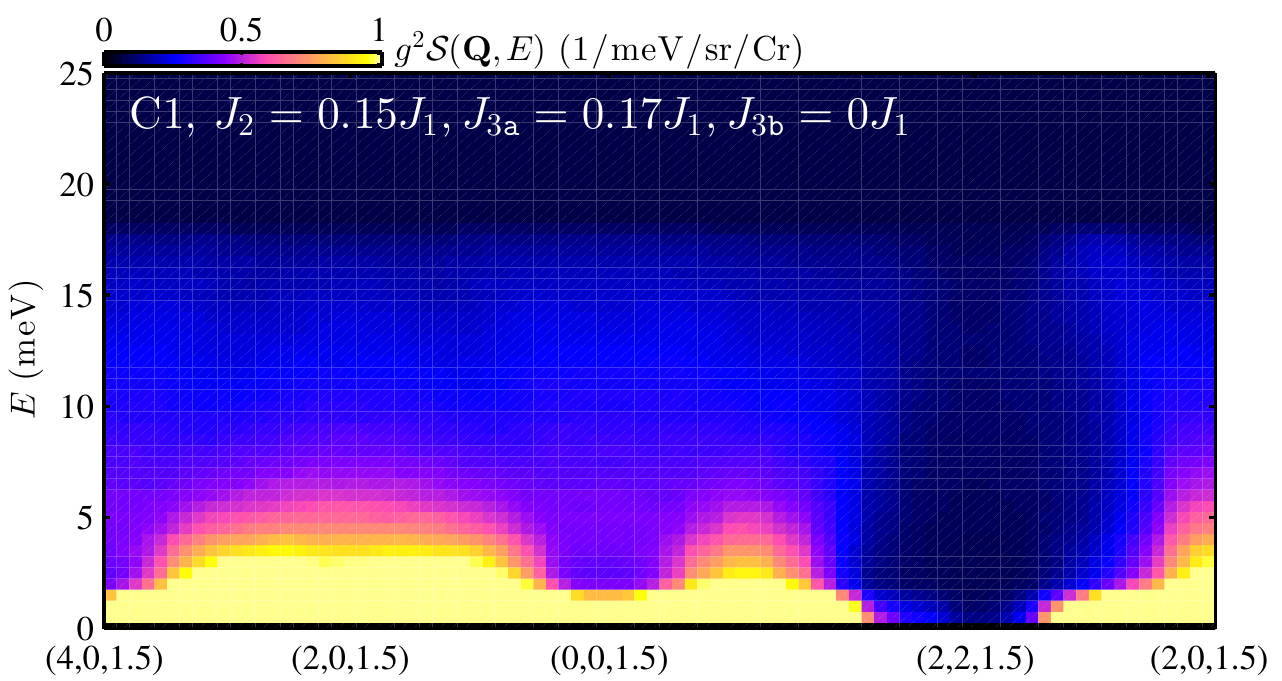}
\includegraphics[width=0.41\columnwidth]{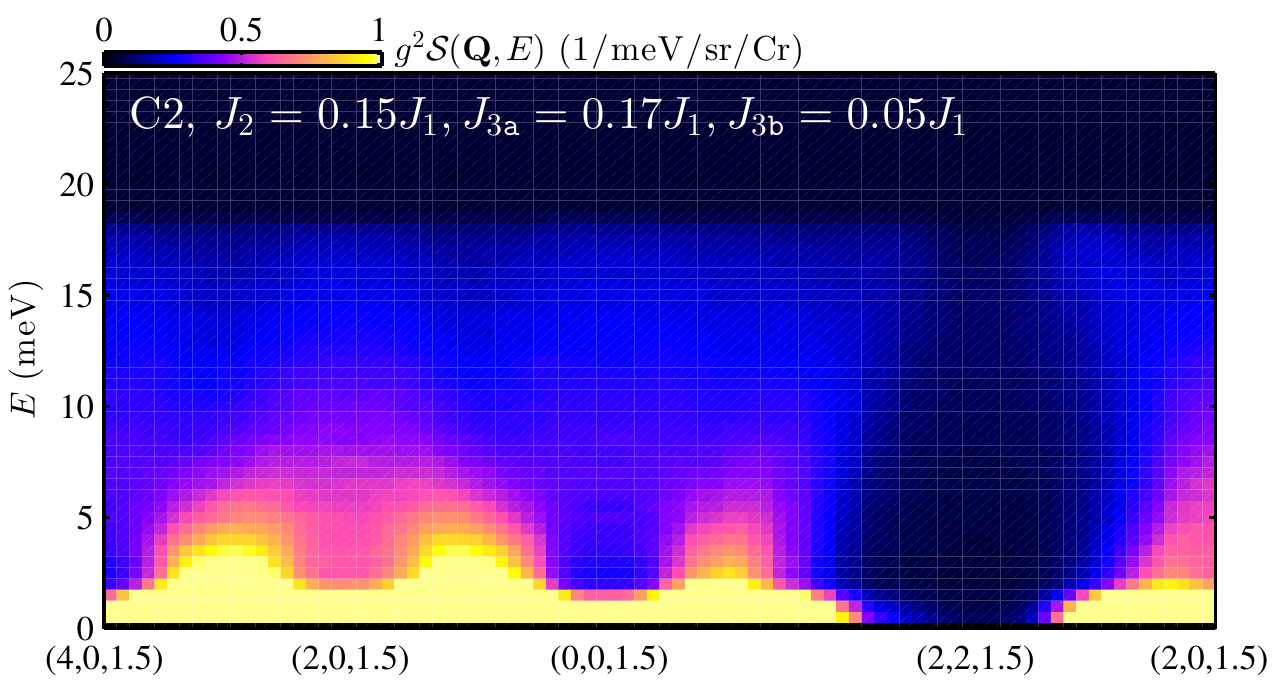}\\
\includegraphics[width=0.41\columnwidth]{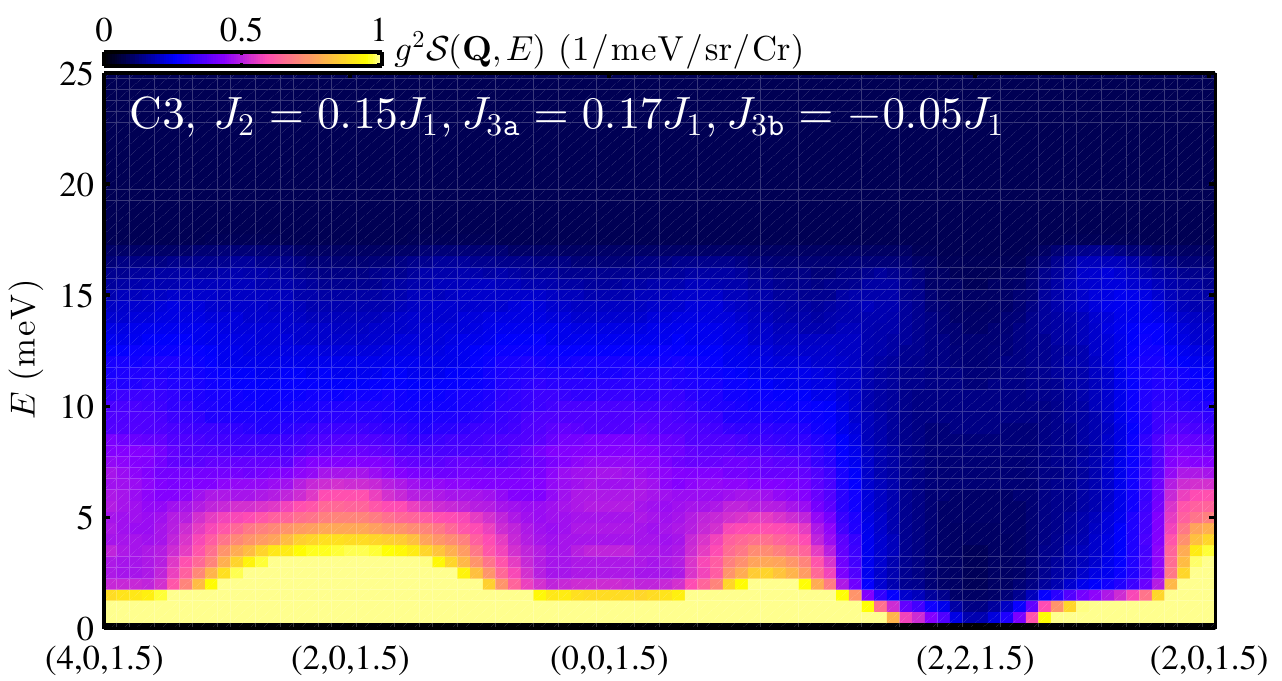}
\includegraphics[width=0.41\columnwidth]{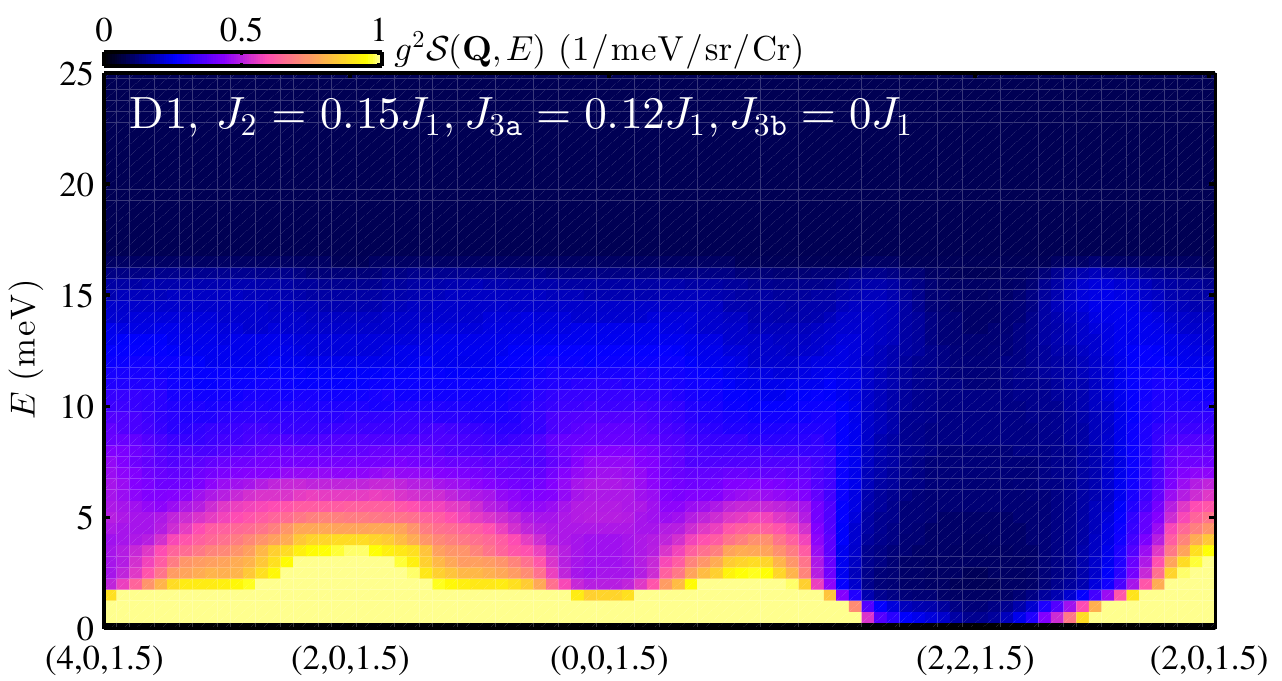}\\
\includegraphics[width=0.41\columnwidth]{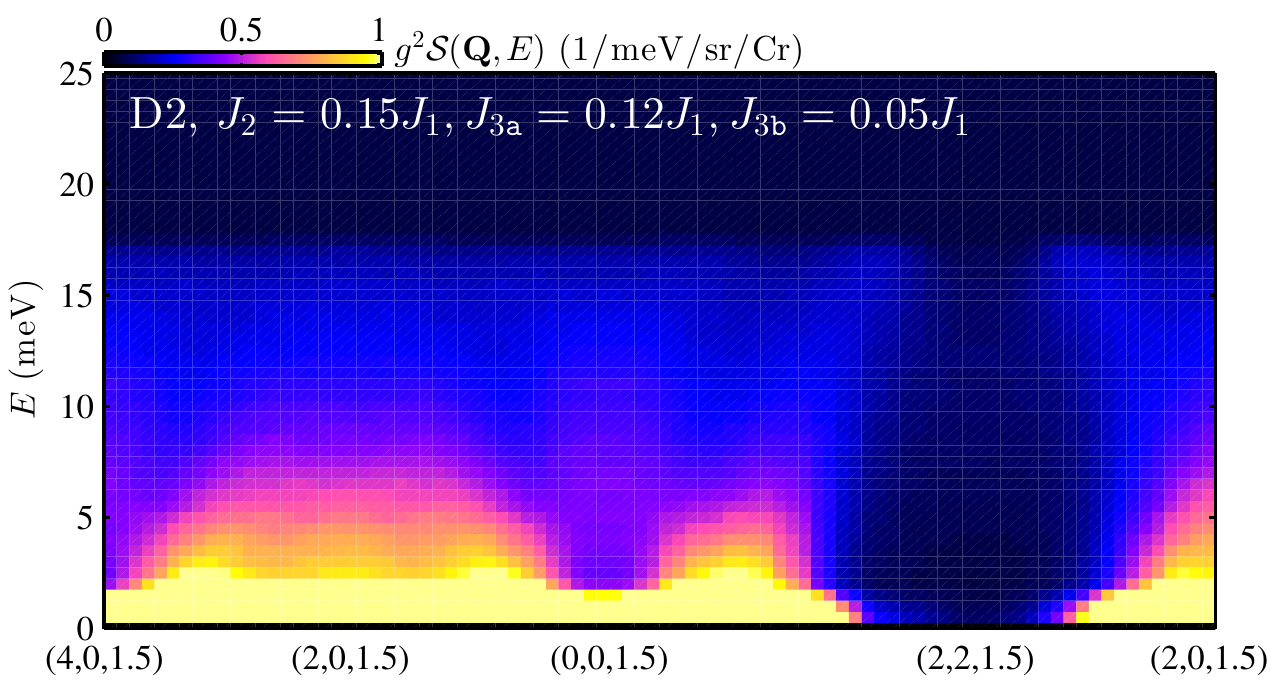}
\includegraphics[width=0.41\columnwidth]{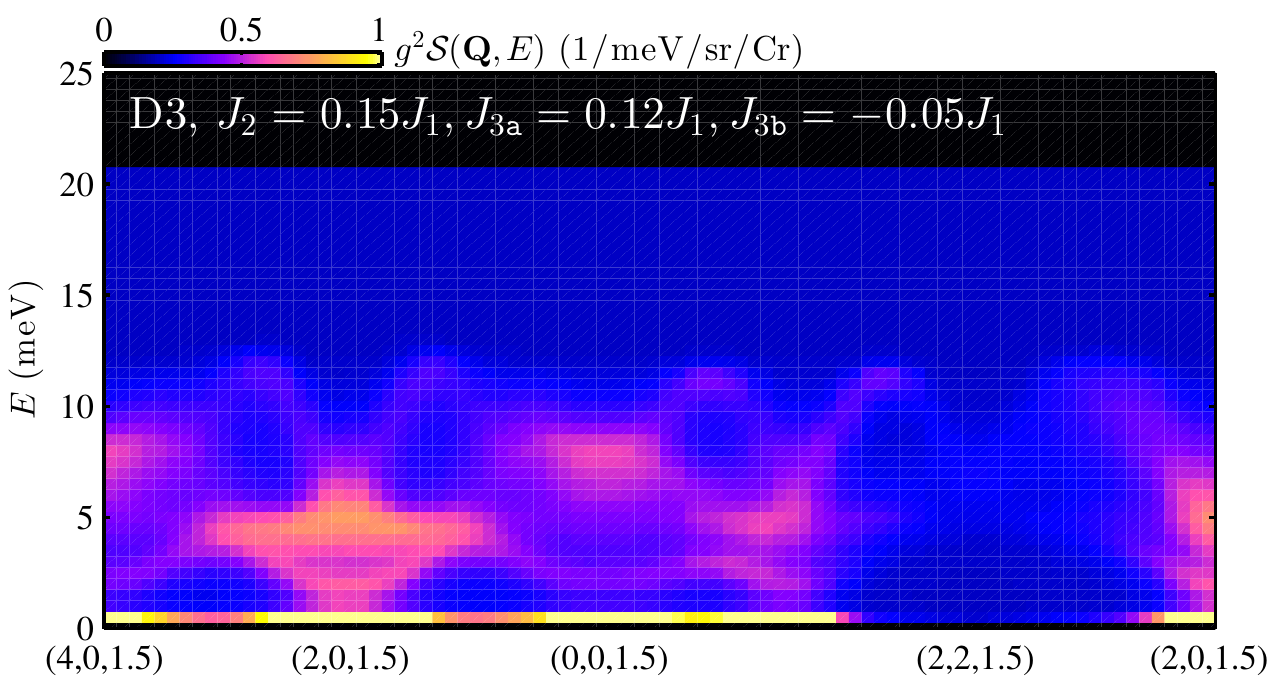}\\
\includegraphics[width=0.41\columnwidth]{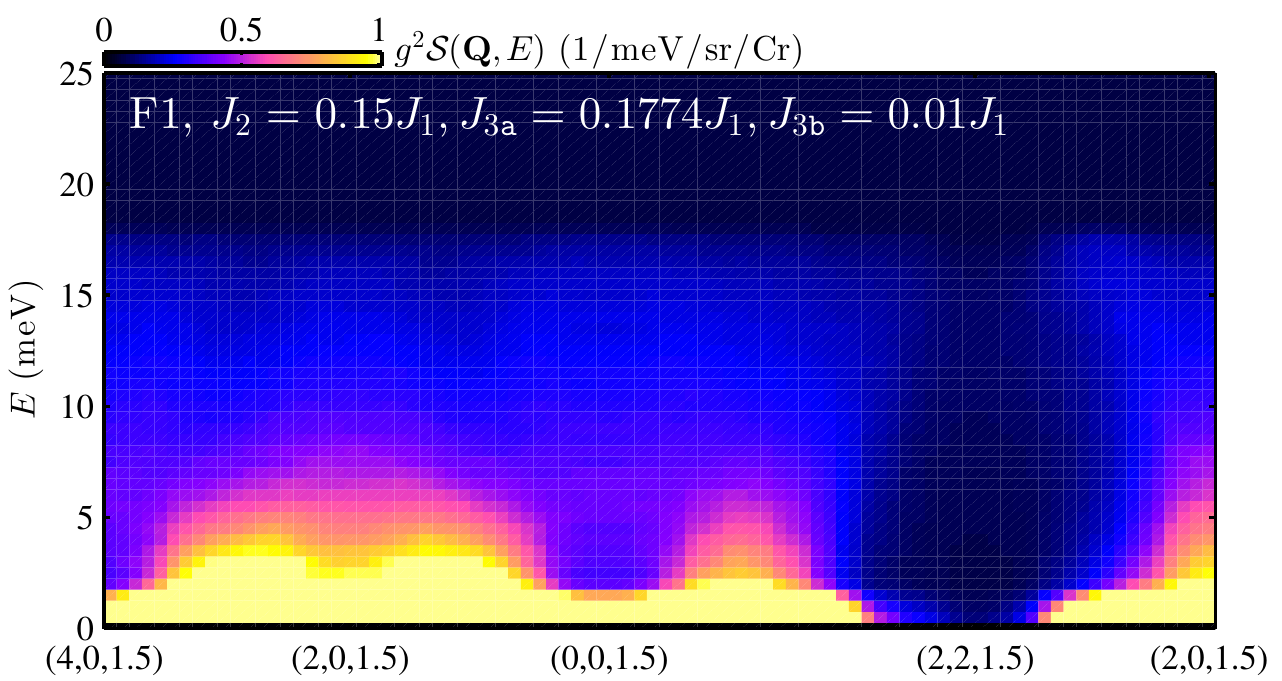}
\includegraphics[width=0.41\columnwidth]{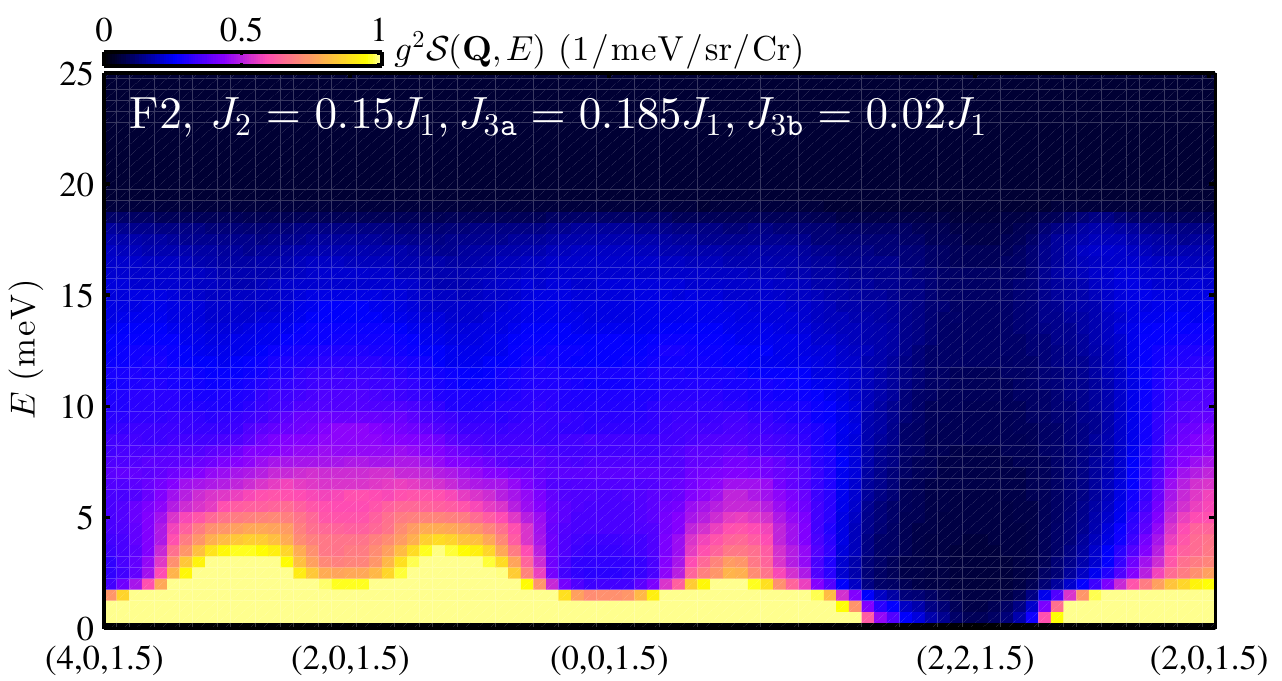}\\
\includegraphics[width=0.41\columnwidth]{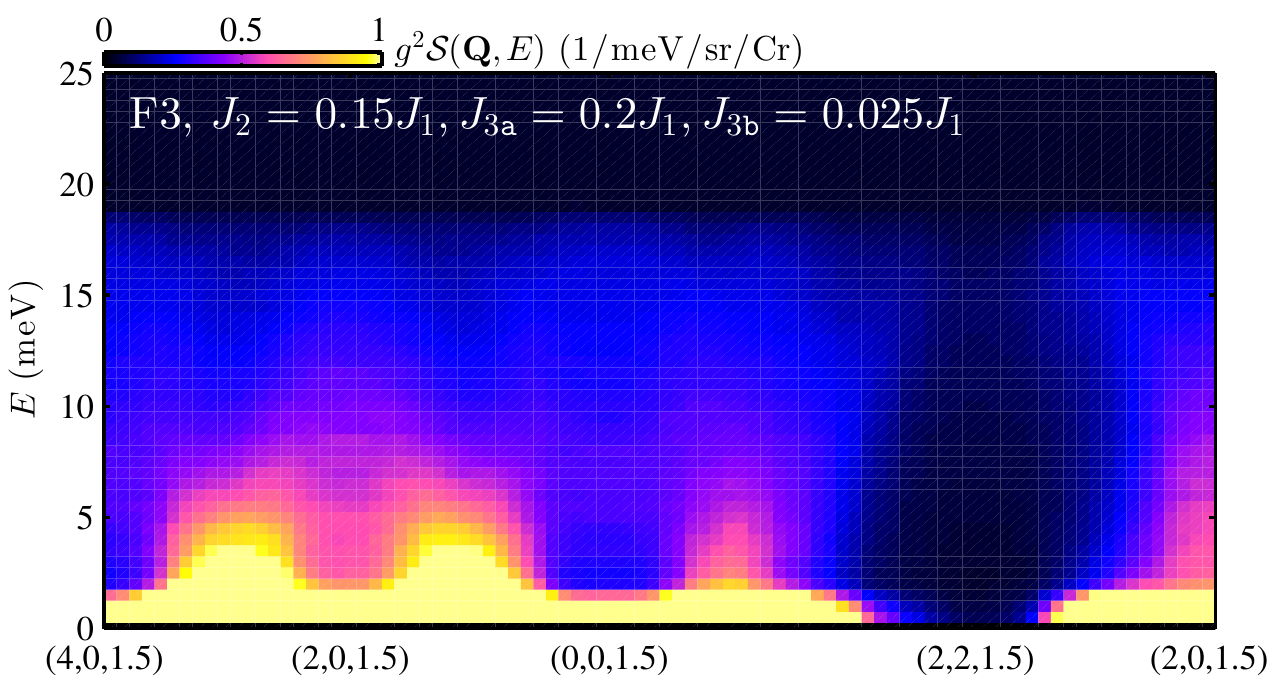}
\includegraphics[width=0.41\columnwidth]{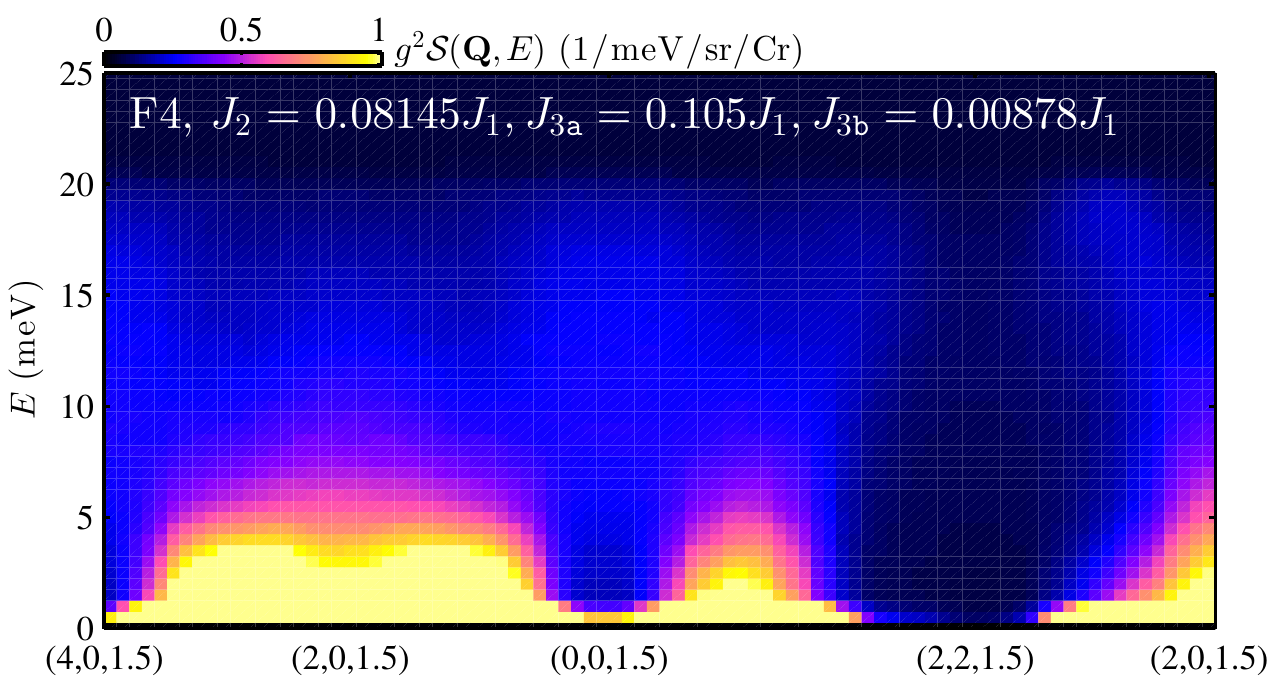}
\caption{{Calculated inelastic spectra along the path $(4,0,1.5)\rightarrow(2,0,1.5)\rightarrow(0,0,1.5)\rightarrow(2,2,1.5)\rightarrow(2,0,1.5)$.}}
\label{SI_Inelastic4}
\end{figure} 
\clearpage

\newpage
\section{S8. Mean-field phase diagram for FN exchange interactions}

The Hamiltonian is 
\begin{align}
\mathcal{H} = \dfrac{1}{2}\sum_{mn}\sum_{\mu\nu} \sum_i\mathcal{J}^{(i)}_{\mu\nu}({\bf R}_{m}-{\bf R}_{n}){\bf S}_{m\mu}\cdot{\bf S}_{n\nu}
\end{align}
where $\mathcal{J}^{(i)}_{\mu\nu}({\bf R}_{m}-{\bf R}_{n})$ is the exchange matrix with $i = 1,2,3\text{a}$ and $3\text{b}$ indexing the level of interactions and ${\bf S}_{m\nu}$ is the spin at the unit cell $m$ and sublattice $\nu$. 
The Fourier transform of the exchange matrix is given by
\begin{align}
\mathcal{J}_{\mu\nu}({\bf Q}) = \dfrac{1}{N}\sum_i\sum_{{\bf R}_{m}-{\bf R}_{n}}\mathcal{J}^{(i)}_{\mu\nu}({\bf R}_{m}-{\bf R}_{n})\,e^{\text{i} {\bf Q}\cdot ({\bf R}_{m\mu}-{\bf R}_{n\nu})}\,,
\end{align}
where ${\bf R}_{m\mu}={\bf R}_{m}+{\bf c}_{\mu}$, ${\bf c}_1 =(0,0,0)$, ${\bf c}_2 =(0,1/4,1/4)$, ${\bf c}_3 =(1/4,0,1/4)$ and ${\bf c}_4 =(1/4,1/4,0)$.
The explicit formula can be found in \cite{Conlon2010a}. 

For ${\bf Q}=2\pi(h,1,0)$, the interaction matrix  reduces to
\begin{align}
{ \bf J} =
\left(
\begin{array}{cc}
 -2 (J_{3\text{b}}+J_{3\text{a}}){\bf I}_{2\times2} & 2 (J_{1}-2 J_{2}){\bf \Lambda} \\
 2 (J_{1}-2 J_{2}){\bf \Lambda}^{\text{T}} & -2 (J_{3\text{a}}+J_{3\text{b}}){\bf I}_{2\times2} \\
\end{array}
\right)\,\quad {\bf \Lambda} = \left(
\begin{array}{cc}
\cos \left(\frac{h \pi }{2}\right) & -\sin \left(\frac{h \pi }{2}\right) \\
 \sin \left(\frac{h \pi }{2}\right) & \cos \left(\frac{h \pi }{2}\right) \\
\end{array}
\right)
\end{align}
which can be diagonalized by a unitary matrix
\begin{align}
{\bf U}=\dfrac{1}{2}\left(
\begin{array}{cccc}
 \text{i} e^{\frac{\text{i} h \pi }{2}} & -\text{i} e^{-\frac{\text{i} h \pi }{2} } & -\text{i} e^{\frac{\text{i} h \pi }{2}} & \text{i} e^{-\frac{\text{i} h \pi }{2}}  \\
 e^{\frac{\text{i} h \pi }{2}} & e^{-\frac{\text{i} h \pi }{2} } & -e^{\frac{\text{i} h \pi }{2}} & -e^{-\frac{\text{i} h \pi }{2} } \\
 \text{i} & -\text{i} & \text{i} & -\text{i} \\
 1 & 1 & 1 & 1 \\
\end{array}
\right)
\end{align}
\begin{align}
{\bf U}\cdot {\bf J}\cdot {\bf U}^{\dagger}=
\left(
\begin{array}{cccc}
 \epsilon_+ & 0 & 0 & 0 \\
 0 & \epsilon_+ & 0 & 0 \\
 0 & 0 & \epsilon_- & 0 \\
 0 & 0 & 0 & \epsilon_- \\
\end{array}
\right)\,,\quad
\epsilon_\pm = \pm2 (J_1-2 J_2)-2(J_{3\text{a}}+J_{3\text{b}}).
\end{align}
The spin configuration for minimal eigenvalue $\epsilon_-$ is 
\begin{align}
{\bf S}_{n\nu}=\hat{\bf x} \sin \left({{\bf Q}}\cdot {\bf R}_{n\nu}+\phi _{\nu}\right)+\hat{\bf y} \cos \left({\bf Q}\cdot {\bf R}_{n\nu}+\phi _{\nu}\right)
\end{align}
where 
\begin{align}
\phi_1 = \dfrac{h\pi}{2}-\dfrac{\pi}{2},\quad
\phi_2 = \dfrac{h\pi}{2}+\pi,\quad
\phi_3 = \dfrac{\pi}{2},\quad
\phi_4 = 0\,.
\end{align}

\end{document}